\documentclass{aa}  

\usepackage{graphicx}
\usepackage{natbib}
\usepackage{multirow}
\usepackage[super]{nth}
\usepackage{listings}
\usepackage{enumitem}
\usepackage{soul}
\usepackage{color}
\sethlcolor{cyan}
\usepackage{cprotect}
\usepackage{caption}
\usepackage{subcaption}
\usepackage{float}
\usepackage{txfonts}
\usepackage[export]{adjustbox}
\usepackage{hyperref}
\usepackage{xcolor}
\hypersetup{
    colorlinks,
    citecolor=blue,
    linkcolor=magenta,
    urlcolor=cyan}

\begin{document}

   \title{The Pristine survey XXIV: The Galactic underdogs}
   
   \subtitle{Dynamic tales of a Milky Way metal-poor population}

   \author{Isaure Gonz\'alez Rivera de La Vernhe\inst{\ref{inst1}}\thanks{\email{isaure.gonzalez-rivera@oca.eu}} \and Vanessa Hill\inst{\ref{inst1}} \and Georges Kordopatis\inst{\ref{inst1}} \and Felipe Gran\inst{\ref{inst1}} \and Emma Fern\'andez-Alvar\inst{\ref{inst2},\ref{inst3},\ref{inst1}} \and Anke Ardern-Arentsen\inst{\ref{inst4}} \and Guillaume F. Thomas\inst{\ref{inst3},\ref{inst2}} \and Federico Sestito\inst{\ref{inst5}} \and Camila Navarrete\inst{\ref{inst1}} \and Nicolas F. Martin\inst{\ref{inst6},\ref{inst7}} \and Else Starkenburg\inst{\ref{inst8}} \and Akshara Viswanathan\inst{\ref{inst8}} \and Giuseppina Battaglia\inst{\ref{inst3},\ref{inst2}} \and Kim A. Venn\inst{\ref{inst5}} \and Sara Vitali\inst{\ref{inst9},\ref{inst10}}}
   
   \institute{Université Côte d'Azur, Observatoire de la Côte d'Azur, CNRS, Laboratoire Lagrange, 06304 Nice, France
        \label{inst1}
    \and 
    Departamento de Astrofísica, Universidad de La Laguna, E-38206 La Laguna, Tenerife, Spain
        \label{inst2}
    \and 
    Instituto de Astrofísica de Canarias, E-38200 La Laguna, Tenerife, Spain
        \label{inst3}
    \and
    Institute of Astronomy, University of Cambridge, Madingley Road, Cambridge CB3 0HA, UK
        \label{inst4}
    \and
    Department of Physics and Astronomy, University of Victoria, PO Box 3055, STN CSC, Victoria BC V8W 3P6, Canada
        \label{inst5}
    \and
    Université de Strasbourg, CNRS, Observatoire astronomique de Strasbourg, UMR 7550, F-67000 Strasbourg, France
        \label{inst6}
    \and
    Max-Planck-Institut für Astronomie, Königstuhl 17, D-69117 Heidelberg, Germany
        \label{inst7}
    \and
    Kapteyn Astronomical Institute, University of Groningen, Landleven 12, 9747 AD Groningen, The Netherlands
        \label{inst8}
    \and
    Instituto de Estudios Astrofísicos, Facultad de Ingeniería y Ciencias, Universidad Diego Portales, Av. Ejército Libertador 441, Santiago, Chile
        \label{inst9}
    \and
    Millenium Nucleus ERIS
        \label{inst10}}

    \date{Received April 25, 2024; Accepted June 24, 2024}

 
  \abstract
   {Metal-poor stars hold key information on the early Milky Way.
   Through the identification and characterisation of substructures, one can understand internal mechanisms (including merger and accretion events), which are indispensable to reconstruct the formation history of the Galaxy.}
   {To allow an investigation of a population of very metal-poor stars ([Fe/H] < -1.7) with disc-like orbits (planar and prograde), high angular momenta ($L_z$/$J_{tot}$ > 0.5) and rotational velocities ($V_\phi$ > 180 km.s$^{-1}$) proposed in the literature, we used a sample of \\ $\sim$ 3M giant stars with \textit{Gaia} DR3 BP/RP information and \textit{Pristine-Gaia} metallicities down to -4.0 dex that we aimed to decontaminate. To achieve this, we constructed a sample as free as possible from spurious photometric estimates, an issue commonly encountered for high $V_\phi$ metal-poor stars.}
   {We created a statistically robust sample of $\sim$ 36 000 \textit{Pristine-Gaia} very metal-poor ([Fe/H] < -1.7) giant stars, using APOGEE and LAMOST data (adding GALAH and GSP-spec for verification) to estimate and remove contamination. We investigated the spatial and kinematic properties of the decontaminated sample, making use of $V_\phi$ as well as the action space, which are both powerful tools to disentangle stellar populations.}
   {The global distribution of very metal-poor stars in our sample shows the typical kinematics, orbital properties, and spatial distributions of a halo; however, as in previous works, we found a pronounced asymmetry in the $L_z$ and $V_\phi$ distributions, in favour of prograde stars. We showed that this excess is predominantly due to prograde-planar stars (10 $\%$ of the very metal-poor population), which can be detected down to [Fe/H] = -2.9 at a 2$\sigma$ confidence level. This prograde-planar population contains stars with $V_\phi$ > 180 km.s$^{-1}$ and $Z_{\text{max}}$ < 1.5\,kpc. While the overall orbital configurations ($Z_{\text{max}}$ - $R_{\text{max}}$ or action space distributions) of our sample match that of a halo, the highly prograde and planar subset (2 $\%$ of the very metal-poor population) also bears characteristics classically associated with a thick disc: (i) a spatial distribution compatible with a short-scaled thick disc, (ii) a $Z_{\text{max}}$ - $R_{\text{max}}$ distribution similar to the one expected from the thick disc prediction of the \textit{Gaia} Universe Model Snapshot, and (iii) a challenge to erase its signature assuming a stationary or prograde halo with $\overline{V_\phi}$ $\sim$ 30-40 km.s$^{-1}$. Altogether, these results seem to rule out that these highly prograde and planar stars are part of a thin disc population and, instead, support a contribution from a metal-weak thick disc. Higher resolution spectra are needed to fully disentangle the origin(s) of the population.}
    %
  {}


   \titlerunning{The Galactic underdogs}
   \authorrunning{I. Gonz\'alez Rivera de La Vernhe et al.}
   \maketitle
%

\section{Introduction}
\label{intro}

Our galaxy, the Milky Way (MW), is a bridge between near and far-field cosmology. Within the scope of $\Lambda$CDM, a hierarchical formation scenario of galaxies can be constrained and compared to high-redshift observations \citep{freeman_bland02, frebelnorris15}. In addition, there is an opportunity to unveil the early history of the MW through the study of its metal-poor ([Fe/H] < -1.0) stars. The atmospheres of the most metal-poor stars, reflective of limited chemical enrichment within their birth gas clouds, is the reason why they are commonly associated with the oldest populations of the Universe \citep{beerschristlieb05,frebelnorris13, frebelnorris15}. 

Until recently, it was widely accepted that the most metal-poor (i.e. oldest) stars would be primarily found on pressure-supported orbits \citep{els62,tumlinson10} - either within the Galactic halo or the bulge \citep{whitespringel00,starkenburg_apostle17,elbadry18}. This idea follows the expectations of a hierarchical assembly process \citep{searlezinn78, schusternissen89, ibata94, chibayoshii98, zolotov09, belokurov22, maosheng22}, and also highlights the contribution of the halo to a metal-poor inner Galaxy \citep{arentsen20, lucey21, arentsen24}.
The search for metal-poor stars in our galaxy goes back to the 1950s (see \cite{beerschristlieb05} for a complete review). The pioneering work of \cite{schwarzschild50} and \cite{roman50} paved the way for the discovery of the first very metal-poor (VMP\footnote{In the following sections, we refer to VMP stars as stars with [Fe/H] < -1.7.}) stars ([Fe/H] < -2.0, \citealt{bond70,bond80}). However, most works from that epoch emphasise the rare nature of this population, making their investigation a rather tedious task in terms of sample selection, [Fe/H] validation, and abundance determination. During the late 1990s to early 2000s, the development of digital photometric and spectroscopic instruments marked the beginning of the era of large ground-based surveys dedicated to the search of metal-poor stars. These include the HK survey \citep{beers85,beers92} and the Hamburg/ESO survey (HES, \citealt{wisotzki96,wisotzki00}), which are responsible for the large samples of stars being uncovered with even lower metallicities, namely extremely metal-poor (EMP) stars ([Fe/H] < -3.0). In particular, HES is at the origin of the class of ultra metal-poor (UMP) stars ([Fe/H] < -4.0) with the discovery of HE 0107-5240 \citep{christlieb04}. 

The mid-2010s represent a turning point in galactic archaeology. With the advent of the space-based \textit{Gaia} mission \citep{perryman01, gaiadr1,gaiadr1bis}, one could get easy access to the 3D positions of millions of celestial objects within $\sim$ 20 kpc and with unprecedented accuracy, leading to the serendipitous detection of metal-poor, VMP, and EMP stars. 
\textit{Gaia}'s third data release \citep{gaiadr3} exploits the medium resolution spectra (R $\sim$ 11 500) obtained by the Radial Velocity Spectrometer (RVS, \citealt{cropper18,katz22}) to provide one of the largest samples of stellar parameters and elemental abundances available \citep{gspspec_firstpaper,gspspec_rvs,recioblanco23}. This combination of astrometric and chemical parameters has offered new insights into the identification and validation of the most metal-poor stars of the Galaxy, and eventually on their chemodynamical characterisation (\citealt{matsuno22, viswanathan24}, see also \citealt{venn20, kielty21}). In parallel, this database can be complemented with recent ground-based spectroscopic surveys, whether high resolution, such as APOGEE DR17 (R $\sim$ 22 500, \citealt{apogee_firstpaper, apogeedr17}), GALAH DR3 (R $\sim$ 28 000, \citealt{galah_firstpaper,galahdr3}), and \textit{Gaia}-ESO (R $\sim$ 19 800 - 21 500 with GIRAFFE and R $\sim$ 47 000 with UVES for MW targets, \citealt{gaiaeso_firstpaper}), or low resolution, such as RAVE DR6 (R $\sim$ 7500, \citealt{ravedr6}) and LAMOST (R $\sim$ 1800, \citealt{lamost_firstpaper, lamost_dr8}). As part of the Sloan Digital Sky Survey (SDSS, \citealt{sdss_firstpaper}), we can also cite SEGUE (R $\sim$ 1800, \citealt{segue_firstpaper}), BOSS, and eBOSS (R $\sim$ 1560-2270 in the blue channel, \citealt{boss_firstpaper,eboss_firstpaper}).
Consequential progress in the identification of metal-poor stars is being made thanks to the development of photometric surveys such as SkyMapper \citep{skymapper_firstpaper}, S-PLUS \citep{splus_firstpaper}, and more specifically with the Pristine survey, a photometric survey at the CFHT using a narrow-band filter centred on the CaHK doublet ($\lambda$ $\sim$ 3900 $\AA$).

Interestingly, several authors are now challenging the formation history of the Galactic disc \citep{sestito19,dimatteo20,mardini22,bellazzini23}.
It has been established for a long time that the disc is a combination of two main components \citep{gilmore83}: the thin disc (scale height $\sim$ 300 pc), with circular orbits and a metallicity distribution peaking close to solar values \citep{hayden14,hayden15,buder19, mackereth19}, and the thick disc, vertically more extended ($\sim$ 1000 pc) with slightly lower metallicities down to $\sim$ -1.5 dex \citep{norris85,morrison90,chiba00,ruchti11,kordo11,kordo13}.
Very recently, several studies identified a significant number of metal-poor stars with kinematics typical of the thin and thick discs \citep{beers14,sestito19,dimatteo20,cordoni21,zhang23,dovgal24,nepal24,gallart24,sestito24}, in particular through the eyes of Pristine \citep{sestito20,venn20,fernandez21,fernandez24, viswanathan24_pgs}.
This population, as well as its prograde nature, is supported by recent cosmological simulations \citep{santistevan21,sestito21,sotillo23,carollo23}, but its full characteristics are not yet well established.

However, one would expect very ancient stars rotating in a disc to have heated up towards halo-like kinematics over time, and not to be found on disc-like orbits.  
A first explanation for such a divergence between observations and theory is that the sample of Pristine low-metallicity candidates is contaminated by a fraction of the overwhelmingly larger population of metal-rich stars ([Fe/H] > -0.5).
In fact, the metallicity distribution function (MDF) of the MW in different Galactic locations shows a dominant metal-rich population related to the thin disc \citep{hayden15,kordo15}. Inevitably, the contamination of metal-poor samples by metal-rich counterparts is higher at rotational velocity ($V_\phi$) values typical of the thin disc. Indeed, metal-poor samples with prograde and cold kinematics (small velocity dispersion), that is, disc-like, are considerably more polluted by metal-rich stars compared to metal-poor samples located at higher latitude \textit{b} and, or with kinematically hotter orbits. Such an effect, adding to the fact that extinction biases on Pristine photometric metallicities are larger along the disc area, represents a serious concern, as misclassifying only a very small fraction of the metal-rich sample could compromise the detection of a kinematically cold metal-poor population. 
Leaving aside doubts related to the quality of photometric metallicities, the origin of this disc-like metal-poor population is still highly debated. 
Some authors relate it to a primordial disc, with divergent opinions on its in situ or ex situ origin \citep{sestito21,mardini22,bellazzini23}, some to a prograde halo \citep{dimatteo20,belokurov22,zhang23}, while others involve the Galactic bar in the migration of metal-poor stars from the bulge or the halo towards the disc \citep{yuan23,li23,dillamore23}.

In this paper, we aim at characterising a population of VMP stars confined on prograde-planar orbits, specifically with high angular momenta and high $V_\phi$, similarly to MW disc stars. Using the \textit{Pristine-Gaia} synthetic photometric metallicity catalogue (PGS), which is part of the first Pristine data release \citep{martin23}, we conscientously selected a sample of about 3M giant stars with available radial velocities (RVs) from \textit{Gaia} \citep{katz22}, in the range -4.0 < [Fe/H] < 0. The main emphasis of this work is to obtain a sample that is statistically significant on the metal-poor end to confidently assess the existence of a VMP disc population, while being as free as possible from metal-rich outliers at high $V_\phi$. This step is the most important to be able to discuss the potential origins of this peculiar population.

This paper is organised as follows. Section 2 presents the relevant catalogues needed to construct a sample with astrometric and chemodynamical information. Section 3 describes the quality cuts and the filtering method applied to the sample, and Section 4 details the validation of the latter. Section 5 is dedicated to the decontamination of the sample using spectroscopic surveys. Finally, in Sections 6 and 7 we discuss and summarise our results.

\section{The data: PGS metallicities and \textit{Gaia} RVS orbits}
\label{desc_pgs}

The Pristine survey \citep{starkenburg17} is a narrow-band photometric survey, observed with the Canada-France-Hawaii Telescope (CFHT), and focused on the metallicity-sensitive Ca H $\&$ K doublet lines (located at 3933.7 and 3968.5 $\AA$), which are of great interest to infer photometric metallicities down to [Fe/H] $\sim$ -4.0 dex. This combination of narrow-band and broad-band photometry has led to the successful discovery of large samples of EMP, VMP and UMP stars in recent years \citep{starkenburg18, aguado19, bonifacio19, venn20, lardo21, kielty21, lucchesi22, caffau23, viswanathan24}.

For this work, we used the publicly available PGS metallicity catalogue, an all-sky metallicity catalogue based on \textit{Gaia} DR3 BP/RP spectro-photometry published by \cite{martin23} as part of Pristine DR1. Relevant columns for this paper are listed in Table~\ref{pgs_var_desc}. 
In this catalogue, 52 million \textit{Gaia} DR3 objects with BP/RP coefficients\footnote{In the \textit{Gaia} archive, BP/RP spectra are stored as coefficients with their corresponding covariance matrix (projections on a set of basis functions), which can be reconstructed as spectral energy distributions by a linear combination.} have their de-reddened synthetic CaHK magnitudes, \verb|CaHK_0|, computed after integrating their BP/RP spectra under the transmission curve of the same CaHK filter as the one used in the Pristine survey. Methods are described in \cite{martin23} following \cite{montegriffo23} for the integration of photometric filters, and in \cite{starkenburg17} for the metallicity estimation. 

Compared to Pristine CFHT observations (6500 $\text{deg}^{2}$ mapped since 2015), PGS allows for a complete sky coverage: the catalogue is a great opportunity to gather a statistically significant sample of stars with a net prograde motion, close to the Galactic plane, while displaying very low metal contents. Such a population was highlighted in \cite{fernandez21} from CFHT Pristine observations. In order to confirm these results, especially at low metallicities, we dedicated a large part of this work to the preliminary verification of the catalogue parameterisation. 

To be able to investigate both the kinematical (velocities) and dynamical (actions) characteristics of this population, we complemented our sample with orbital parameters derived using \textit{Gaia} RVs and the geometric distances of \cite{bj21} as our distance estimator. Galactocentric cartesian positions \textit{X, Y, Z}, Galactocentric cylindrical radius \textit{R}, and Galactocentric cylindrical velocities $V_R$, $V_\phi$, $V_Z$ were computed for all stars with available RVs. Orbital parameters were obtained following \cite{kordo23}, who used the Stäckel fudge method for their computations \citep{binney12b,bovyrix13,mackereth_bovy_2018}, along with the Galpy\footnote{\url{http://github.com/jobovy/galpy}} code \citep{bovy2015}. The adopted MW potential is the \cite{mcmillan17}; it is axisymmetric, with a modelling of cold gas discs, a thin and a thick disc, a bar, a bulge, and a halo. The potential was adapted to recent values of Solar positions, that is, $(R, Z)_{\odot}$ = (8.249, 0.0208) kpc \citep{gravity20, bennettbovy19}, and Solar velocities, that is, $(V_{R}, V_{\phi}, V_{Z})_{\odot} = (-9.5, 250.7, 8.56)$ km.s$^{-1}$ (\citealt{gravity20,reidbrunthaler20}). Further details on the computation are provided in Sec. 3.1 of \cite{kordo23}.

The following sections describe the different quality cuts applied. The number of stars removed from each individual cut is summarised in Table \ref{cuts_pgs_recap}.

\begin{table*}
\centering
\caption{Description of key columns from the published \textit{Pristine-Gaia} synthetic photometric metallicity catalogue.}
\begin{tabular}[t]{ccc}
\hline
Column & Description & Unit \\
\hline
\verb|G_0| & de-reddened G magnitude & mag\\
\verb|BP_0| & de-reddened BP magnitude & mag\\
\verb|RP_0| & de-reddened RP magnitude & mag\\
\verb|CaHK_0| & synthetic CaHK magnitude & mag\\
\verb|d_CaHK| & uncertainty on synthetic CaHK magnitude & mag\\
\verb|FeH_CaHKsyn| & photometric metallicity obtained directly from the Pristine model & dex\\
\verb|FeH_CaHKsyn_50th| & photometric metallicity obtained after 100 Monte-Carlo draws of the Pristine model & dex\\
\verb|mcfrac_CaHKsyn| & fraction of Monte-Carlo draws falling into the [Fe/H] grid & --\\
\verb|Pvar| & probability for a source to be variable & --\\
\verb|E(B-V)| & \cite{schlegel98} extinction value & mag\\
\hline
\end{tabular}
\label{pgs_var_desc}
\end{table*}

\section{The sample: Data selection}

\subsection{\textit{Gaia} DR3 quality cuts}
\label{gaiaqualcut}
 
 After cross-matching PGS with the \textit{Gaia} DR3 source catalogue, a filtering based on astrometry has been applied: 
\begin{itemize}[leftmargin=*]
\item[--] $\varpi$ > 0 mas to avoid negative parallaxes\footnote{Quality cuts relative to parallax do not take into account the correction from the parallax zero point of -17 $\mu$as \citep{lindegren20}.};
\item[--] relative parallax error, that is, $\delta \varpi /\varpi$ < 0.20 to avoid large errors propagated onto velocities\footnote{When applying a stricter cut on the parallax (e.g. $\delta \varpi /\varpi$ < 0.10), we restrict the accessible volume, therefore our sample selection sharply decreases (by 24 $\%$), which hampers our ability to probe the population of disc-like metal-poor stars we focus on. However, the conclusions of this work remain unchanged.};
\item[--] \verb|sqrt(phot_g_n_obs)/phot_g_mean_flux_over_error| < 0.015 to minimise the presence of variable stars;
\item[--] Renormalised Unit Weight Error (RUWE) < 1.4 to avoid astrometric binaries, following \cite{lindegren21};
\item[--] we removed stars with \verb|source_id| in \verb|gaiadr3.vari_summary|, \textit{Gaia} DR3's variable sources catalogue (complementary to Pristine's \verb|Pvar|, see Table \ref{pgs_var_desc}).
\end{itemize}

\subsection{Pristine DR1 quality cuts}
\label{pgsqualcut}

We adopted the following cuts relative to the PGS catalogue (see also Sec. 7.3 of \cite{martin23} and Table~\ref{pgs_var_desc}): 

\begin{itemize}[leftmargin=*]
\item[--] \verb|mcfrac_CaHKsyn| > 0.9 to remove stars with a large fraction of their Monte Carlo samplings falling beyond the edges of the resulting metallicity grid (\cite{martin23} advise  \verb|mcfrac_CaHKsyn| > 0.8, we adopted a stricter cut since 80 $\%$ of the catalogue's \verb|mcfrac_CaHKsyn| lies above 0.9);
\item[--] \verb|FeH_CaHKsyn_50th| > -4.0 to avoid stars at the lower edge of the metallicity grid adopted for the training sample;
\item[--] \verb|Pvar| < 0.3 to avoid photometrically variable stars\footnote{As discussed in Sec. 5 of \cite{martin23}, the variability of stars is complex on the bright end of the G band, when obtained from \textit{Gaia} data. It is necessary to develop a model that is dependent on the noise budget of the source to determine which \textit{Gaia} stars are intrinsically variable.};
\item[--] \verb|d_CaHK| < 0.05 mag to remove noisy photometric measurements;
\item[--] \cite{schlegel98} extinction value $E(B-V)$ < 0.5 mag\footnote{Applying a stricter cut on $E(B-V)$ (e.g. < 0.2) results in depleting the final metal-poor sample statistically, but in a homogeneous way, preserving the distributions in kinematics discussed in the following sections.} following Sec. 7.5 of \cite{martin23}.
\end{itemize}

\textit{Pristine-Gaia} synthetic provides two metallicity estimates: \verb|FeH_CaHKsyn_50th|, the median photometric estimate obtained after 100 Monte-Carlo draws of the giant Pristine model\cprotect\footnote{The Pristine model focusses on giant stars, thus the generic output for the photometric metallicities is the one computed with the giant model (\verb|FeH_CaHKsyn_50th|). However, we note that we chose to assign to dwarf stars the photometric estimate obtained with the dwarf model (\verb|FeH_CaHKsyn_dw_50th|).}, and \verb|FeH_CaHKsyn|, the direct estimate (see Table~\ref{pgs_var_desc}). We found that, irrespective of the location in the PGS colour-magnitude diagram (CMD, in this context, refers to a colour-absolute magnitude diagram), there is an overall good agreement between median and direct metallicity estimates, with an absolute difference consistently below 0.04 dex, except for a few sources with $(BP - RP)_0 < 0.8$ and $M_{G_0}\footnote{We used the geometric distances of \cite{bj21} to \\ compute $M_{G_0}$.} \geqslant 4$ mag, for which the difference is greater than 0.1 dex (0.83 $\%$ of the total sample).
For the remainder of this work, we define \verb|FeH_CaHKsyn_50th| as our generic photometric metallicity, that we dub $\text{[Fe/H]}_{\text{PGS}}$, to be compared with spectroscopic metallicities. 

\subsection{Stellar population quality cuts and isochrone filtering method}
\label{isofit}

\begin{figure*}
    \centering
    \includegraphics[scale=0.6,center]{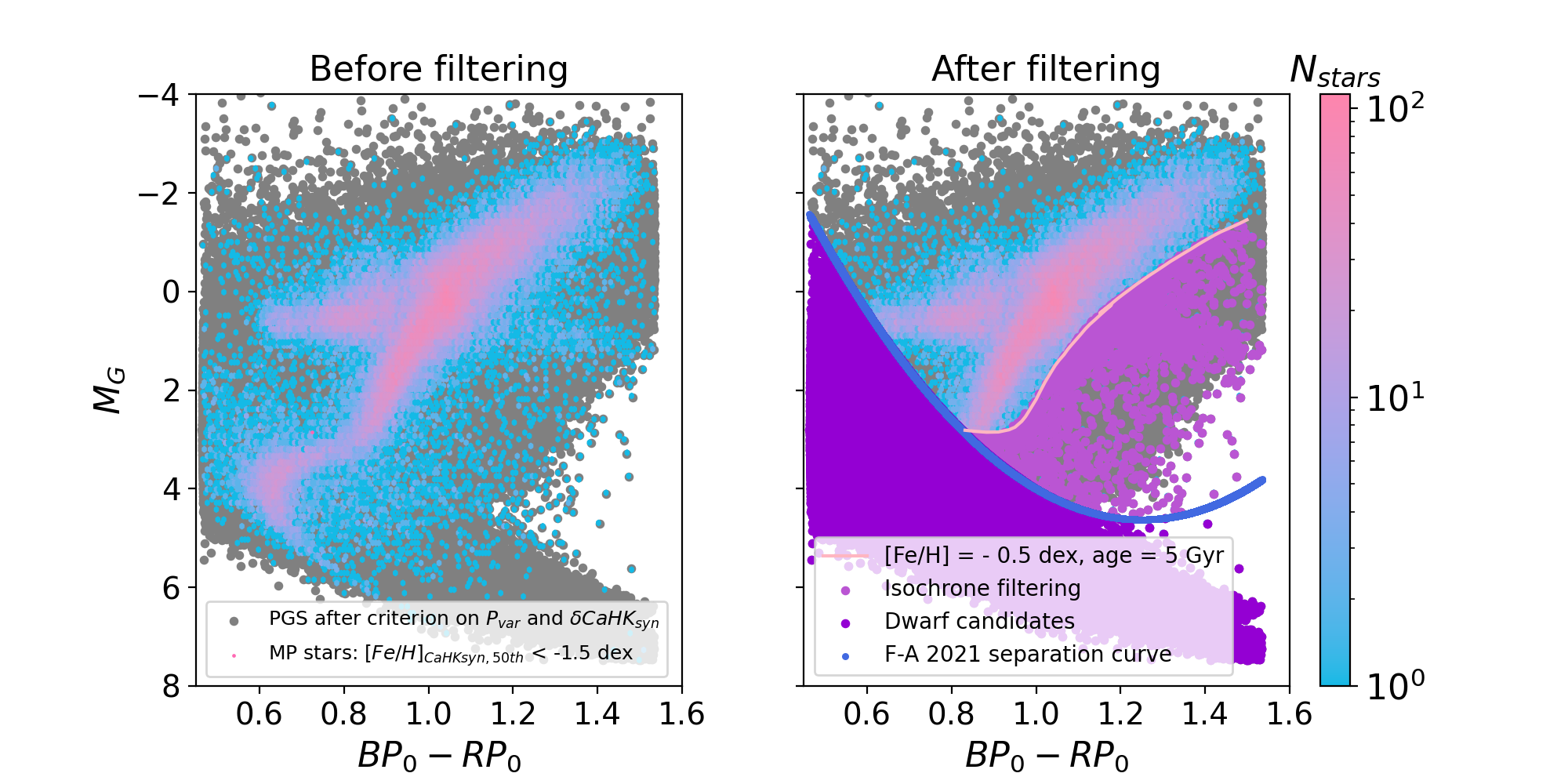}
    \caption{Colour - magnitude diagram (CMD) of the \textit{Pristine-Gaia} synthetic (PGS) metallicity catalogue. The sample, after applying the quality cuts described in Sec.~\ref{gaiaqualcut} and \ref{pgsqualcut} is colour-coded in grey. \textit{Pristine-Gaia} synthetic stars with $\text{[Fe/H]}_{\text{PGS}}$ < -1.5 dex are shown with a density plot. There are two colour codes for the interlopers: dwarf stars are displayed in deep purple and isochrone-filtered sources in light purple. Left panel: before filtering, 9 432 286 stars. Right panel: after, 2 880 338 stars (Sec.~\ref{isofit}). The blue curve corresponds to the dwarf and giant separation of \cite{fernandez21}.}
    \label{cmd_fullpgs}
\end{figure*}

Once our data have been curated based on \textit{Gaia} DR3 astrometry (Sec.~\ref{gaiaqualcut}) and PGS criteria (Sec.~\ref{pgsqualcut}), additional filters were applied:
\begin{itemize}[leftmargin=*]
\item[--] we removed stars identified in recent censuses of globular \citep{vasiliev21} and open clusters \citep{hunt23};
\item[--] we selected giant stars using the separation curve in the CMD of \cite{fernandez21} (blue curve in the right panel of Fig.~\ref{cmd_fullpgs}). As advised in \cite{starkenburg17} and \cite{martin23}, the Pristine survey was designed to target mostly metal-poor giant stars of the MW halo and is consequently less precise for main sequence stars (smaller training sample). 
\end{itemize}

An additional filter was created to remove stars with underestimated PGS metallicities, based on their location in the CMD. Indeed, it is expected that genuine metal-poor giants (intrinsically old) occupy the bluest part of the red giant branch (RGB). 
For that reason, one can use theoretical isochrones in order to define areas in which genuine metal-poor stars (in that context, stars with estimated $\text{[Fe/H]}_{\text{PGS}}$ < -1.5) should not reside in. However, on the one hand, metal-poor stars can have a wide range of ages; on the other hand, PGS metallicities can have relatively large uncertainties ($\delta_{\text{[Fe/H]}} \sim$ 0.3 dex). So, the definition of this area needs to be quite generous.
That is why we adopted a PARSEC isochrone\footnote{\url{http://stev.oapd.inaf.it/cgi-bin/cmd}} \citep{bressan12} of [Fe/H] = -0.5 and $\tau$ = 5 Gyr to define the physical boundaries into which a PGS star with [Fe/H] < -1.5 can reside (see Fig.~\ref{cmd_fullpgs}).

We used an argument based on age and metallicity to justify the physicality of the location of our sample in the CMD.
If a star within the metal-poor subsample appears redder than the boundary defined by the isochrone, then either:
\begin{itemize}[leftmargin=*]
\item[--] its metallicity is underestimated, and is therefore higher than the metallicity we are targeting for the metal-poor subsample; 
\item[--] its isochrone age is older than that of the Universe ($\tau$ = 13.5 Gyr);
\item[--] the distance or the de-reddening procedure is inaccurate, causing erroneous metallicity and, or orbits.
\end{itemize}

With this method, we filtered out 15 $\%$ (9811 sources) of the PGS metal-poor subsample. Eventually, the PGS sample comprises a total of 2 880 338 giant stars, including 54 009 metal-poor stars according to our definition. In particular:
\begin{itemize}[leftmargin=*]
\item[--] stars with $\text{[Fe/H]}_{\text{PGS}}$ < -1.5 were filtered according to Sec.~\ref{gaiaqualcut}, ~\ref{pgsqualcut} and ~\ref{isofit};
\item[--] stars with $\text{[Fe/H]}_{\text{PGS}}$ > -1.5 were filtered similarly without the isochrone filtering;
\item[--] we note that this sample does contain horizontal branch (HB) stars; however, they only make up for 11 $\%$ of the total size.
\end{itemize}

\begin{figure*}
    \centering
    \includegraphics[width=\linewidth]{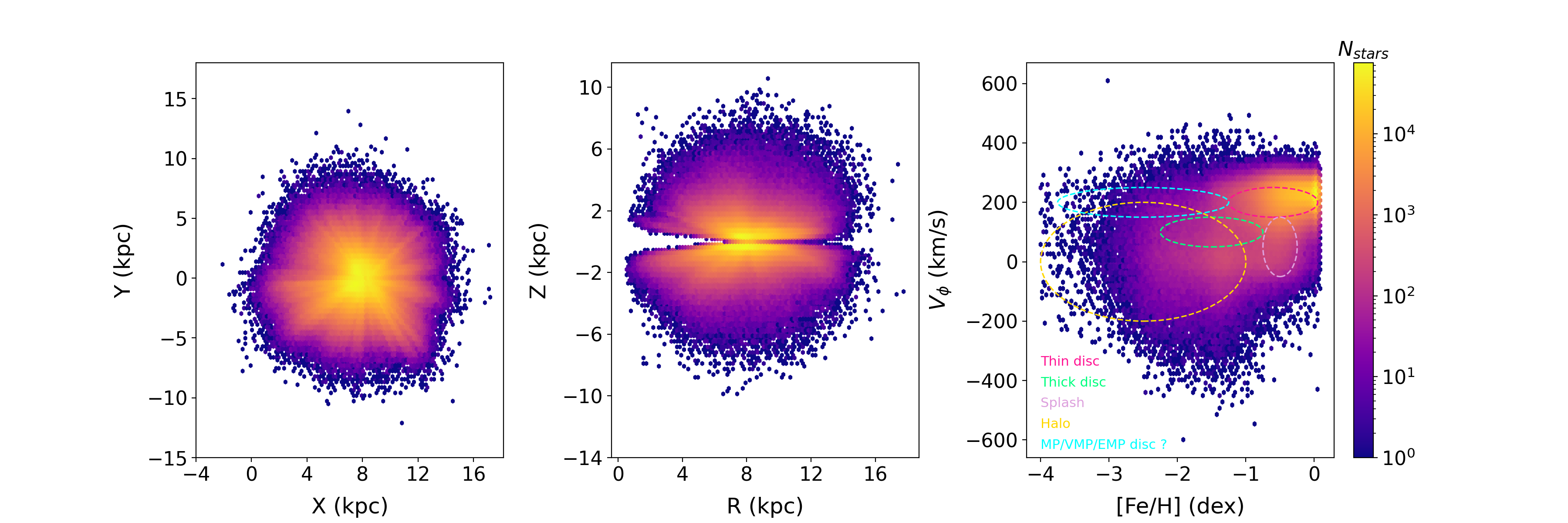}
    \caption{Spatial density distribution of the isochrone-filtered PGS in logarithmic scale (only giant stars). Left panel: Galactocentric Y versus Galactocentric X. Middle panel: Galactocentric Z versus Galactocentric cylindrical radius R. Right panel: $V_\phi$ versus [Fe/H]; the different ellipses qualitatively indicate the location of several identified Galactic components, following Fig. 8 of \cite{fernandez21}.}
    \label{spatialdistpgs}
\end{figure*}

The spatial distribution of our PGS sample is displayed in Fig.~\ref{spatialdistpgs}. It is essentially distributed within the extended Solar neighbourhood, but covers quite well the X-Y space (left panel), and also between 2 < R < 14 kpc and -6 < Z < 6 kpc (middle panel). As seen from the R-Z distribution, most stars lie close to the plane, although due to Pristine extinction cuts, there are very few stars below |Z| = 2 kpc at large R. As expected, we can clearly see the thin and thick disc stand out in the $V_\phi$ - [Fe/H] space (right panel), but we do see a tail of distribution extending down to the lowest metallicities, as discussed in \cite{fernandez21}. Additionally, we note the presence of a 'pencil-beam' structure in X-Y that could be caused by the \textit{Gaia} scanning law (see Sec. 5.2 of \citealt{gaiadr1bis}), in particular because of our sample selection, which only comprises stars with \textit{Gaia} RVS RVs (RVS selection effects can also be a cause for the structure). We discuss these asymmetries in Sec.~\ref{trends_spatialdist}.

\section{Validations using spectroscopic surveys}

\subsection{Validation of the isochrone filtering procedure}
\label{validapo}

To confirm the reliability of PGS, and to evaluate the amount of metal-rich contaminants, we used the publicly available spectroscopic metallicity catalogues of APOGEE DR17 \citep{apogee_firstpaper,apogeedr17}, GALAH DR3 \citep{galah_firstpaper,galahdr3}, LAMOST DR8 \citep{lamost_firstpaper,lamost_dr8}, and \textit{Gaia} RVS GSP-spec \citep{gspspec_firstpaper,gspspec_rvs}. 
It is important to note that in this section, we kept both dwarfs and giants, although white dwarfs were discarded for all surveys (stars with $M_{G_0} \geqslant 7.5$).
The orbital parameters of the spectroscopic catalogues were computed similarly to the method described in Sec.~\ref{desc_pgs}. For APOGEE and GALAH, we used their spectroscopic RVs as input, whereas we used \textit{Gaia} RVs for LAMOST and GSP-spec. Table \ref{cuts_spectro_recap} gives the sample size of each cross-match before and after performing the isochrone filtering method.
The following quality cuts were applied to each cross-match between PGS and the spectroscopic catalogues:
\begin{itemize}[leftmargin=*]
\item[--] APOGEE: following \cite{apogeedr17} and \cite{hegedus22}, we removed stars with any of its parameter flagged as \verb|_BAD|\footnote{\url{https://www.sdss4.org/dr17/irspec/apogee-bitmasks/##APOGEEBitmasks}};
\item[--] GALAH: we followed \cite{galahdr3} and \cite{hegedus22} and the best practices for the use of GALAH DR3\footnote{\url{https://www.galah-survey.org/dr3/using_the_data/}}, that is, columns and flags recommendations;
\item[--] LAMOST: following \cite{luo15} and \cite{tsantaki2022}, we selected stars with S/N > 15 in the \textit{g}- and -\textit{i}- bands at all [Fe/H] ranges (see \citealt{li18} for limitations), and excluded stars with negative RV errors; 
\item[--] GSP-spec: following Table 2 of \cite{gspspec_rvs}, we selected stars with null values for flags 1-13 (related to atmospheric parameters except abundances)\cprotect\footnote{\verb|flags_gspspec LIKE ‘0000000000000%’|}, except flags 7 (\verb|fluxNoise|) and 8 (\verb|extrapol|) for which values $\leqslant$ 2 were allowed. Table~\ref{gspspec_flags} summarises the different flags applied to this sample. Additionally, some hot metal-poor stars were discarded as they are likely badly parameterised (i.e, with $T_{\text{eff}}$ > 6000 K and $\text{[Fe/H]}_{\text{GSP-spec}}$ < -1, see Sec.~10.5 of \citealt{gspspec_rvs}). We note that for GSP-spec, we use $\text{[M/H]}_{\text{calibrated}}$ as our metallicity estimator.
\end{itemize}

In the following, we focus on comparisons with APOGEE since it is of high resolution and highly reliable for giant stars. Besides, APOGEE contains a large number of higher-metallicity disc stars, our primary source of contaminants. Validation using all spectroscopic catalogues can be found in Appendix~\ref{isofit_other}.

The left panel of Fig.~\ref{fehfeh_pgs_apogee} shows the cross-match between PGS and APOGEE in metallicity, where the broad majority of the sample follows a one-to-one relation. 
Between -2.5 < $\text{[Fe/H]}_{\text{APOGEE}}$ < -1, we see a relatively good agreement. APOGEE metallicities do not go below -2.5 because of the lower boundary of their adopted training grid \citep{aspcap16}, whereas some PGS stars can be as metal-poor as -4.0. Thus, for these stars, APOGEE values are most likely biased, while PGS values can still be reliable. 
However, a fraction of PGS metal-poor candidates ($\text{[Fe/H]}_{\text{PGS}}$ < -1.5) have nearly solar APOGEE metallicities. This is manifesting as a horizontal sequence in the left panel of Fig.~\ref{fehfeh_pgs_apogee} between -4.0 < $\text{[Fe/H]}_{\text{PGS}}$ < -1.5, and makes up for 8.04 $\%$ of the total number of PGS metal-poor candidates within the PGS-APOGEE cross-match. In what follows, we dub these contaminants 'interlopers'.
Although in small numbers, the metal-rich interlopers evidenced most likely belong to the disc and might contaminate highly prograde VMP stars. For this paper that aims to characterise VMP disc stars, the number of interlopers therefore needs to be minimised, which is why they are removed from our final sample.

To infer whether a certain type of star causes the bias observed between both metallicity estimates, we looked at the CMD of the PGS-APOGEE cross-match (left panel of Fig.~\ref{cmd_pgs_apogee}).
The majority of the metal-poor subsample is located on the blue side of the RGB, while interlopers can be found either on the main sequence or on the red side of the RGB. We note that interlopers are unlikely to be APOGEE stars with a wrong colour estimation (therefore, a wrong $T_{\text{eff}}$), as no clear dependency on the extinction coefficient $E(B-V)$ was identified.
The isochrones with different combinations of age and metallicity visually translate the arguments used to construct the isochrone filtering method (Sec.~\ref{isofit}).
After applying the isochrone filtering method, 94 $\%$ of APOGEE interlopers are removed (see middle panel of Fig.~\ref{fehfeh_pgs_apogee} and right panel of Fig.~\ref{cmd_pgs_apogee}). For LAMOST, GALAH and GSP-spec, the percentages are detailed in Appendix~\ref{isofit_other}.
The strength of this method can be seen in the middle panel of Fig.~\ref{fehfeh_pgs_apogee}, where the majority of the horizontal sequence caused by interlopers was successfully removed.

We note that this method does prevent some interlopers from being discarded because they occupy the region where metal-poor stars physically exist, and conversely, it removes badly parameterised stars that are not interlopers but that still occupy the region where metal-poor stars should not exist. The latter represents an excess of 2.4 $\%$ of the PGS metal-poor sample that might have been removed. Finally, it is also possible that outliers from APOGEE make up for a fraction of the undiscarded percentage of interlopers. In the right panel of Fig.~\ref{fehfeh_pgs_apogee}, the aftermath of removing dwarf stars from our sample using the separation criterion of \cite{fernandez21} is the disappearance of the overdensity at $\text{[Fe/H]}_{\text{PGS}}$ $\sim$ -0.3, and an overall lower density along the [1:1] diagonal.

\begin{figure*}
    \centering
    \includegraphics[width=\linewidth, center]{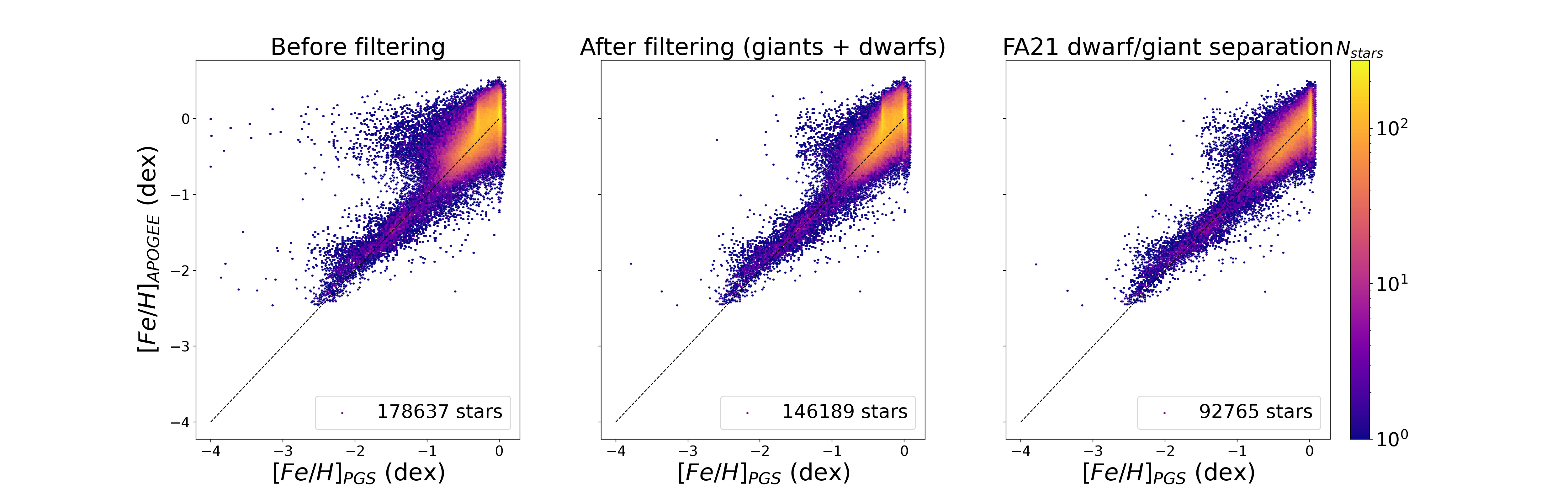}
    \caption{APOGEE versus PGS metallicities, colour-coded in number density. Left panel: before applying the isochrone filtering method (dwarfs + giants). Middle panel: after applying the isochrone filtering method (giants + dwarfs). Right panel: after applying the dwarf and giant separation criterion of \cite{fernandez21} (only giants, see Sec.~\ref{validapo}). The black dashed line represents the [1:1] relation.}
    \label{fehfeh_pgs_apogee}
\end{figure*}

\begin{figure}
    \centering
    \includegraphics[width=1.3\linewidth, center]{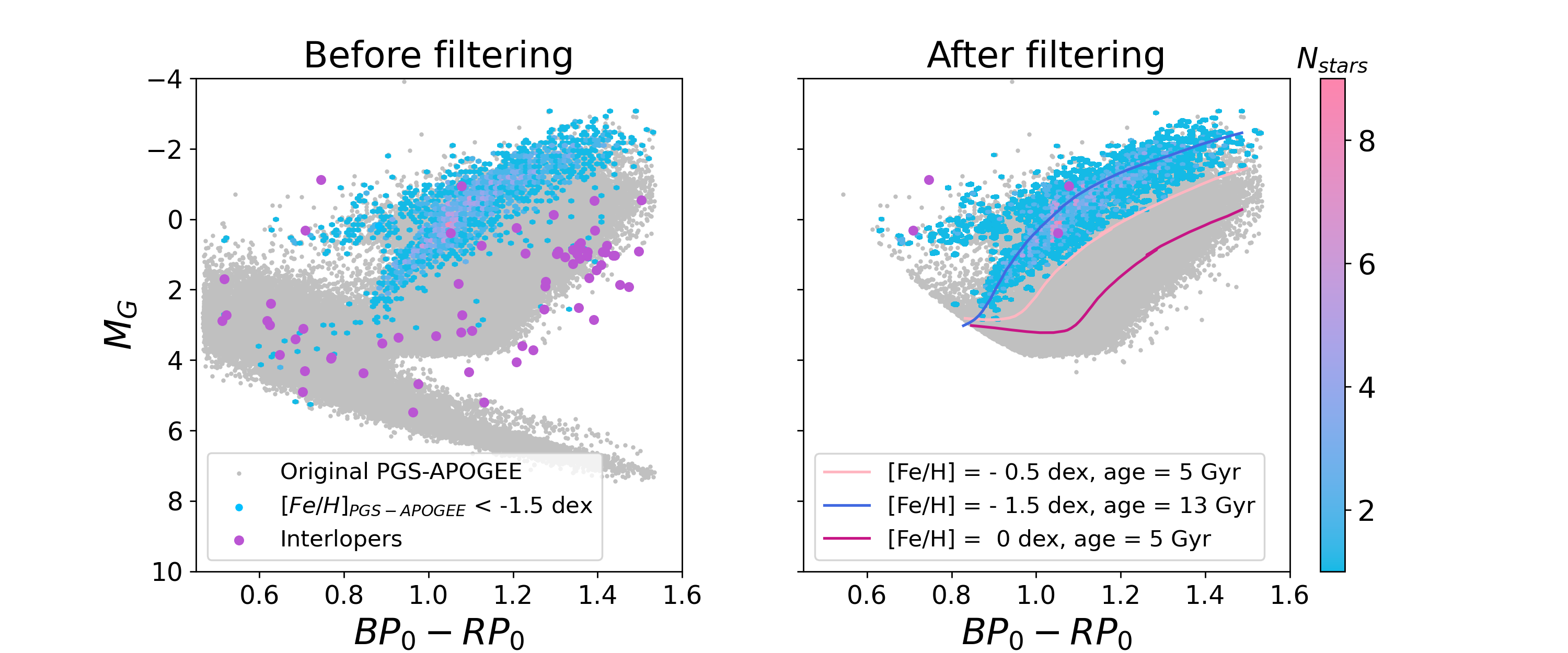}
    \caption{CMD of the PGS-APOGEE cross-match (grey). The metal-poor subsample ($\text{[Fe/H]}_{\text{PGS}}$ < -1.5 dex), is colour-coded in cyan. Interlopers are colour-coded in purple (Sec.~\ref{validapo}). PARSEC isochrones of different age-metallicity combinations are also displayed on the right panel.}
    \label{cmd_pgs_apogee}
\end{figure}

\subsection{Estimating contamination in the isochrone filtered PGS}
\label{spectroconts}

After applying the isochrone filtering method to our PGS sample, we statistically assessed the likely remaining contamination using scaled comparisons with the spectroscopic catalogues. Here, we investigated the PGS-APOGEE cross-correlation. Other surveys are discussed in Appendix~\ref{choice_spectro_cont}. We adopted the following procedure:

\begin{itemize}[leftmargin=*]
\item[--] for a given $\text{[Fe/H]}_{\text{PGS}}$ range, we counted and flagged as 'spectroscopic contaminant' every star (already isochrone-filtered) located beyond the $2\sigma$ threshold of the \\ $\Delta_{\text{[Fe/H]}} = \text{[Fe/H]}_{\text{PGS}} - \text{[Fe/H]}_{\text{APOGEE}}$ distribution (see Fig.~\ref{sigma_feh_pgs_apogee});
\item[--] we then rescaled the contaminants: for a given $\text{[Fe/H]}_{\text{PGS}}$ range, we multiplied the number of contaminant counts by a factor, corresponding to the total number of PGS stars in the given $\text{[Fe/H]}_{\text{PGS}}$ range, divided by the number of PGS stars with spectroscopic estimates (including the contaminants), in the same $\text{[Fe/H]}_{\text{PGS}}$ range.
\end{itemize}

This definition of contaminants is conservative because it includes a significant number of true metal-poor stars. Specifically, it may include stars with underestimated spectroscopic metallicities as well as stars with only slightly overestimated photometric metallicities, which still belong to the metal-poor regime and cannot be mistaken for metal-rich disc contamination.

\subsection{Impact of the spectroscopic contamination}
\label{filteredpgs}

In this section, we analyse the $V_\phi$ distribution and the action space of the isochrone-filtered PGS after estimating the spectroscopic contamination using APOGEE. 

\subsubsection{In the $V_\phi$ distribution}

\begin{figure*}
    \centering
    \includegraphics[width=1.1\textwidth,center]{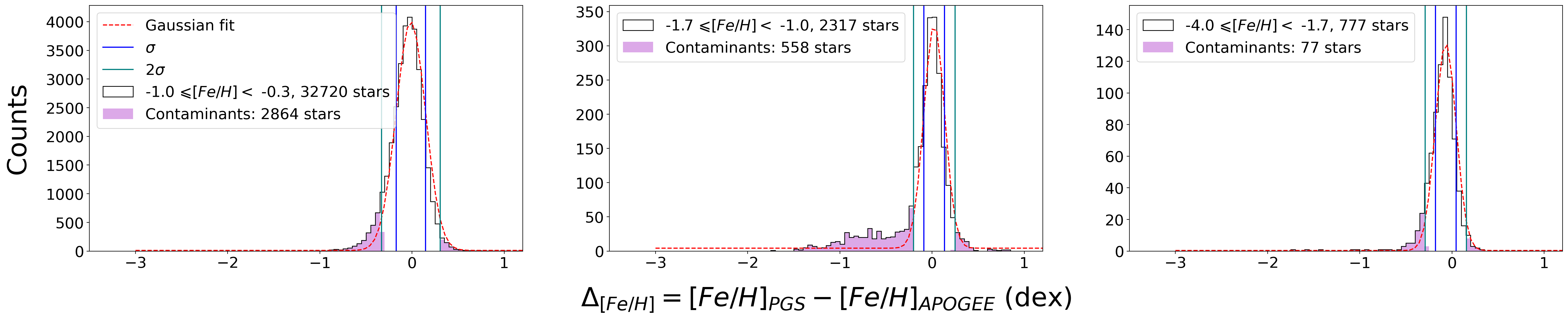}
    \caption{$\Delta_{\text{[Fe/H]}} = \text{[Fe/H]}_{\text{PGS}} - \text{[Fe/H]}_{\text{APOGEE}}$ distribution (black). The Gaussian fit to the distribution is overplotted in red dashed lines. $\sigma$ and 2$\sigma$ limits are respectively indicated in blue and teal vertical lines. The PGS-APOGEE contaminants, i.e. sources with $\Delta_{\text{[Fe/H]}} = \text{[Fe/H]}_{\text{PGS}} - \text{[Fe/H]}_{\text{APOGEE}} > 2\sigma$ are colour-coded in purple. 
    Left panel: -1.0 < $\text{[Fe/H]}_{\text{PGS}}$ < -0.3. Middle panel: -1.7 < $\text{[Fe/H]}_{\text{PGS}}$ < -1.0. Right panel: -4.0 < $\text{[Fe/H]}_{\text{PGS}}$ < -1.7 (Sec.~\ref{spectroconts}).}
    \label{sigma_feh_pgs_apogee}
\end{figure*}

\begin{figure}
    \centering
    \includegraphics[width=\linewidth]{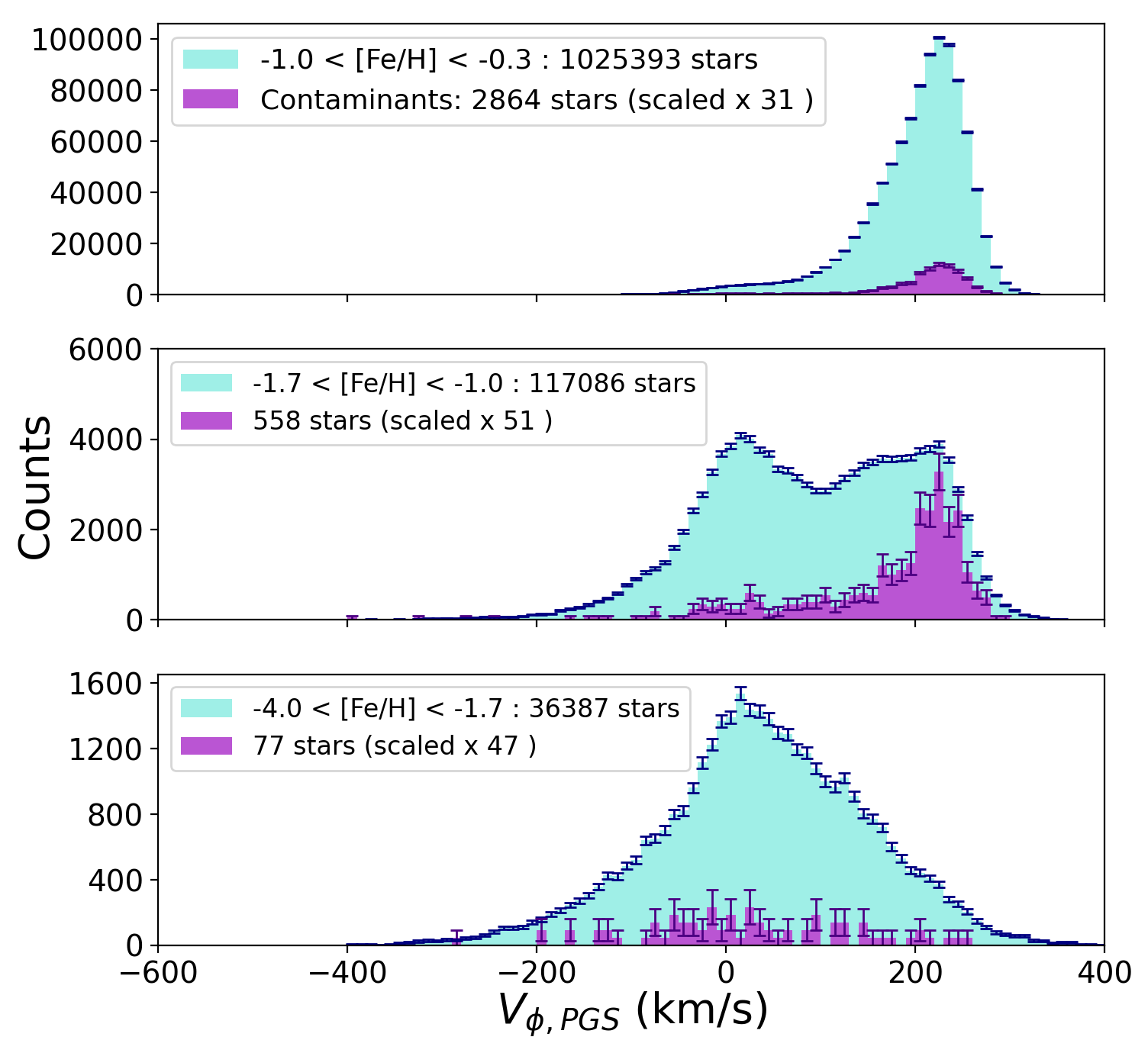}
    \caption{$V_\phi$ distribution of PGS. Counts with associated error bars (Poisson noise) are displayed in blue while PGS-APOGEE spectroscopic contaminants are displayed in purple. Contaminants are scaled by a factor equal to the size of PGS in each $\text{[Fe/H]}_{\text{PGS}}$ interval, divided by the size of APOGEE in the same intervals. Top panel: -1.0 < $\text{[Fe/H]}_{\text{PGS}}$ < -0.3. Middle panel: -1.7 < $\text{[Fe/H]}_{\text{PGS}}$ < -1.0. Bottom panel: -4.0 < $\text{[Fe/H]}_{\text{PGS}}$ < -1.7.}
    \label{vphidist_allbins}
\end{figure}

Figure~\ref{vphidist_allbins} shows the $V_\phi$ distribution of PGS with associated spectroscopic contaminants in three [Fe/H] intervals.
In the top panel, for the most metal-rich interval (-1.0 < [Fe/H] < -0.3), the bulk of the PGS sample (in cyan) is compatible with the circular velocity measurement in the solar vicinity (mostly thin disc stars) of $\sim$ 230\,km.s$^{-1}$ \citep{eilers18,vieira22,poder23}. The population is mildly affected by contaminants (in purple). 
The middle panel, with -1.7 < [Fe/H] < -1.0, is interesting because the PGS distribution seems to be compatible with three components: one centred at $\sim$ 0-10 km.s$^{-1}$, in line with a stationary and, or slightly prograde halo \citep{tian19}, a second one at $\sim$ 180 km.s$^{-1}$, compatible with the thick disc value of \cite{fernandez21}, and a third one centred at $\sim$ 240-250 km.s$^{-1}$, similar to the one identified in the top panel and dominated by metal-rich contamination, as expected at high $V_\phi$ \citep{hayden15,kordo15}.  
Finally, in the bottom panel, where the most metal-poor stars reside (-4.0 < [Fe/H] < -1.7), the PGS sample is centred at 0 km.s$^{-1}$, although a prograde-retrograde asymmetry is definitely visible, with a skewed distribution towards prograde values. Contaminants are spread homogeneously along the entire $V_\phi$ range, and are not discarding PGS VMP stars with high $V_\phi$. 
Out of all the metallicity intervals, the one between -1.0 and -1.7 is the most contaminated. The reason for the presence of higher contamination rates at high $V_\phi$ in this particular interval can be seen in Fig.~\ref{sigma_feh_pgs_apogee}. The middle panel displays a strong asymmetry in the distribution of contaminants with respect to the two side panels. In fact, this contamination is mostly made of metal-rich (solar) stars shifted in [Fe/H] down to almost -1.5; as explained in Sec.\ref{intro}, this is because the metal-rich population is dominant in the thin disc and is more likely to pollute VMP prograde samples. 
Therefore, it is not surprising to find contamination at MW thin disc $V_\phi$. 

\subsubsection{In the action space}

\begin{figure*}
    \centering
    \includegraphics[width=\textwidth]{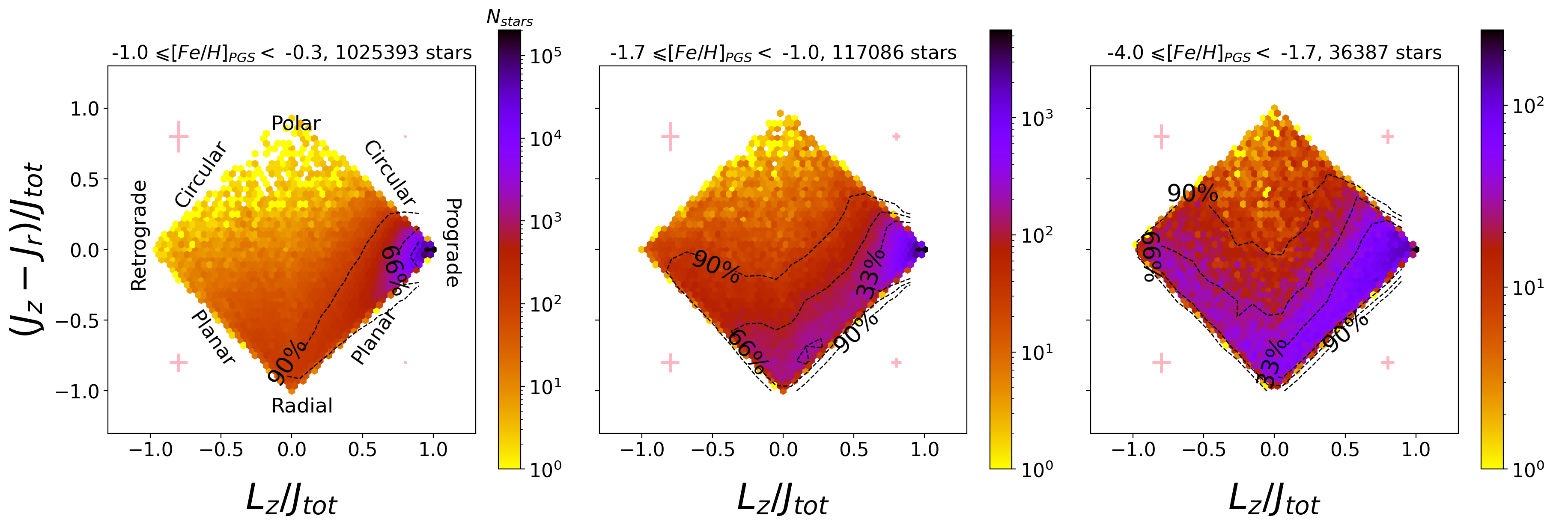}
    \caption{Action space: ($J_z - J_r$) as a function of $L_z$. Axes are normalised by $J_{tot} = |L_z| + J_r + J_z$. Left panel: -1.0 < $\text{[Fe/H]}_{\text{PGS}}$ < -0.3. Middle panel: -1.7 < $\text{[Fe/H]}_{\text{PGS}}$ < -1.0. Right panel: -4.0 < $\text{[Fe/H]}_{\text{PGS}}$ < -1.7. We note that the colour bar range changes for each interval. We add mean error bars for each quadrant of the action space in light pink.}
    \label{diamondplot}
\end{figure*}

Figure~\ref{diamondplot} shows the action space of our PGS sample in the same [Fe/H] intervals chosen in Fig.~\ref{vphidist_allbins}. It is colour-coded in density, and contour lines of the 33, 66 and 90 $\%$ of the sample are also displayed. 

The action space is defined as ($J_z$ - $J_r$)/$J_{\text{tot}}$ versus $J_\phi$/$J_{\text{tot}}$, where $J_r$ is the radial action, $J_\phi$ is the azimuthal action ($J_\phi$ = $L_z$ since we assume an axisymmetric potential), $J_z$ is the vertical action, and $J_{\text{tot}}$ is $J_r$ + |$L_z$| + $J_z$. The action space allows us to identify different families of orbits \citep{binney12b,binney12a}: 
\begin{itemize}[leftmargin=*]
\item[--] radial: $L_z$/$J_{\text{tot}}$ = 0 and ($J_z$ - $J_r$)/$J_{\text{tot}}$ = -1; the radial component dominates
\item[--] polar: $L_z$/$J_{\text{tot}}$ = 0 and ($J_z$ - $J_r$)/$J_{\text{tot}}$ = +1; the vertical component dominates
\item[--] prograde: $L_z$/$J_{\text{tot}}$ $\rightarrow$ 1
\item[--] retrograde: $L_z$/$J_{\text{tot}}$ $\rightarrow$ -1
\item[--] circular: $J_z$/$J_{\text{tot}}$ > $J_r$/$J_{\text{tot}}$
\item[--] planar: $J_z$/$J_{\text{tot}}$ < $J_r$/$J_{\text{tot}}$
\end{itemize}

In the left panel of Fig.~\ref{diamondplot}, 66 $\%$ of the sample is labelled as circular-prograde-planar, which undoubtedly corresponds to a thin disc population with high $V_\phi$ (hence high angular momenta $L_z$), low eccentricities (circular orbits) and low $Z_{\text{max}}$ (close to the plane). More specifically, the mean eccentricity of the 66 $\%$ of the sample is 0.20 and the mean $Z_{\text{max}}$ is 1.3 kpc. The rest of the sample extends along the planar down to radial regions of the action space with null $L_z$; this structure could come from different contributions, mainly the \textit{Splash} \citep{belokurov20} or the thick disc. 
In the middle panel, 33 $\%$ of the PGS sample is prograde-planar between -1.7 < [Fe/H] < -1.0, and can be associated with the thin disc. However, it is difficult to completely trust this distribution in the action space, due to the high contamination rates at high $V_\phi$ (middle panel of Fig.~\ref{vphidist_allbins}). In the right panel, the bulk of the sample extends towards the planar and radial regions. The 90 $\%$ contour line expands more towards retrograde values than for the two previous metallicity ranges; this is in line with the second peak in the $V_\phi$ distribution that we attribute to a halo contribution. 
The prograde-planar population is found again in the right panel, although the retrograde and prograde regions are more evenly distributed. 
Altogether, the error bars in the action space are small with respect to the signatures that we seek, especially since they are even smaller for prograde stars. Errors are driven mostly by distance errors, which stem from the maximum errors on parallaxes allowed in our selection. We note that the depletion of data when applying a stricter cut on the parallax (e.g. $\delta \varpi /\varpi$ < 0.10) is localised in the polar area of the action space; however, we recovered the same trends as the study conducted with the original cut of 0.20.

To get a better understanding of what to expect in terms of dynamics, we compared our sample with the \textit{Gaia} Universe Model Snapshot (GUMS, \citealt{robin12}), a model based on the Besançon Galaxy Model (BGM, \citealt{robin03}). This model generates a synthetic catalogue at a given static time (snapshot) to simulate the environment in which \textit{Gaia} objects are or will be observed. It is semi-empirical as it is partly based on stellar evolution theory and constrained by observations. GUMS contains the main Galactic components (the thick and thin discs, the halo, the bulge and the bar), but no stellar streams, nor any other known substructures. The parameters chosen for the Galactic components are constrained by the BGM\footnote{\url{https://gea.esac.esa.int/archive/documentation/GEDR3/Data_processing/chap_simulated/sec_cu2UM/ssec_cu2starsgal.html}}. 
For this mock model, we considered GUMS up to G = 15.3 mag, for |\textit{b}| > 7 deg, and for giant stars (defined here as $T_{\text{eff}}$ < 6000 K and log \textit{g} < 3.9 dex) to be able to compare with our catalogue of PGS giants. We also removed very young stars (< 0.15 Gyr) since, as indicated in the \textit{Gaia} documentation, such stars have had their kinematics badly implemented into the simulated catalogue. We computed stellar orbits using the \cite{mcmillan17} potential, in the same fashion as in Sec.~\ref{desc_pgs}; we note that, to remain coherent with the GUMS dynamical model, we adopted the GUMS solar velocities (12.75, 0.93, 7.10) km.s$^{-1}$ \citep{robin17}, as well as $V_{\text{LSR}}$ = 230.6 km.s$^{-1}$ at $R_{\odot}$ = 8 kpc \citep{sofue15}. In this first selection function of GUMS stars, we verified that the main Galactic components were retrieved as expected in the BGM. The thin disc contains stars with a maximum scale length of 2011 pc and a mean rotation of 221.85 km.s$^{-1}$. The thick disc is modelled as a 'young' and 'old' thick disc of respectively 10 and 12 Gyr, in respective proportions of 68 and 32$\%$; we found the expected scale heights of 400 and 795 pc, and scale lengths of 2040 and 2919 pc, and a mean rotation for the entire population 186.52 km.s$^{-1}$. Finally, halo stars have an eccentricity of 0.774, and a mean rotation of -0.34 km.s$^{-1}$. 
We stress that although GUMS does not capture the fine details of Galactic structure that would be needed to properly model our metal-poor population, it is a very helpful guide to understand the behaviour of classical MW stellar populations in various kinematical spaces, since it has been shown to reproduce the overall star counts and kinematics of the Galaxy.

\begin{figure*}
    \centering
    \includegraphics[width=\linewidth]{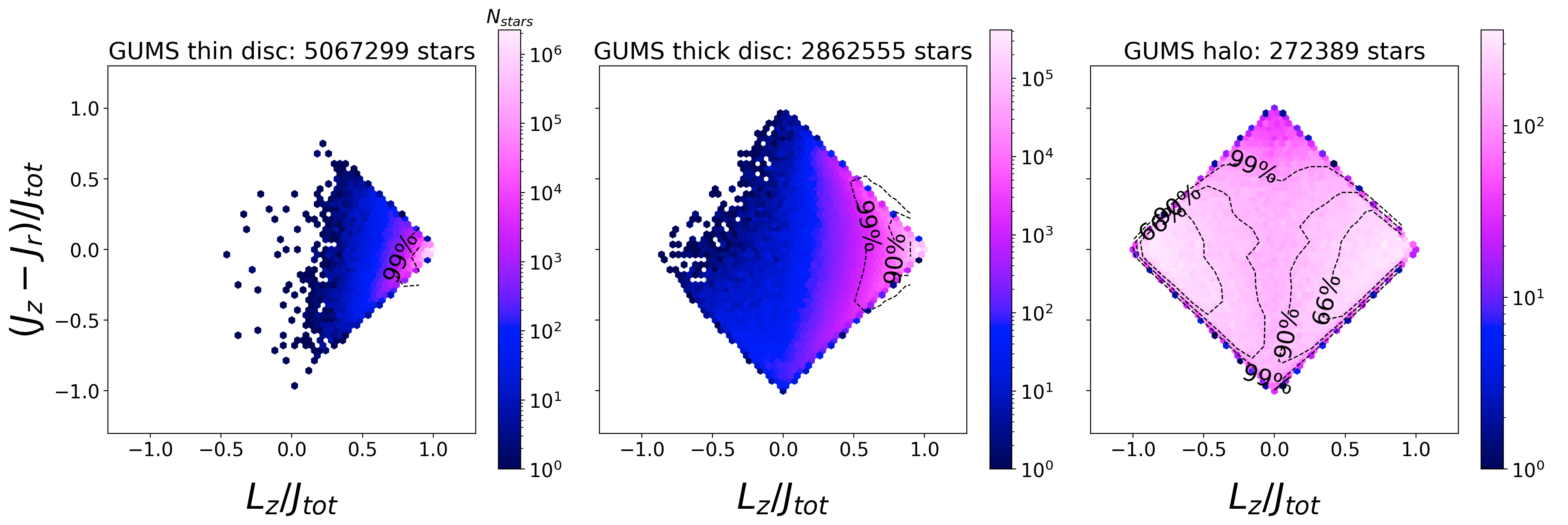}
    \caption{\textit{Gaia} Universe Model Snapshot (GUMS) action space. Left panel: thin disc stars. Middle panel: thick disc stars. Right panel: halo stars. The contour lines at 33, 66, 90 and 99 $\%$ are plotted in black dashed lines. We note that the colour bar range changes for each panel.}
    \label{gums_diamondplot}
\end{figure*}

We display the GUMS action space in Fig.~\ref{gums_diamondplot}. Out of the sample of GUMS thin disc stars, 90 $\%$ are located on highly prograde and circular orbits; the GUMS thick disc is essentially prograde with equal amounts of radial and polar orbits. The GUMS halo is very symmetrical with respect to $L_z$/$J_{\text{tot}}$ = 0. 
As expected, the left panel of Fig.~\ref{gums_diamondplot} resembles the distribution of our data in the left panel of Fig.~\ref{diamondplot}. According to the distribution of the GUMS thin and thick disc, we suspect that our sample at intermediate [Fe/H] is a mix of thin and thick disc stars. As opposed to the symmetry observed for the GUMS halo, our sample at the lowest [Fe/H] interval (right panel of Fig.~\ref{diamondplot}) is clearly asymmetric in favour of the prograde component. 
Given the high contamination of PGS at intermediate [Fe/H] (between -1.7 and -1.0) by disc stars, it would be risky to interpret the high $V_\phi$ PGS VMP stars ([Fe/H]<-1.7) without considering contamination. Although it is low overall in that range (see right panel of Fig.~\ref{sigma_feh_pgs_apogee}), it may contain a higher proportion of stars with disc kinematics than the truly metal-poor parent sample. In the following section, we statistically decontaminate our PGS sample using spectroscopic estimates.

\section{Statistical decontamination of PGS}
\label{decontamination}

In this section, we detail how we decontaminated PGS for -4.0 < [Fe/H] < -1.7, using spectroscopic estimates of contamination. 
We worked with APOGEE and LAMOST in the $V_\phi$ - [Fe/H] space and in the action space. As discussed in Appendix~\ref{choice_spectro_cont}, due to completeness (GSP-spec) and statistical (GALAH) issues, we investigated the decontamination with APOGEE because of its high resolution, and with LAMOST because of its statistics, despite its low resolution. To remain within the range of 0.3 dex in [Fe/H] uncertainties, we computed the mean of contamination in different areas of the $V_\phi$ - [Fe/H] space and of the action space. In each area, the scaling factor for contamination changes with the density of stars in PGS, and the density of stars in the survey, in designated areas of the $V_\phi$ - [Fe/H] space and of the action space.  
For both surveys, we checked : 
\begin{itemize}[leftmargin=*]
\item[--] the density of stars in the contaminants samples;
\item[--] PGS minus individual contamination (pixel-by-pixel);
\item[--] PGS minus the mean of contamination.
\end{itemize}

\subsection{$V_\phi$ - [Fe/H] space}
\label{decont_vphi}

\begin{figure*}
    \centering
    \begin{subfigure}{0.5\textwidth}
        \includegraphics[width=0.9\textwidth]{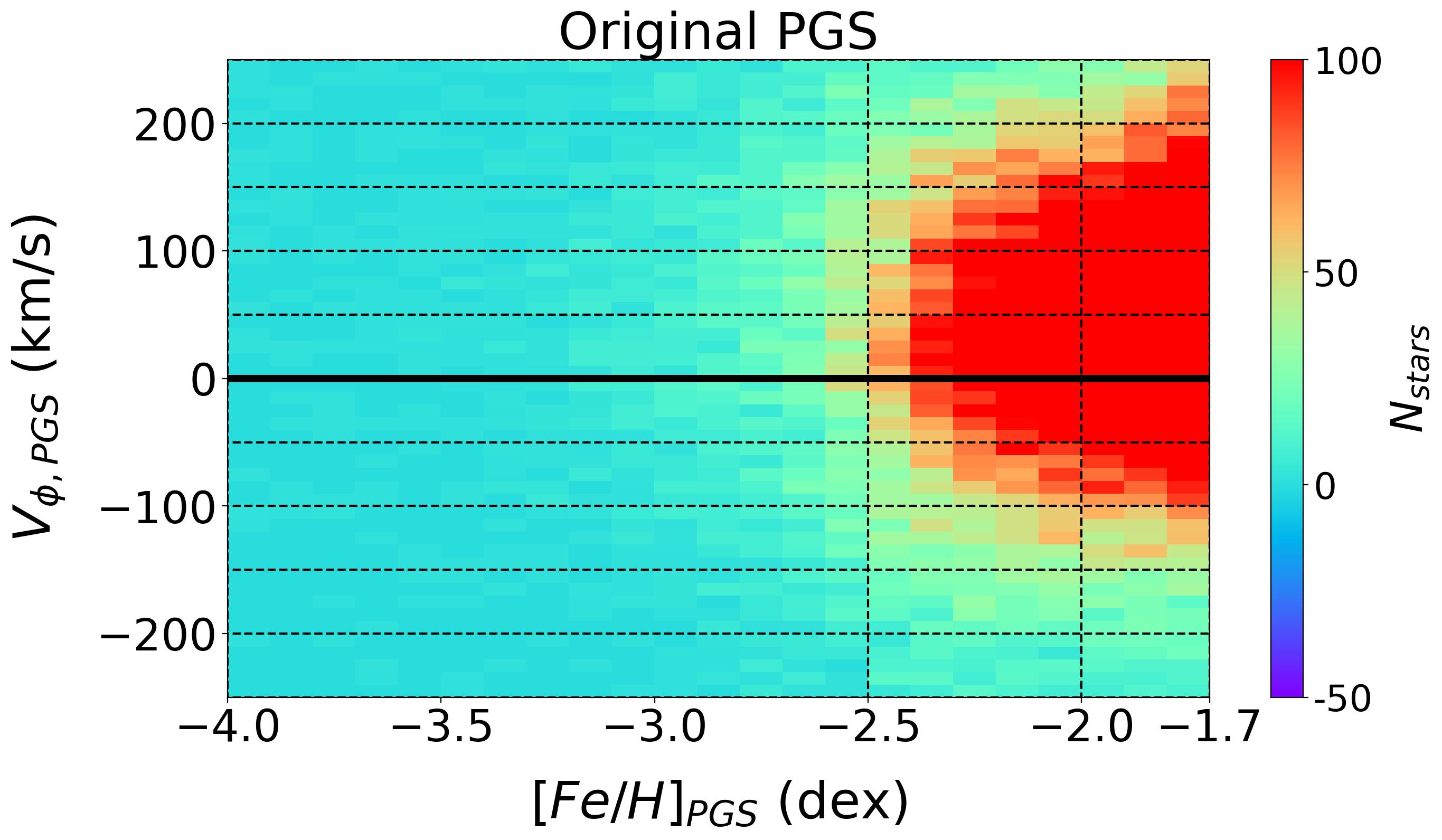}
        \end{subfigure}
    \hfill
    \begin{subfigure}{\textwidth}
        \centering
        \includegraphics[width=0.9\textwidth]{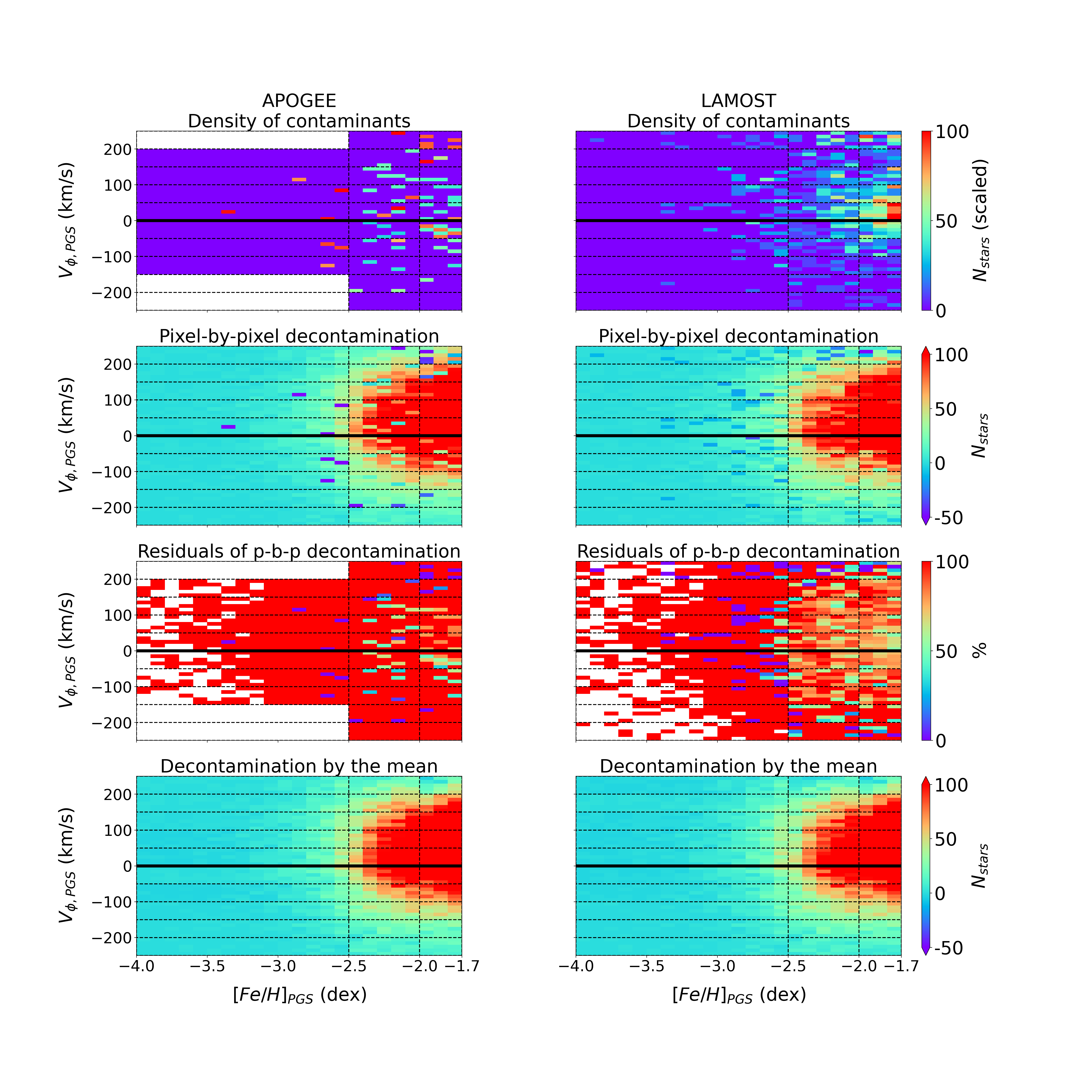}
    \end{subfigure}
    \hfill
    \caption{Densities of stars in the $V_\phi$ - [Fe/H] space. The grid corresponds to areas delimited in $V_\phi$ and [Fe/H]. The scaling of contaminants varies from one area to another with the density of PGS in each area. The dimension of a pixel is 0.1 dex $\times$ 10 km.s$^{-1}$. First panel (top): PGS only. In the next rows of panels: left = APOGEE, right = LAMOST. Second row: density of spectroscopic contaminants, scaled to the sample size of PGS in each pixel. Third row: PGS decontaminated pixel-by-pixel, i.e. subtraction between a pixel of PGS density and a pixel of spectroscopic contamination density. Fourth row: residuals percentages of the pixel-by-pixel decontamination of PGS. Fifth row: PGS decontaminated by the mean of the spectroscopic contamination, i.e. by the mean of the spectroscopic contamination density in a designated $V_\phi$ - [Fe/H] area.}
    \label{thebigplot}
\end{figure*}

Fig.~\ref{thebigplot} summarises the main steps of decontamination in the $V_\phi$ - [Fe/H] space.
Here, we plotted 30 individual areas, ranging from -250 to 250 km.s$^{-1}$ vertically (spaced by 50 km.s$^{-1}$), and from -1.7 to -4.0 horizontally (spaced at -2.0 and -2.5). We note that the vertical bin size is 10 km.s$^{-1}$ and the horizontal bin size is 0.1 dex.  
The first panel (top) shows the density of the original PGS data. We notice an asymmetry in $V_\phi$ which was already observed in the bottom panel of Fig.~\ref{vphidist_allbins}. For [Fe/H] < -2.5, there is a sharp decrease in density, 
reflecting the shape of the MDF of the MW; even with these low counts, an asymmetry in the $V_\phi$ distribution is visible. 
The second row of panels shows the density of the spectroscopic contamination. Starting with the left panel (APOGEE), we notice blank areas at [200, 250] and [-150, -250] km.s$^{-1}$; we have no information on the contamination there since APOGEE does not provide spectroscopic metallicity measurements below -2.5. Hence, only contaminants with $\text{[Fe/H]}_{\text{PGS}}$ < -2.5 and $\text{[Fe/H]}_{\text{APOGEE}}$ > -2.5 can appear in that range. When there are none, no contamination can be measured. Nonetheless, for the stars that verify this condition, we have evidence of a low contamination rate (at [150, 200] and [0, -50] km.s$^{-1}$).
Overall, the contamination is lower with decreasing [Fe/H], but it is also very high in the three upper right areas, corresponding to the population we are aiming to characterise.
In the right panel, as expected, we are able to follow the contamination down to -4.0 with LAMOST because of the more complete sample available. However, this time contamination does not decrease with decreasing [Fe/H], but is rather concentrated between -1.7 and -2.5 dex. It is not restricted to high $V_\phi$ values but also extends to the [0, -50] km.s$^{-1}$ area. 
The third row of panels shows the pixel-by-pixel decontamination of PGS. We note that the blank APOGEE contamination areas were replaced by PGS densities, since there is in practice no contamination to account for.
In both cases, the decontamination is causing pixels with negative values, due to statistical fluctuations of the low contamination rates (the contaminant pixel removed is too dense in comparison with the PGS pixel). This is highly dependent on the sample sizes of PGS, APOGEE and LAMOST since they rule the weight of the scaling factor. However, this view is useful to identify the areas most affected by the decontamination. In particular, we note that regardless of the negative structure, the density of PGS in the upper right areas of the $V_\phi$ - [Fe/H] space ($V_{\phi}$ > 150 km.s$^{-1}$, -2.5 < [Fe/H] < -1.7) is still remarkably high when decontaminating with APOGEE, and non-negligible when decontaminating with LAMOST.

Another way to visualise the effect of the pixel-by-pixel decontamination is to show the residuals, after removing spectroscopic contaminants, as seen in the fourth row of panels. After both decontaminations, we notice that between 50 and 100 $\%$ of the sample is conserved, especially at [-100; 100] km.s$^{-1}$ for all metallicities. We do see more clearly in the upper right areas of the $V_\phi$ - [Fe/H] space mentioned before that a large percentage of data is preserved.

To overcome the statistical fluctuations, we also display the result in the fifth row of panels, when the decontamination is made by subtracting the mean value of contamination in each area, delimited the dashed lines corresponding to different combinations of $V_\phi$ and [Fe/H].
Impressively, there is not much difference between both decontaminations, and also between the decontaminations and the original PGS. Again, we do see a change in the densities of the upper right areas, but the population of VMP planar-prograde stars we suspected could be erased after decontaminating is in fact still present.

\subsection{Action space}

\begin{figure}
    \centering
    \begin{subfigure}{0.5\textwidth}
        \centering
        \includegraphics[width=0.8\linewidth]{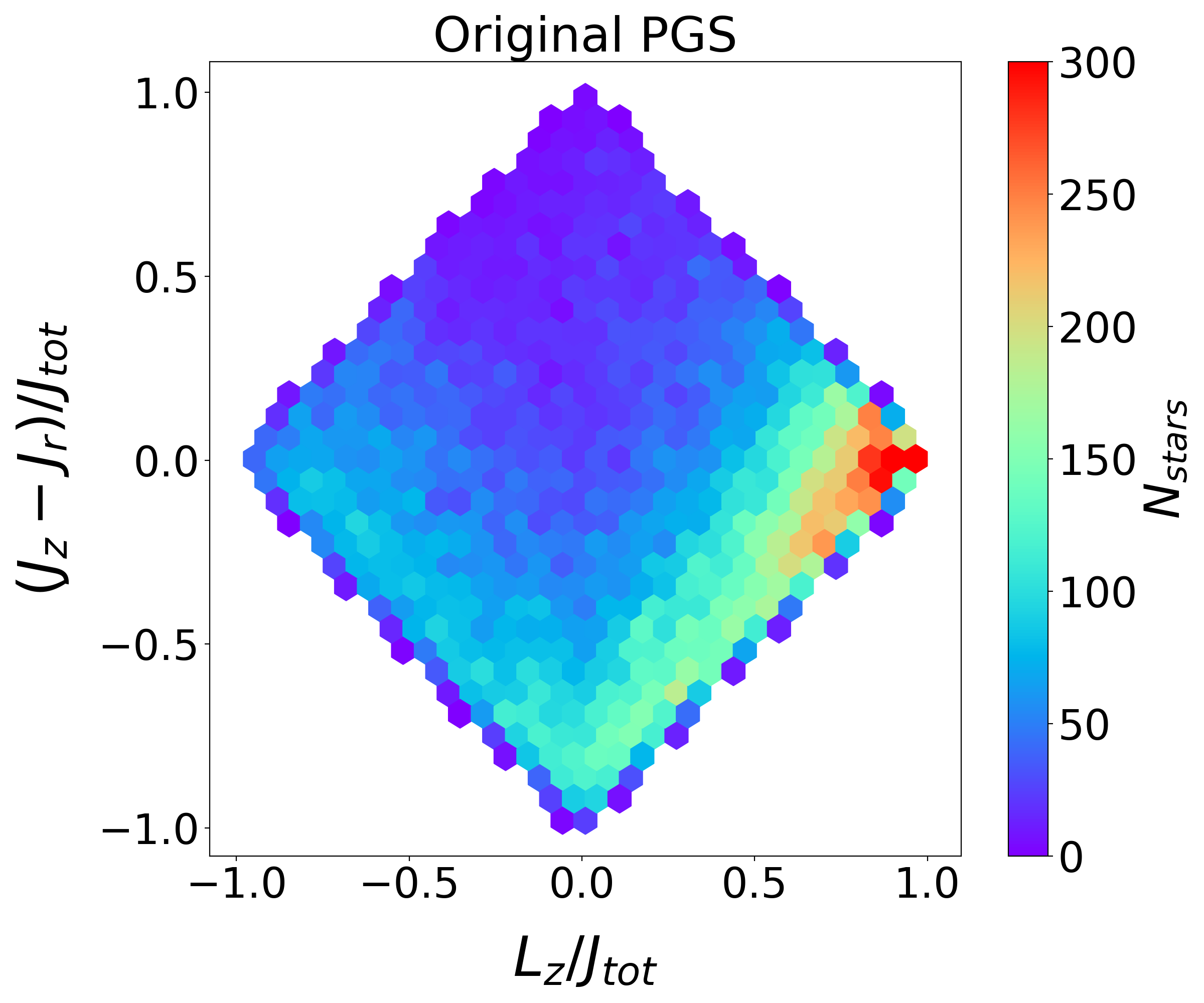}
    \end{subfigure}
    \begin{subfigure}{0.5\textwidth}
        \centering
        \includegraphics[width=0.8\linewidth]{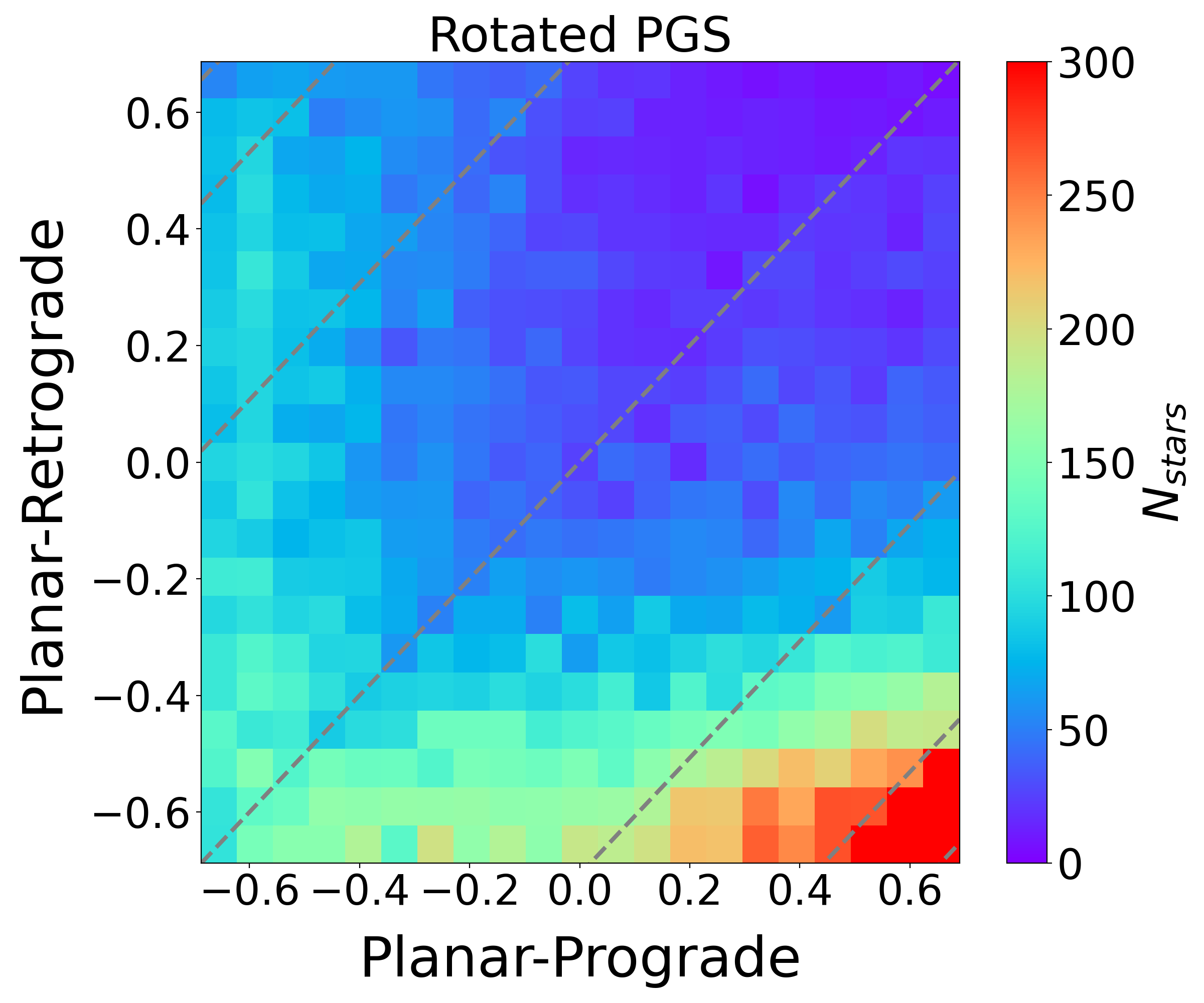}
    \end{subfigure}
    \caption{Densities in the action space. Top: original PGS after applying the isochrone filtering method, in the most metal-poor interval (-4.0 $\leqslant$ [Fe/H] < -1.7). Same plot as the right panel of Fig.~\ref{diamondplot}. Bottom: same as the top panel, after rotating the (central) coordinates of each pixel by + 45 degrees. The limits of the plot in that space correspond to the rotated coordinates of the four corners of the original action space. The grey dashed lines of the bottom panel correspond to different values of $L_z$/$J_{\text{tot}}$: [-0.95, -0.8, -0.5, 0, 0.5, 0.8, 0.95]. }
    \label{actspace_oriandrot}
\end{figure}

The principle of decontamination in the action space is the same as in the $V_\phi$ - [Fe/H] space. The main difference is that we manipulated the action space to resemble the layout of Fig.~\ref{thebigplot}, so that comparisons between both spaces are more straightforward.  
To do so, we rotated the (central) coordinates of each pixel composing the original action space (top panel of Fig.~\ref{actspace_oriandrot}) by + 45 degrees. Then, we assigned a weight to each pixel, equal to the scaling factor defined in Sec.~\ref{spectroconts}. The limits of this newly defined action space correspond to the rotated coordinates of each corner of the original action space; the resulting space is displayed in the bottom panel of Fig.~\ref{actspace_oriandrot}.
The rotated action space is subsequently divided into nine even areas of length $\sim$ 0.4 in both directions.

\begin{figure}
    \centering
    \includegraphics[width=0.77\linewidth]{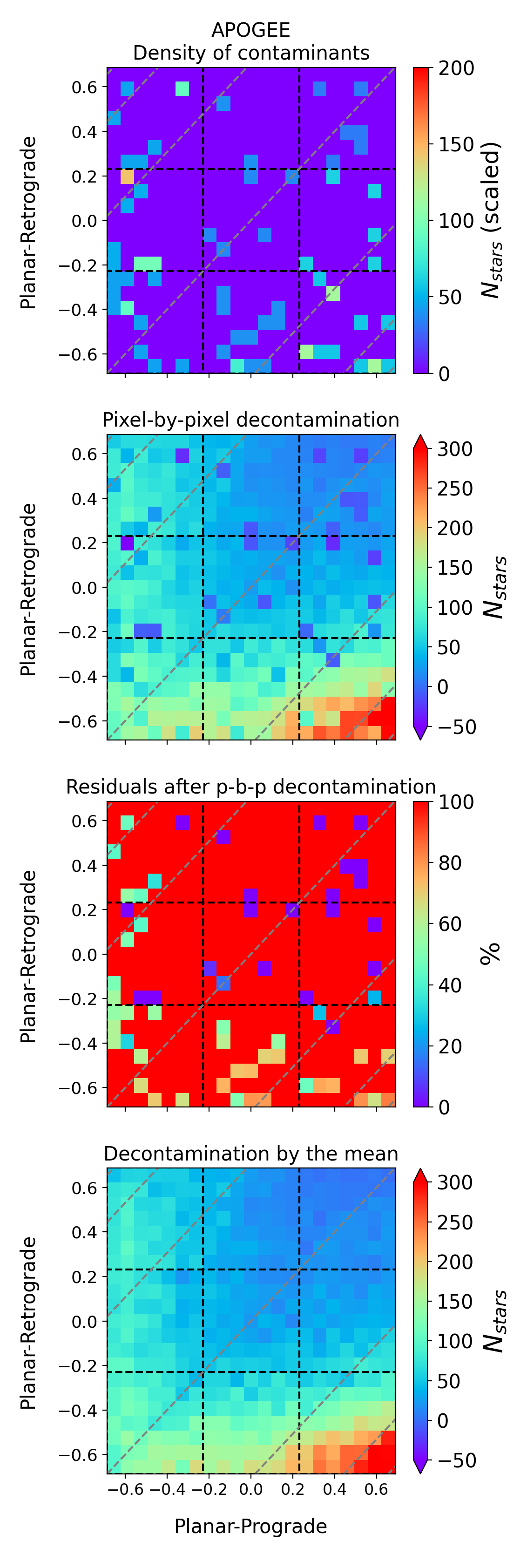}
    \caption{Densities in the action space, in the most metal-poor interval (-4.0 $\leqslant$ [Fe/H] < -1.7). The grid corresponds to areas in the action space delimited in ($J_z - J_r$) and $L_z$. From top to bottom: first panel: density of APOGEE contaminants. Second panel: PGS decontaminated pixel-by-pixel, using APOGEE. Third panel: residuals percentages of the pixel-by-pixel PGS decontamination using APOGEE. Fourth panel: PGS decontaminated by the mean of the APOGEE contamination. The grey dashed lines correspond to different values of $L_z$/$J_{\text{tot}}$: [-0.95, -0.8, -0.5, 0, 0.5, 0.8, 0.95].}
    \label{actspacedecont}
\end{figure}

Figure~\ref{actspacedecont} summarises the decontamination in the action space with APOGEE. The results of the decontamination with LAMOST, which are very similar to APOGEE, are displayed in Fig.~\ref{actspacedecont_lamost}. Comparably to Fig.~\ref{thebigplot}, regardless of the method employed and the survey used to decontaminate PGS, we notice a high fraction of stars with circular-prograde-planar orbits, and also prograde-planar orbits in the decontaminated PGS. 

\section{VMP stars on disc-like orbits}

After decontaminating PGS in the $V_\phi$ - [Fe/H] space and in the action space, regardless of the survey used there are strong hints indicating the presence of a population with the following characteristics:
\begin{itemize}[leftmargin=*]
\item[--] high $V_\phi$ (> 180 km.s$^{-1}$);
\item[--] prograde-planar orbits (10 $\%$ of the PGS VMP sample in the rotated action space, between 0.2 and 0.7, regardless of the survey used for decontamination);
\item[--] metallicities below -1.7 dex.
\end{itemize}

In this section, we aim to provide a kinematical and dynamical characterisation of the highly prograde\footnote{In the following, we refer to highly prograde stars as stars with $V_\phi$ > 180 km.s$^{-1}$.} and planar population ($Z_{\text{max}}$ < 1.5 kpc), which makes up for 2$\%$ of the PGS VMP sample, and to constrain its lower limit in [Fe/H], with special attention to test whether this is a tail of the MDF of known stellar populations (thin and thick disc, prograde halo). 

\subsection{Trends in the spatial distribution}
\label{trends_spatialdist}

\begin{figure*}
    \sidecaption
        \includegraphics[width=12cm]
        {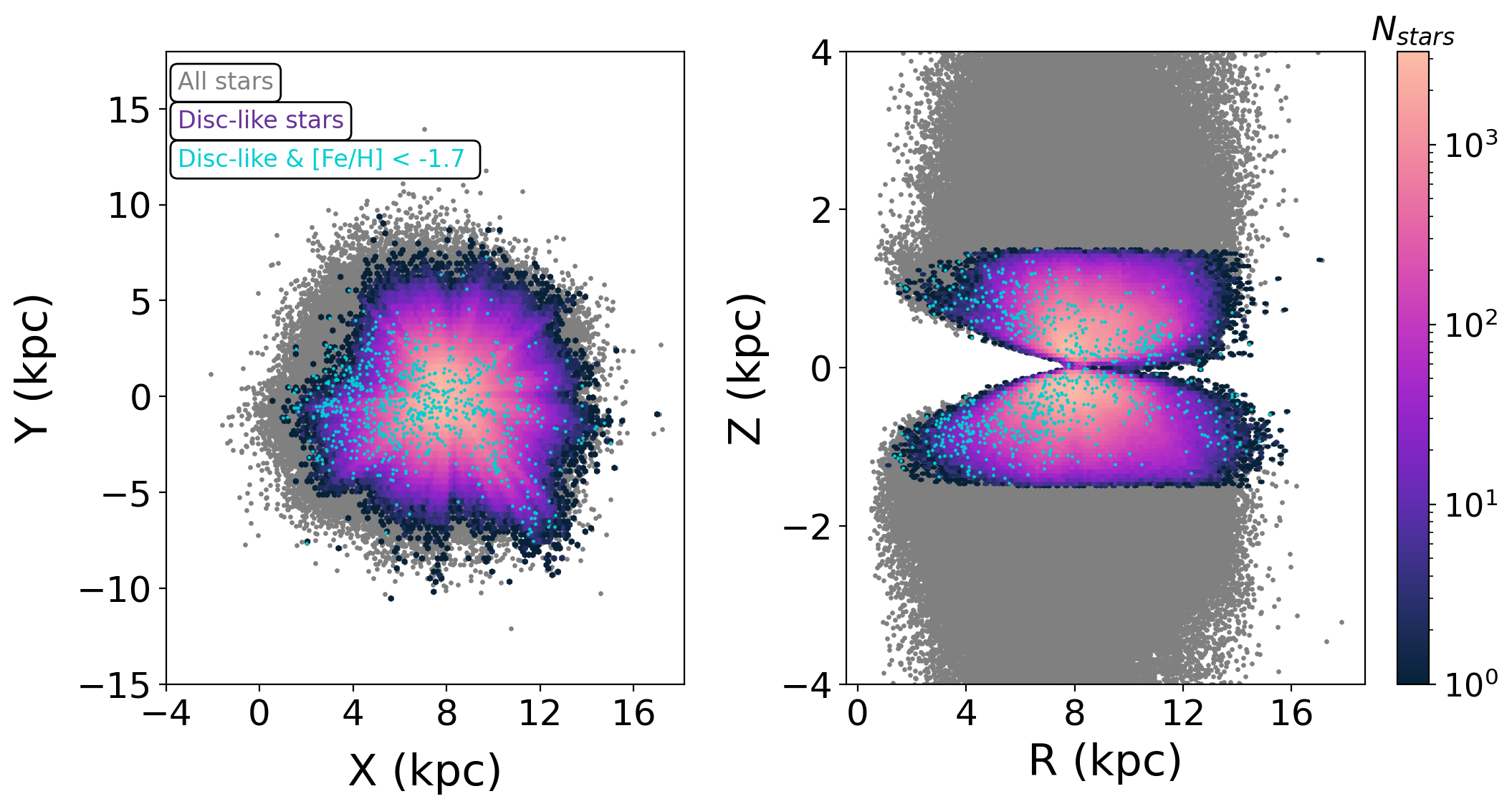}
        \caption{Spatial distributions. Grey scatter plot: same as in Fig.~\ref{spatialdistpgs}. Density plot: disc-like stars ($V_\phi$ > 180 km.s$^{-1}$ and $Z_{\text{max}}$ < 1.5 kpc). Teal scatter plot: disc-like VMP stars. Left panel: X versus Y. Right panel: R versus Z.}
        \label{spatialdist_pgs_subsamples}
\end{figure*}

\subsubsection{X-Y and R-Z}
\label{xyrz}

First, we investigated the spatial distribution of the isochrone-filtered PGS\footnote{In Sec.~\ref{xyrz} and ~\ref{sec_rmax_zmax}, we neglected the contamination of the PGS metal-poor sample, as we have shown that for metallicities below -1.7, it is not a dominant effect on our sample, and because we do not expect contamination to impact directly the spatial and $Z_{\text{max}}$ - $R_{\text{max}}$ distributions examined in these sections.} in Fig.~\ref{spatialdist_pgs_subsamples}. As previously noted in Sec.~\ref{isofit} with Fig.~\ref{spatialdistpgs}, there is a visual asymmetry in the X-Y and R-Z distribution of PGS, where stars are mostly located in the inner parts of the Galaxy. 
In Fig.~\ref{spatialdist_pgs_subsamples}, we show the full PGS with a selection of disc-like stars, that is, stars with $V_\phi$ > 180 km.s$^{-1}$ and $Z_{\text{max}}$ < 1.5 kpc (density plot). We also plot a subsample of disc-like stars with [Fe/H] < -1.7 dex (teal scatter).
We note that, as expected from our cut in $Z_{\text{max}}$, the R-Z distribution of the disc-like population is concentrated very close to the plane.

We can divide the R(X)-Z plane (right panel) into four quadrants to evidence any asymmetry in the full PGS sample, the disc-like subset and the VMP disc-like subset. First, we looked for asymmetries in Y (left panel); overall, the fraction of stars with Y > 0 kpc is relatively constant for the full PGS sample, its disc-like subset (both 43 $\%$), and the disc-like subset (38 $\%$) for which we evidenced a non-negligible asymmetry with negative Y stars.

In the R-Z plane, we did not evidence strong asymmetries in R for the full PGS sample (52 $\%$ of the sample lies below R = 8 kpc) nor Z (45 $\%$ above 0 kpc.) The disc-like subset also does not show strong asymmetries in R (47 $\%$ of the subset below 8 kpc) nor Z (also 45 $\%$ above 0 kpc.). We conclude that the asymmetries identified in Fig.~\ref{spatialdistpgs} are simply probable visual effects from the \textit{Gaia} scanning law. 
We notice a pronounced asymmetry in R and Z for the VMP disc-like subset, with 65 $\%$ located below R = 8 kpc, and 42 $\%$ with positive Z. When checking R and Z simultaneously, 29 $\%$ of disc-like stars have R < 8 kpc and Z < 0 kpc, while for the VMP disc-like subsample it represents 41 $\%$, a value consistent with \cite{thomas19}. 
These observations could be hints of a disc population with a shorter scale length than the MW thin disc, thus reminiscent of the thick disc \citep{bensby11,bovy12,hayden15}, or the flaring of the disc \citep{thomas19}. It could also correspond to halo stars, for which the density is higher in the inner Galaxy \citep{rix22}.

\begin{figure*}
    \centering
    \includegraphics[width=\textwidth, center]{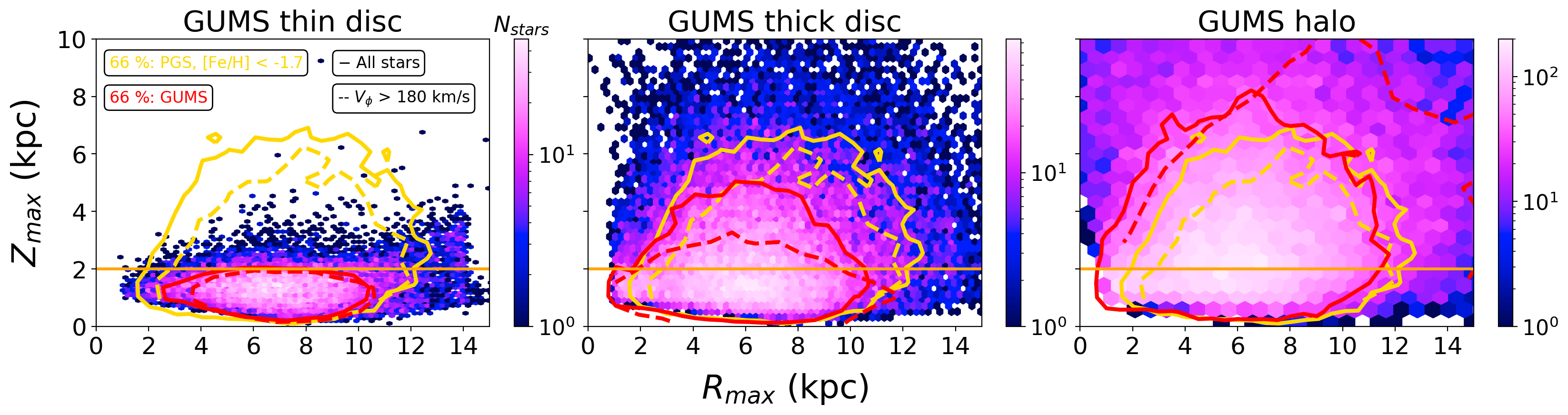}
    \caption{$Z_{\text{max}}$ as a function of $R_{\text{max}}$. The density plot corresponds to GUMS stars that follow the selection function of the PGS VMP sample (below [Fe/H] = -1.7) in $\ell$, \textit{b}, $\varpi$ and apparent G magnitude. The red solid line corresponds to the 66 $\%$ contour line of each GUMS sample, while the yellow solid line is the 66 $\%$ contour line of PGS VMP stars ([Fe/H] < -1.7). The red dashed line corresponds to the 66 $\%$ contour line of each highly prograde ($V_\phi$ > 180 km.s$^{-1}$) GUMS subsample; the yellow dashed line is the 66 $\%$ contour line of PGS VMP highly prograde stars. The orange solid horizontal line corresponds to $Z_{\text{max}}$ = 2 kpc. Left panel: GUMS thin disc. Middle panel: GUMS thick disc. Right panel: GUMS halo.}
    \label{rmax_zmax}
\end{figure*}

\subsubsection{$Z_{\text{max}}$ - $R_{\text{max}}$}
\label{sec_rmax_zmax}

Following \cite{haywood18}, \cite{dimatteo20}, and \cite{koppelman21}, we then probed the $Z_{\text{max}}$ - $R_{\text{max}}$ plane. Here, $R_{\text{max}}$ corresponds to the square root of the difference between the apocenter cylindrical radius, $R_{\text{apo}}$, and $Z_{\text{max}}$: \\ $R_{\text{max}}$ = $\sqrt{R_{\text{apo}}^{2} - Z_{\text{max}}^{2}}$. We used this space to better constrain the bulk of the distribution of each GUMS Galactic component (thin disc, thick disc, halo), in order to compare it with PGS, below [Fe/H] = -1.7. 
To achieve this, we selected for each PGS star, the nearest neighbour from GUMS in longitude $\ell$, latitude \textit{b}, $\varpi$ and apparent G magnitude simultaneously. We allowed the procedure to select several times the same GUMS star if it is the closest match. We obtained:
\begin{itemize}[leftmargin=*]
\item[--] 13 $\%$ of nearest neighbour repeats for the GUMS halo sample;
\item[--] 12 $\%$ of repeats for the GUMS thick disc sample;
\item[--] 53 $\%$ of repeats for the GUMS thin disc sample.
\end{itemize} 

Interestingly, the amount of repeats in each component reflects the difference between the spatial distribution of our PGS metal-poor sample, and that of a pure halo, a pure thin or thick disc. 
In fact, the observed spatial distribution of our PGS metal-poor sample does not resemble a classical thin disc, as seen in Fig.~\ref{rmax_zmax}; thus, forcing a matching selection in $\ell$, \textit{b}, $\varpi$ and G magnitude results in a larger fraction of duplicates. In the following, we only worked with unique stars from our GUMS samples, therefore the GUMS halo, thick and thin disc samples are respectively 13, 12 and 53 $\%$ smaller than our PGS VMP sample.

Figure~\ref{rmax_zmax} compares the distribution of PGS stars below \\-1.7\,dex with the three GUMS Galactic components convolved by the selection function as described above. For legibility, the 66 $\%$ contour lines of GUMS are plotted in red solid lines, and the 66 $\%$ contour line of the PGS sample is plotted in yellow solid lines. We also added the respective contour lines of each highly prograde subsample ($V_\phi$ > 180 km.s$^{-1}$) in red and yellow dashed lines. The orange solid horizontal line corresponds to a constant $Z_{\text{max}}$ = 2 kpc.
To add more meaning to this representation, Table~\ref{frac_rmax_zmax} shows the fraction of stars located below $Z_{\text{max}}$ = 2 kpc, for each GUMS component as well as PGS (in different [Fe/H] intervals) for the full samples and their highly prograde subsets.

The left panel displays the density of the GUMS thin disc. 
As indicated by the orange solid line defined as $Z_{\text{max}}$ = 2 kpc, the bulk of the distribution is located below 2 kpc\footnote{$Z_{\text{max}}$ = 2 kpc may seem high for a classical thin disc component such as the one assumed in GUMS; we verified that this is largely due to applying the PGS VMP sample selection function, preferentially selecting stars at low R and high Z, which in turn enter together with the orbits of the stars, to produce $Z_{\text{max}}$ and $R_{\text{max}}$, eventually affecting the distribution in that space.}, whether we check the full sample (84 $\%$) or solely highly prograde stars (89 $\%$). The overall $Z_{\text{max}}$ - $R_{\text{max}}$ distribution of PGS shows that the main 66 $\%$ of the sample is constrained on scale heights $Z_{\text{max}}$ < 7 kpc. Similarly, $R_{\text{max}}$ is contained within 2 and 13 kpc. If we consider the full PGS sample, 22 $\%$ fall below $Z_{\text{max}}$ = 2 kpc. As for the highly prograde subsample, this fraction increases to 33 $\%$ while the 66 $\%$ contour line still rises to $Z_{\text{max}} \sim$ 6 kpc, showing that the PGS distribution is distinct (more extended in $Z_{\text{max}}$) from that of the GUMS thin disc.

The middle panel shows the density plot of the GUMS thick disc. The 66 $\%$ GUMS contour lines are higher in $Z_{\text{max}}$ than in the previous panel. Interestingly, the 66 $\%$ PGS contour line for all PGS VMP stars seems to match the outer contours ($\sim 90 \%$) of the GUMS thick disc. We also notice that the contour line of the highly prograde PGS VMP subsample coincides with the contour line of the GUMS total thick disc. In fact, we found that 38 $\%$ of GUMS thick disc stars lie below $Z_{\text{max}}$ = 2 kpc, which is only 5 $\%$ more than the PGS VMP prograde subsample, a hint that the thick disc might significantly contribute to this population.

Finally, the right panel shows the density plot of the GUMS halo. By contrast to the previous panels, both 66 $\%$ contour lines match with each other when considering the full samples. However, we note that the contour lines of the GUMS halo extend to higher values of $Z_{\text{max}}$, in particular for $R_{\text{max}} \leqslant$ 8 kpc, where PGS VMP stars have significantly lower $Z_{\text{max}}$ than GUMS halo stars; furthermore, the highly prograde GUMS halo extends to $R_{\text{max}}$ > 15 kpc (presumably including stars with very large $R_{\text{apo}}$), which is not observed in the PGS sample. 

In summary, within the limitation of GUMS (i.e. a mock model with no additional substructures other than a prograde or dual halo, a thick disc and a thin disc), when no selection on $V_\phi$ is made, the comparison between GUMS components and PGS leads to the conclusion that the best match in fraction is the halo. 
Simulating self-consistently a prograde halo (or a dual halo with one prograde component) is beyond the scope of this paper, although we have verified that a simple shift of $V_\phi$ in GUMS does not modify the $Z_{\text{max}}$ - $R_{\text{max}}$ distribution.
We therefore confirm that the majority of the PGS sample would be composed of stars with a halo $Z_{\text{max}}$ - $R_{\text{max}}$ distribution, and halo kinematics, as we show in the following sections. However, when focusing only on the highly prograde subsample, the fraction of PGS stars below $Z_{\text{max}}$ = 2 kpc regardless of the [Fe/H] interval, is twice higher than the corresponding GUMS halo fraction, and resembles more the fraction seen in the GUMS thick disc. This is a hint that the spatial distribution and orbital configuration of the highly prograde PGS metal-poor population is consistent with that of a kinematically warm Galactic disc (possibly even slightly warmer than the canonical GUMS thick disc). In the next section, we investigate the $V_\phi$ distribution and the action space of the most metal-poor PGS stars, after decontaminating with spectroscopic surveys, to further characterise the kinematics and orbital configuration of this population.

\begin{table*}[h]
    \caption{Fraction of stars below $Z_{\text{max}}$ = 2 kpc, for each GUMS component, and for PGS in different [Fe/H] intervals (all < -1.7 dex).}
    \centering
    \begin{tabular}[t]{ccc}
    \hline
    Sample & $Z_{\text{max}}$ < 2 kpc & $Z_{\text{max}}$ < 2 kpc out of the highly\\
     & out of all stars ($\%$) & prograde ($V_\phi$ > 180 km.s$^{-1}$) sample ($\%$)\\
    \hline
    GUMS thin disc & 84 &  89 \\ 
    GUMS thick disc &  38 &  57 \\ 
    GUMS halo & 18 &  14 \\ 
    PGS, [Fe/H] < -1.7 & 22 &  33 \\
    PGS, -2.0 < [Fe/H] < -1.7 & 24 &  38 \\
    PGS, -2.5 < [Fe/H] < -2.0 & 21 &  30 \\
    PGS, -4.0 < [Fe/H] < -2.5 &  18 &  26 \\
    \hline
    \end{tabular}
    \tablefoot{Each GUMS subset follows the selection function of the PGS VMP sample (below [Fe/H] = -1.7) in $\ell$, \textit{b}, $\varpi$ and apparent G magnitude. The fractions are given for full samples, and for highly prograde subsets ($V_\phi$ > 180 km.s$^{-1}$).}
    \label{frac_rmax_zmax}
\end{table*}

\subsection{Asymmetries in the decontaminated PGS}
\label{discussion_folding}

\subsubsection{$V_\phi$ distribution}

To identify and quantify a potential contribution from the halo to our population of interest, an empirical test is to fold the $V_\phi$ distribution after decontaminating PGS. 
This enables to visualise any remaining prograde count after subtracting retrograde counts. But, folding along an axis of symmetry centred on 0 amounts to assuming that the halo has zero net rotation, which is still an open debate \citep{morrison90,deason11,deason17,tian19,fernandez21}. 

\begin{figure}
    \centering
    \includegraphics[width=0.8\linewidth]{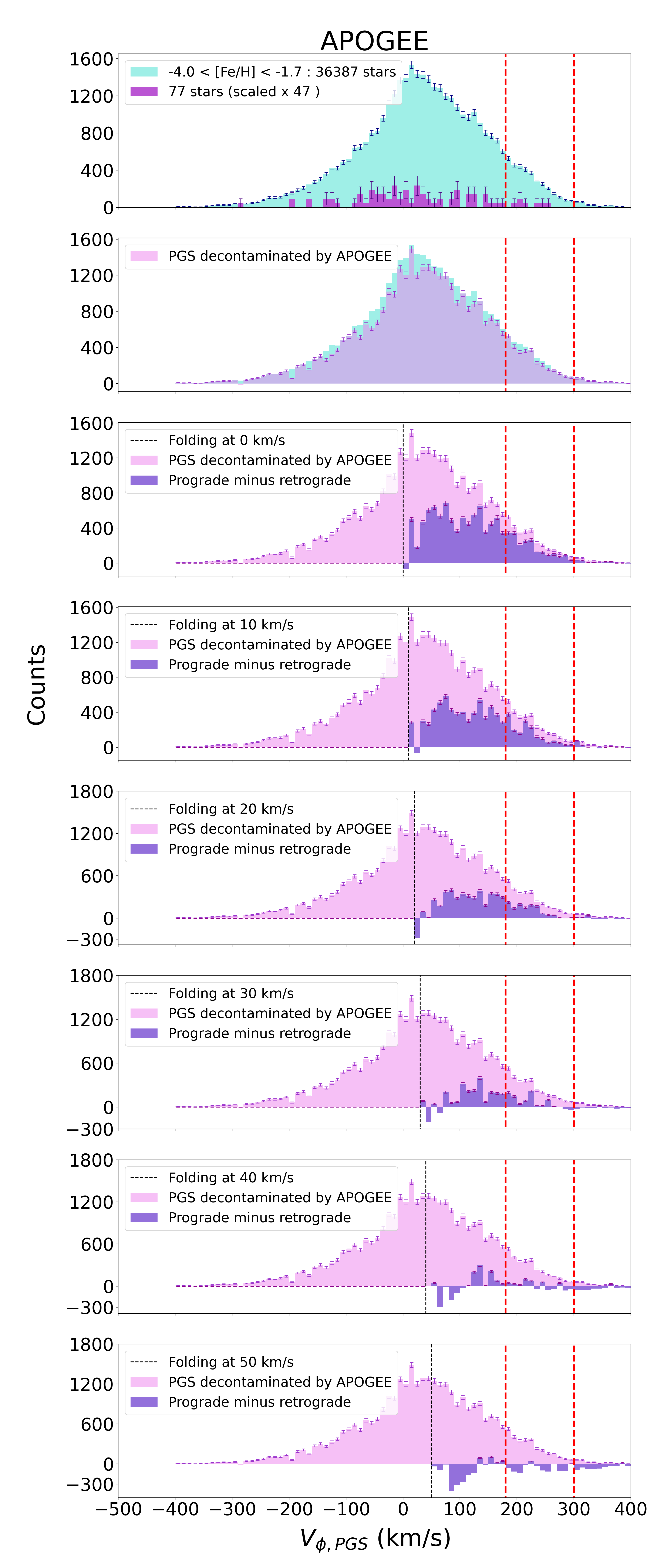}
    \caption{$V_\phi$ distribution for stars with PGS metallicities between -1.7 and -4.0. From top to bottom: panels 1 and 2: PGS decontamination by APOGEE. Panels 3-8: folding of the retrograde counterparts over the prograde counterparts of the decontaminated PGS. The black dashed line indicates the location of our presumed halo. In panel 3, the folding is done assuming a theoretical halo. In the remaining panels, it is done assuming a prograde halo shifted by 10 km.s$^{-1}$ at every panel. The red dashed lines delimit the location of the population we aim to characterise.}
    \label{folding_vphi}
\end{figure}

Therefore, we investigated the folded $V_\phi$ distribution below -1.7 dex with increasing halo velocity central value, with the aim of finding the limit for which the highly prograde population is entirely removed; this is summarised in Fig.~\ref{folding_vphi}. Because the results for LAMOST are very similar, we show them in Fig.~\ref{folding_vphi_lamost}. 
The first two (top) panels show the decontamination by APOGEE in the $V_\phi$ distribution, similar to the pixel-by-pixel decontamination detailed in Sec.~\ref{decontamination}.  
The third panel shows the folding in the case where we assume a stationary halo, centred at 0 km.s$^{-1}$. The red dashed lines delimit the $V_\phi$ at which we expect our population to be (180 - 300 km.s$^{-1}$). 
In both decontaminated and folded PGS $V_\phi$ distributions, high $V_\phi$ stars are present in non-negligible numbers (maximum counts $\sim$ 250 for APOGEE and 200 for LAMOST). This means that removing the entire contribution from halo stars when the halo is stationary is not sufficient to suppress highly prograde VMP stars. We note that stars with thick disc-like $V_\phi$ ($\sim$ 100 km.s$^{-1}$) are also present, in high numbers. 
In the following fourth to the eighth panels, we progressively shift the central value of the halo by 10 km.s$^{-1}$ until we reach 50 km.s$^{-1}$. The value of 30 km.s$^{-1}$ (panel 6) simultaenously minimises the global counts and does not over-correct from the prograde component (negative structure) at high $V_\phi$ (180 - 300 km.s$^{-1}$), both for APOGEE and LAMOST. If we do not account for negative structure at high $V_\phi$, the value of 40 km.s$^{-1}$ (panel 7) is the one that globally minimises the prograde population of our sample for both spectroscopic surveys\footnote{We verified that the results regarding the folding in the action space were not affected by a choice of either value.}. Therefore, we conclude that a single-component halo would need to be prograde with $\overline{V_\phi}$ of at least 30 - 40 km.s$^{-1}$ for the population of VMP highly prograde stars to be considered as the high-rotation tail of such a halo.

This value is in qualitative agreement with \cite{zhang23}, who work with a sample of stars with \textit{Gaia} RVs and \textit{Gaia} DR3 XP spectra-based metallicities from \cite{andrae23}. Their approach is to fit three-dimensional Gaussian Mixture Models (GMM) in a ($V_\phi$, $V_R$, $V_Z$) space in bins of decreasing [Fe/H], 
from which they consistently find a dual halo; one that is stationary and one that rotates. If we combine the mixture and associated $V_\phi$, the qualitative agreement becomes quantitative at -2.0 < [Fe/H] < -1.6 (24 $\%$ of stationary halo rotating at -8 km.s$^{-1}$ and 60 $\%$ of prograde halo rotating at $\sim$ 72 km.s$^{-1}$ give an average rotation of 40 km.s$^{-1}$) and can also be verified below -2.0 dex, where only 28 $\%$ of their population rotates. Taken individually, their prograde halo below -1.6 dex (two most metal-poor bins) rotates $\sim$ 30 - 40 km.s$^{-1}$ faster (72 - 80 km.s$^{-1}$) than the one we infer here (30 - 40 km.s$^{-1}$). 
This difference in mean rotation is most probably driven by the different approaches of our works. More specifically, our approach has been purely empirical; it is agnostic on the functional form (no assumption of gaussianity) and does not rely on any modelling of the velocity distributions. However, it is not free of hypothesis. First, we assumed that the highly retrograde and prograde tails of the velocity distribution are symmetrical; second, we assumed that the halo is a single component. This latter hypothesis is different from other works, including \cite{belokurov22} and \cite{zhang23} in a similar sample, or \cite{arentsen24} in the inner Galaxy. The difference in the mean rotation of the halo needed to account for the population of highly prograde VMP stars may stem from this: a faster rotation of a less numerous sub-population (i.e. a second halo population) would be needed to erase that same signature. 

\begin{figure*}
    \sidecaption
        \includegraphics[width=12cm]
        {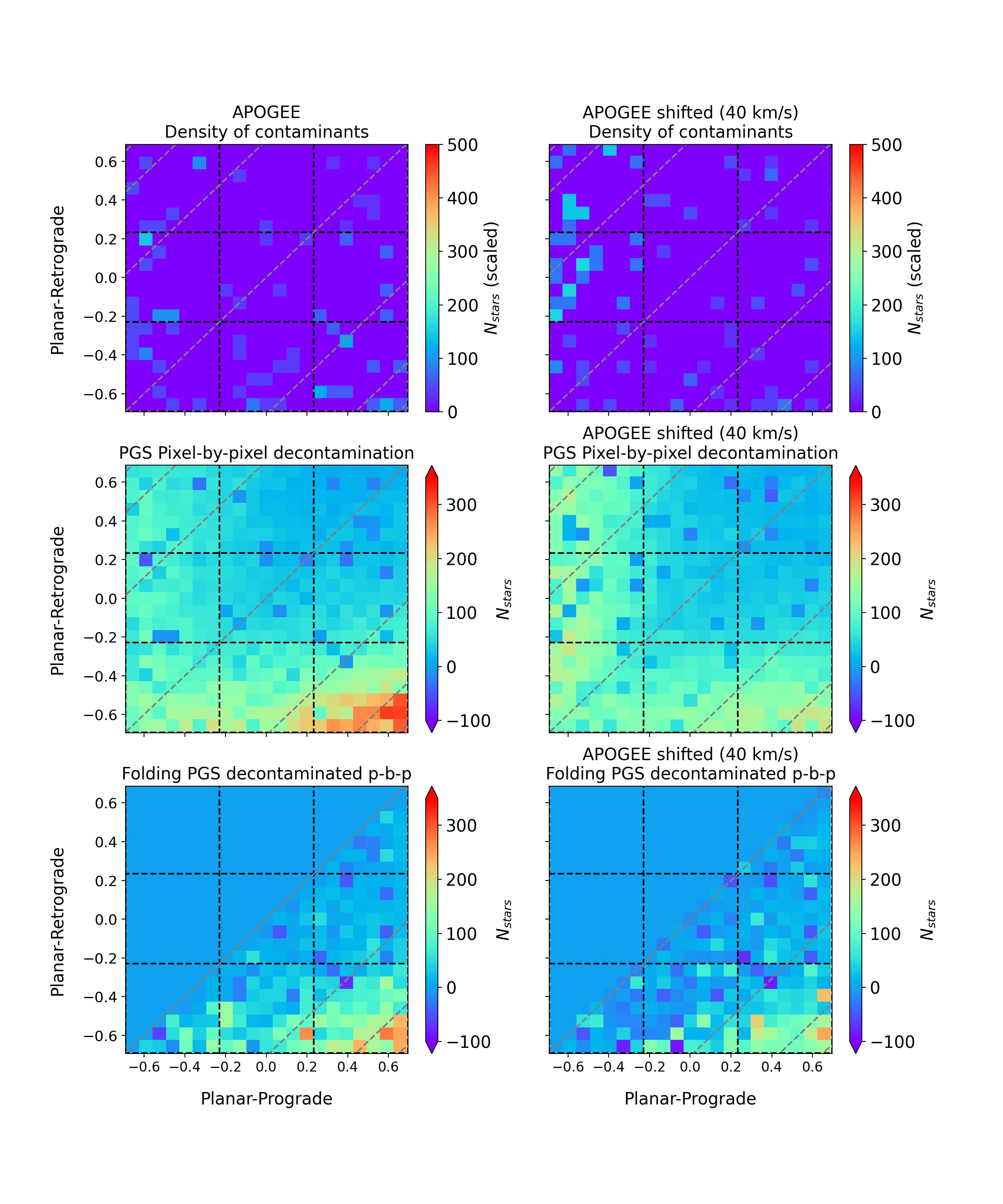}
        \caption{Rotated action space between -4.0 $\leqslant$ [Fe/H] < -1.7. The grey dashed lines correspond to different values of $L_z$/$J_{\text{tot}}$: [-0.95, -0.8, -0.5, 0, 0.5, 0.8, 0.95]. Left column: decontamination process and folding using APOGEE contamination. Top row: density of spectroscopic contaminants. Middle row: pixel-by-pixel decontamination. Bottom row: folding of the retrograde components over the prograde components, for PGS decontaminated pixel-by-pixel. In this space, the folding axis of symmetry is the grey dashed line at $L_z$/$J_{\text{tot}}$ = 0. Right column: same as the left panels, but the three actions were computed shifting the observed $V_\phi$ by 40 km.s$^{-1}$.
        }
        \label{folding_actspace}
\end{figure*}

\subsubsection{Action space}

To visualise how these results translate dynamically, we folded the rotated action space in Fig.~\ref{folding_actspace}. The trends obtained for LAMOST are displayed in Fig.~\ref{folding_actspace_lamost}; they are very close to the observations we make with APOGEE. 
The middle panel of Fig.~\ref{folding_actspace} shows PGS decontaminated by APOGEE, pixel-by-pixel. The decontaminated PGS has a high density of stars with prograde-planar-circular orbits according to the grey dashed lines corresponding to $L_z$/$J_{\text{tot}}$ = 0.8 and 0.95. But this structure is statistically significant along the prograde-planar region, down to the radial area. 
When folding the action space (prograde minus retrograde) with respect to the central diagonal (dashed line at $L_z$/$J_{\text{tot}}$ = 0), we obtain the plot at the bottom left panel of Fig.~\ref{folding_actspace}, which is the equivalent of the third panel of Fig.~\ref{folding_vphi}. 
After folding, pixels with negative values appear below $L_z$/$J_{\text{tot}}$ = 0.5. However, the prograde-planar region is still populated in relatively high counts. In particular, above $L_z$/$J_{\text{tot}}$ = 0.8 the counts are the highest. This is in line with the observations of Fig.~\ref{folding_vphi}. 

We now focus on the right column of Fig.~\ref{folding_actspace}.
To model the effect of a shifted prograde halo in the action space, we recomputed the three action components assuming a shifted $V_\phi$, while $V_R$ and $V_Z$ remain unchanged. To be consistent with the limit value for the presence of VMP highly prograde stars found empirically, we shifted it by 40 km.s$^{-1}$. 
It is important to note that we folded the shifted retrograde counterparts onto the original prograde PGS counts. 
Since our main motivation is to get an idea of how the original action space evolves with a shifted halo component, we compared:
\begin{itemize}[leftmargin=*]
\item[--] the folding of the original action space, that is, the subtraction of the original retrograde component from the original prograde component;
\item[--] the folding of the original action space, but replacing the original retrograde component by the shifted (by 40 km.s$^{-1}$) retrograde component (the prograde component remains the same).
\end{itemize}

That is, we subtracted the upper triangle matrix (with respect to the diagonal at $L_z$/$J_{\text{tot}}$ = 0) representing the halo shifted by 40 km.s$^{-1}$ (middle right panel of Fig.~\ref{folding_actspace}) from the lower triangle matrix (with respect to the diagonal at $L_z$/$J_{\text{tot}}$ = 0) representing the decontaminated PGS prograde counts (middle left panel of Fig.~\ref{folding_actspace}).
Therefore, the result of the folding in the lower right panel should be treated very cautiously at $L_z$/$J_{\text{tot}}$ = 0 because there is no physical meaning in matching the original PGS prograde counts with counts that are also prograde but were shifted to 0.
As expected from the $V_\phi$ distribution of Fig.~\ref{folding_vphi}, the overall star counts in the residuals, in the bottom right panel of Fig.~\ref{folding_actspace}, are close to 0, which verifies our methodology.
In spite of that, subtracting retrograde stars from our prograde sample has left a significant signature in the action space. While VMP stars with polar orbits are subtracted in excess (at all $L_z$/$J_{\text{tot}}$ < 0.8 or even higher), stars with planar and circular orbits remain (where $L_z$/$J_{\text{tot}}$ > 0.95 and stars with $L_z$/$J_{\text{tot}}$ > 0.8 which are slightly polar). This indicates that although the rotational signature of these VMP stars with high $V_\phi$ can be explained by a prograde halo, their orbits are more radial and less polar than those of our empirical rotating halo. However, we might reach the limits of our method, since only the $V_\phi$ are arbitrarily shifted, which is not the case for the other velocities (E is not conserved anymore).
From this dual view on the $V_\phi$ distribution and the action space, we conclude that if the halo is shifted by 40 km.s$^{-1}$, the remaining population is mostly planar-prograde, with $V_\phi$ of the order of 100-150 km.s$^{-1}$, which is compatible with a metal-poor thick disc \citep{kordo13}. Besides, we can highlight a clear asymmetry in the distribution of the prograde and retrograde stars at low [Fe/H], in favour of the prograde counterparts.

\subsection{Insights on the [Fe/H] lower boundary}

\begin{figure*}
    \centering
    \includegraphics[width=\linewidth]{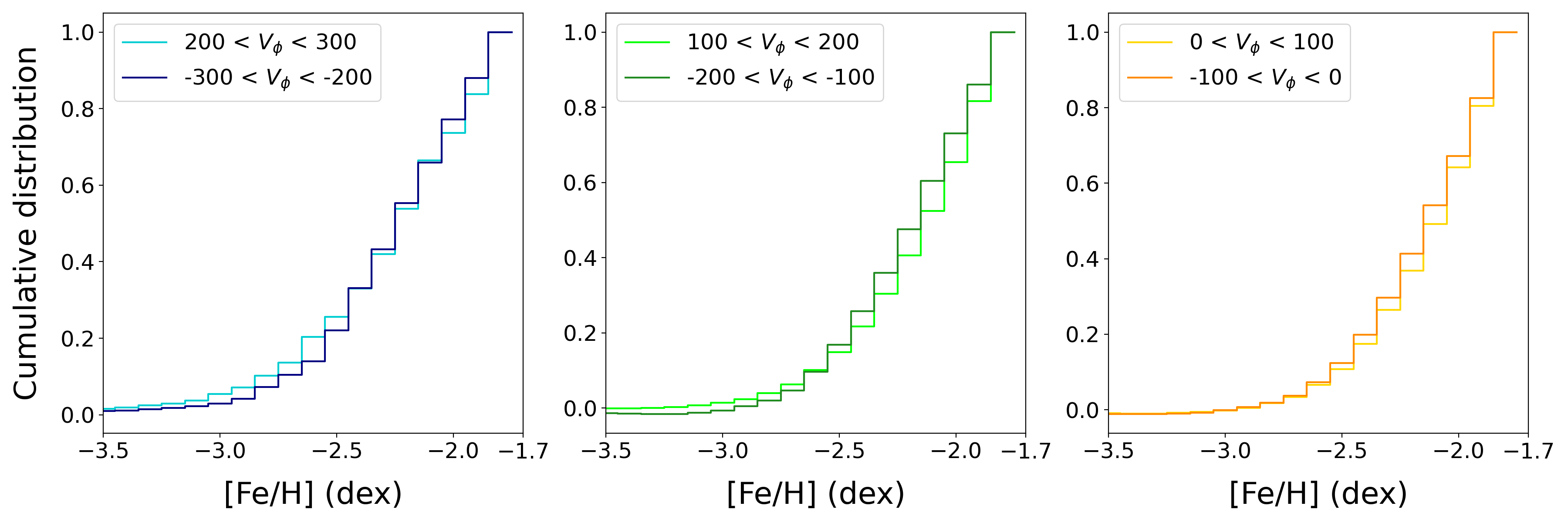}
    \caption{Cumulative [Fe/H] distribution of the normalised counts of the decontaminated PGS, in different $V_\phi$ intervals. Left panel: 200 < |$V_\phi$| < 300 km.s$^{-1}$. Middle panel: 100 < |$V_\phi$| < 200 km.s$^{-1}$. Right panel: 0 < |$V_\phi$| < 100 km.s$^{-1}$.}
    \label{histvphi_projfeh_cumul}
\end{figure*}

\begin{figure}
    \centering
    \includegraphics[width=\linewidth]{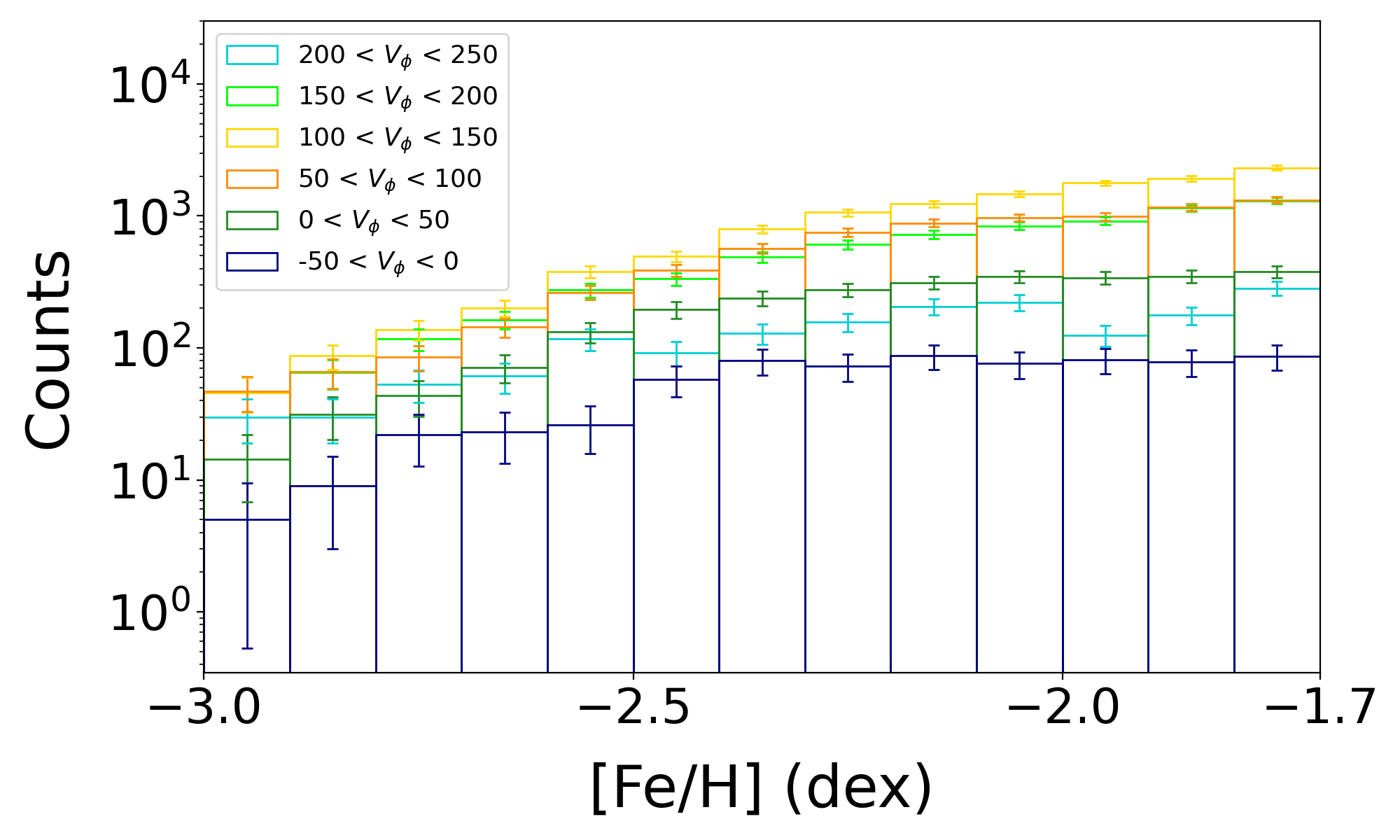}
    \caption{[Fe/H] distribution of the decontaminated PGS, below -1.7 dex for different ranges of $V_\phi$.}
    \label{histvphi_projfeh_zoomin}
\end{figure}

The lower limit in [Fe/H] of our sample can be constrained by comparing its MDF in different $V_\phi$ slices, which is simply another view of the bottom left panel of Fig.~\ref{thebigplot}. We note that this particular panel takes into account PGS decontaminated by the mean of APOGEE contaminants. We plotted the cumulative distribution of metallicities below -1.7 dex in Fig.~\ref{histvphi_projfeh_cumul}. The left panel shows high |$V_\phi$| stars, the middle panel shows intermediate |$V_\phi$| stars and the right panel shows stars with |$V_\phi$| close to 0. 
Regardless of the panels, between -1.7 and -2.5 dex, there are globally as many highly prograde (> 200 km.s$^{-1}$) stars as highly retrograde (< -200 km.s$^{-1}$), if not more in the middle panel. Yet, below \\-2.5 dex the trend changes in favour of the prograde cumulative counts, which seem to dominate down to the lowest metallicities. This is especially visible for high |$V_\phi$| stars (left panel). We draw similar conclusions when comparing the intermediate prograde and retrograde distributions. 

To get a better sense of the prograde-retrograde asymmetry, we could check their ratios as a function of [Fe/H]. However, below -3 dex we over-decontaminate PGS (as discussed in Sec.~\ref{decont_vphi}); the resulting counts are negative, which is unsuited for determining these ratios. Therefore, we can only confidently check the prograde-retrograde ratio down to -3 at most. We compared three categories of prograde and retrograde populations: the high $|V_\phi|$ (> 200 km.s$^{-1}$), the intermediate $|V_\phi|$ ($\sim$ 150 km.s$^{-1}$), and the small $|V_\phi|$ ($\sim$ 50 km.s$^{-1}$). We found:
\begin{itemize}[leftmargin=*]
\item[--] [0 ; +100] versus [0 ; -100] km.s$^{-1}$: the prograde-to-retrograde ratio is of the order of 1.5 in favour of the prograde;
\item[--] [+100 ; +200] versus [-100 ; -200] km.s$^{-1}$: the ratio is $\sim$ 2.5 in favour of the prograde; 
\item[--][+200 ; +300] versus [-200 ; -300] km.s$^{-1}$: the ratio is $\sim$ 2.6 in favour of the prograde. 
\end{itemize}
If we solely check the original distribution (Fig.~\ref{histvphi_projfeh_zoomin}) below -2.5 dex, assuming a 2$\sigma$ error bar based on Poisson noise, the highly prograde population (> 200 km.s$^{-1}$) is reaching 0 at -2.9 dex, that is, 1.3 dex lower than the lower limit of the disc population of \cite{zhang23}. However, they do detect highly prograde stars below -2 dex, but they associate them to an underlying halo distribution rather than a separate population.
Since their result is based on GMM, it is possible that our population of highly prograde and planar ($Z_{\text{max}}$ < 1.5 kpc) stars is not detected below -1.6 dex, because it represents only 2 $\%$ of our PGS VMP sample.
Nevertheless, our result is in line with \cite{fernandez24}, who also find this population below -1.5 dex.

\section{Summary and conclusions}

Our goal was to investigate and characterise the kinematics and orbital properties of VMP ([Fe/H] < -1.7) MW stars, with special emphasis on stars with disc-like orbits, namely prograde (in particular highly prograde stars, with $V_\phi$ > 180 km.s$^{-1}$) and planar ($Z_{\text{max}}$ < 1.5 kpc). To achieve this, we combined $\textit{Pristine-Gaia}$ synthetic photometry, \textit{Gaia} astrometry and \textit{Gaia} RVS radial velocities. We successfully filtered PGS from metal-rich outliers with a method based on isochrones in the CMD, and constructed a sample of $\sim$ 3M stars with -4 < [Fe/H] < 0.
To further refine the quality of our data, we statistically decontaminated the PGS VMP subsample containing $\sim$ 36 000 stars, using the comparison to the APOGEE and LAMOST spectroscopic surveys, through a criterion related to the difference between photometric and spectroscopic metallicity estimates. We found that contamination has minimal impact on the kinematical properties of the VMP sample.
Globally, we recovered the classical halo trends in orbital $Z_{\text{max}}$ - $R_{\text{max}}$, where the halo prediction of GUMS matches the bulk of our sample distribution, and in the action space, where a large fraction of our data (90 $\%$ between -1.7 < [Fe/H] < -1.0 and 66 $\%$ below -1.7) is distributed along the retrograde area.

Yet, similarly to \cite{sestito20}, \cite{bellazzini23}, and \cite{viswanathan24_pgs} (see their Fig. 7), we evidenced an asymmetry between prograde and retrograde-planar components down to the most metal-poor values. We confirmed that stars on prograde-planar orbits are present well below -1.7 dex, that is, 1.0 dex below the typical limit suggested for that kind of population (e.g. \citealt{bensby14,fuhrmann17}). This prograde-planar subset represents 10 $\%$ of the PGS VMP population, in the rotated action space of the decontaminated PGS between 0.2 and 0.7 (see Fig.~\ref{actspacedecont} and Fig.~\ref{actspacedecont_lamost}). Comparably to \cite{sestito20}, we observed traces of the VMP highly prograde ($V_\phi$ > 180 km.s$^{-1}$) and planar ($Z_{\text{max}}$ < 1.5 kpc) sample (corresponding to a subset of 2 $\%$ of the total VMP sample) down to -2.9 dex with a 2$\sigma$ confidence interval, in a sample four times larger.
Below a metallicity of -3 dex, our analysis of the PGS VMP sample is limited simultaneously by the lack of reference spectroscopic values in APOGEE, LAMOST, GALAH, and GSP-spec needed to properly perform the decontamination, less precise PGS metallicities, and overall low statistics to detect strong signal.

The main findings from the orbital characteristics of this population can be summarised as follows:
\begin{itemize}[leftmargin=*]
\item[--] the GUMS thin disc in the action space (Fig.~\ref{gums_diamondplot}) and the $Z_{\text{max}}$ - $R_{\text{max}}$ distribution (Fig.~\ref{rmax_zmax}) is essentially concentrated towards the Galactic plane and displays circular orbits, which greatly differs with the representation of our VMP highly prograde and planar population; 
\item[--] the PGS VMP highly prograde and planar sample is more concentrated towards the inner MW and at larger Z than the higher-metallicity prograde-planar stars in the spatial distribution (Fig.~\ref{spatialdist_pgs_subsamples}), compatible either with a short-scaled thick disc, or with the metal-poor inner Galaxy population of \cite{rix22} and \cite{arentsen24}; 
\item[--] the PGS VMP highly prograde and planar sample is more concentrated towards the Galactic plane ($Z_{\text{max}}$ < 2 kpc) than the full VMP sample (33 $\%$ versus 22 $\%$), while the classical GUMS halo shows significantly lower concentrations (14 to 18 $\%$). This calls for a contribution from a thick disc-like population, which in GUMS shows a similar concentration below $Z_{\text{max}}$ = 2 kpc (Fig.~\ref{rmax_zmax});
\item[--] the PGS VMP highly prograde and planar sample cannot solely be part of a canonical halo, as shown in the $V_\phi$ distribution (residuals in the sixth panel of Fig.~\ref{folding_vphi} $\sim$ 150 km.s$^{-1}$) and the action space (residuals in the bottom panels of Fig.~\ref{folding_actspace} and Fig.~\ref{folding_actspace_lamost}). To remain agnostic on distribution functional forms and models, we assumed that the halo is a single component, with symmetrical prograde and retrograde tails of velocity distribution. When folding the retrograde over the prograde distribution of our sample at the lowest metallicities, we found that a stationary halo was not enough to suppress highly prograde VMP stars, while a prograde halo centred at 30 or 40 km.s$^{-1}$ potentially could. 
\end{itemize}

These characteristics show that this population is a combination of 
some (possibly prograde) halo contribution and some thick disc-like component, reminiscent of the metal-weak thick disc \citep{morrison90,kordo13,beers14,kordo17,carollo19}, or the Atari disc of \cite{mardini22}.
Different origins can be considered for this population. 
Using ARTEMIS simulations of MW-mass galaxies, \cite{dillamore24} examine the formation of the MW disc. In fact, they show that an early spin-up anti-correlates with the fraction of accreted stars and correlates with high halo mass. Since the MW has an early spin-up, most stars that formed the disc were most likely already inside the MW at the epoch of spin-up. Within the scope of an early spin-up, \cite{semenov24} use the TNG50 cosmological simulation \citep{pillepich18} and show that the 10$\%$ of their MW analogues that formed discs at an early epoch (more than 10 Gyr ago) had a lower metallicity limit located between -1 and -1.5 dex. 
We note, however, that \cite{sotillo23} use the same suite of simulations to constrain the orbital properties of MW-like galaxies, solely based on their circularity, their stellar mass and their environment, and find that 20 $\%$ of stars with [Fe/H] < -2.0 belong to kinematically cold geometrical thin discs, with some UMP analogues reaching $\sim$ 50 $\%$. They also find that such stars were formed ex situ. In our case, the stars we consider to be disc-like (highly prograde and planar) make up for 2 $\%$ of the PGS VMP sample. Such simulations, in light of our results, suggest that we might be just a few instruments away from uncovering a larger population of these singular stars.
\cite{belokurov22} work with APOGEE and \textit{Gaia} EDR3 data and evidenced Aurora, a low-metallicity in situ component with [Fe/H] $\leqslant$ -1.3 and mild net rotation; such population is in agreement with the prograde halo of \cite{zhang23}, who made use of the XGBoost metallicity sample of \cite{andrae23}. 
The test-particle simulations of \cite{li23}, created with \textit{Gaia} RVS and Pristine DR1 data, reveal that a rapidly decelerating bar can transfer inner Galaxy stars towards very prograde-planar orbits, giving an explanation for the presence of stars with very low metallicities along the Galactic disc. With the same method, this time only using \textit{Gaia} RVS data, \cite{dillamore23} find ridges in phase space interpreted as signatures of bar resonances, causing the formation of substructure in the disc. \cite{arentsen24}, using spectroscopic data from the Pristine Inner Galaxy Survey (PIGS, \citealt{arentsen20}) and distances derived with \verb|StarHorse| \citep{queiroz18}, suggest that within 1.5 < R < 3.5 kpc and below [Fe/H] = -2.0, there could be a mix of a stationary halo component (40 $\%$) and a prograde halo component (60 $\%$). Finally, \cite{li24} study the N-body simulations of barred disc galaxies, with spinning and stationary dark matter halos, obtained with the hydrodynamical code GIZMO \citep{hopkins15}. They propose a mechanism to drive metal-poor stars onto disc orbits consisting of cooling the vertical oscillations of halo stars, leading to a flattening of $Z_{\text{max}}$ and $R_{\text{max}}$.
Another possibility could involve accretion, as discussed in \cite{fiorentin21}. One of the substructures identified, dubbed Icarus, is quite similar to the population investigated in this paper; nonetheless, it could be the remnant of a dwarf galaxy progenitor located on a prograde and low-inclination orbit, according to the study of its chemodynamics, in particular $\alpha$-elements, metallicity and eccentricity. Simulations such as that of \cite{sestito21} point out the possibility of later accretions of disrupting galaxies, permeating the disc with low-metallicity stars.

To conclude, we must derive elemental abundances to disentangle the possible origins of the PGS VMP prograde-planar population. \cite{hawkins15} and \cite{feuillet23}, among other studies, show how powerful chemical tagging is, in particular when investigating the [$\alpha$/Fe]-[Fe/H] and the Mg-Mn-Al-Fe planes, on the one hand to locate the canonical Galactic components, and on the other hand to effectively separate in situ from accreted stars. For example, the work of \cite{fernandez24} (see their Fig. 4 and 5) constrains the thin disc at -1.0 < [Fe/H] < -0.7 and shows that high $V_\phi$ stars with [Fe/H] < -1.0 have trends typical of early chemical enrichment, in opposition with thick disc trends. 
Other studies such as that of \cite{dovgal24} underline the importance of clarifying the chemical enrichment history of metal-poor disc stars, paving the way towards efficient use of upcoming spectroscopic facilities such as WEAVE \citep{weave_firstpaper}, 4MOST \citep{4most_firstpaper}, DESI \citep{desi_firstpaper}, and MOONS \citep{moons_firstpaper}. Future prospects involve the spectral analysis of a statistically robust sample composed of canonical thin and thick disc stars as well as Pristine metal-poor and VMP stars, to confirm their metallicities, analyse, and compare their chemical content. We believe that this may be the missing piece needed to reconstruct a clear history of the Galactic disc.

\begin{acknowledgements}
We thank the anonymous referee for their insightful comments that helped clarify the contents of this paper.
We warmly thank Pedro A. Palicio for useful discussions that helped improve the quality of this paper.
NFM, VH, GK and IGR gratefully acknowledge support from the French National Research Agency (ANR) funded project “Pristine” (ANR-18-CE31-0017). 
GK, VH, FG and IGR gratefully acknowledge support from the French National Research Agency (ANR) funded project MWDisc (ANR-20-CE31-0004). 
AAA acknowledges support from the Herchel Smith Fellowship at the University of Cambridge and a Fitzwilliam College research fellowship supported by the Isaac Newton Trust.
GT and GB acknowledge support from the Agencia Estatal de Investigación del Ministerio de Ciencia en Innovación (AEI-MICIN) and the European Regional Development Fund (ERDF) under grant number PID2020-118778GB-I00/10.13039/501100011033 and the AEI under grant number CEX2019-000920-S.
ES acknowledges funding through VIDI grant "Pushing Galactic Archaeology to its limits" (with project number VI.Vidi.193.093) which is funded by the Dutch Research Council (NWO). This research has been partially funded from a Spinoza award by NWO (SPI 78-411). This research was supported by the International Space Science Institute (ISSI) in Bern, through ISSI International Team project 540 (The Early Milky Way).
SV thanks ANID (Beca Doctorado Nacional, folio 21220489) and the Millennium Nucleus ERIS (ERIS NCN2021017). 
This work has made use of data from the European Space Agency (ESA) mission
{\it Gaia} (\url{https://www.cosmos.esa.int/gaia}), processed by the {\it Gaia}
Data Processing and Analysis Consortium (DPAC,
\url{https://www.cosmos.esa.int/web/gaia/dpac/consortium}). Funding for the DPAC
has been provided by national institutions, in particular the institutions
participating in the {\it Gaia} Multilateral Agreement.

Funding for the Sloan Digital Sky 
Survey IV has been provided by the 
Alfred P. Sloan Foundation, the U.S. 
Department of Energy Office of 
Science, and the Participating 
Institutions. 

SDSS-IV acknowledges support and 
resources from the Center for High 
Performance Computing  at the 
University of Utah. The SDSS 
website is www.sdss4.org.

SDSS-IV is managed by the 
Astrophysical Research Consortium 
for the Participating Institutions 
of the SDSS Collaboration including 
the Brazilian Participation Group, 
the Carnegie Institution for Science, 
Carnegie Mellon University, Center for 
Astrophysics | Harvard \& 
Smithsonian, the Chilean Participation 
Group, the French Participation Group, 
Instituto de Astrof\'isica de 
Canarias, The Johns Hopkins 
University, Kavli Institute for the 
Physics and Mathematics of the 
Universe (IPMU) / University of 
Tokyo, the Korean Participation Group, 
Lawrence Berkeley National Laboratory, 
Leibniz Institut f\"ur Astrophysik 
Potsdam (AIP),  Max-Planck-Institut 
f\"ur Astronomie (MPIA Heidelberg), 
Max-Planck-Institut f\"ur 
Astrophysik (MPA Garching), 
Max-Planck-Institut f\"ur 
Extraterrestrische Physik (MPE), 
National Astronomical Observatories of 
China, New Mexico State University, 
New York University, University of 
Notre Dame, Observat\'ario 
Nacional / MCTI, The Ohio State 
University, Pennsylvania State 
University, Shanghai 
Astronomical Observatory, United 
Kingdom Participation Group, 
Universidad Nacional Aut\'onoma 
de M\'exico, University of Arizona, 
University of Colorado Boulder, 
University of Oxford, University of 
Portsmouth, University of Utah, 
University of Virginia, University 
of Washington, University of 
Wisconsin, Vanderbilt University, 
and Yale University.

This work made use of the Third Data Release of the GALAH Survey (Buder et al. 2021). The GALAH Survey is based on data acquired through the Australian Astronomical Observatory, under programs: A/2013B/13 (The GALAH pilot survey); A/2014A/25, A/2015A/19, A2017A/18 (The GALAH survey phase 1); A2018A/18 (Open clusters with HERMES); A2019A/1 (Hierarchical star formation in Ori OB1); A2019A/15 (The GALAH survey phase 2); A/2015B/19, A/2016A/22, A/2016B/10, A/2017B/16, A/2018B/15 (The HERMES-TESS program); and A/2015A/3, A/2015B/1, A/2015B/19, A/2016A/22, A/2016B/12, A/2017A/14 (The HERMES K2-follow-up program). We acknowledge the traditional owners of the land on which the AAT stands, the Gamilaraay people, and pay our respects to elders past and present. This paper includes data that has been provided by AAO Data Central (datacentral.org.au).

Guoshoujing Telescope (the Large Sky Area Multi-Object Fiber Spectroscopic Telescope LAMOST) is a National Major Scientific Project built by the Chinese Academy of Sciences. Funding for the project has been provided by the National Development and Reform Commission. LAMOST is operated and managed by the National Astronomical Observatories, Chinese Academy of Sciences.

\end{acknowledgements}

\bibliographystyle{aa}
\bibliography{biblio}

%

\begin{appendix} 

\section{PGS quality selection path}

Table~\ref{cuts_pgs_recap} describes the chronological cuts applied to PGS in order to build the sample used for this work. As a reminder, the published PGS contains $\sim$ 52 million stars; our selection represents $\sim$ 6 $\%$ of the original catalogue, corresponding to roughly 3 million stars. 

\begin{table*}[ht]
\centering
\caption{Description of the various cuts applied to the \textit{Pristine-Gaia} synthetic catalogue, resulting in the version used for this work.}
\begin{tabular}[t]{ccc}
\hline
Cut & Number of stars & Percentage removed from PGS ($\%$) \\
\hline
 No cut (published PGS) & 52 300 000 & 0\\
 \textit{Gaia} cuts except \verb|gaiadr3.vari_summary| & 41 612 344 & 20.4 \\
 $(BP-RP)_{0}$ > 0.5 mag and \verb|mcfrac_CaHKsyn| > 0 (no MC draw available) & 39 446 013 & 4.1\\
 $M_{G} < 7.5$ &  39 217 953 & 0.44 \\
 \verb|mcfrac_CaHKsyn| > 0.9 & 37 069 113 & 4.11\\
 Sources with available RVS orbits & 12 777 188 & 46.45\\
  $\text{[Fe/H]}$ > -4.0 & 12 776 653 & 0.001 \\
  $E(B-V)$ < 0.5 & 12 737 484 & 0.08 \\
  \verb|Pvar| < 0.3 and \verb|d_CaHK| < 0.05 &  9 548 136 & 6.10 \\
  \verb|gaiadr3.vari_summary| & 9 450 368 & 0.19\\
  Catalogue of globular and open cluster stars & 9 432 286 & 0.03\\
  Dwarf and giant separation + isochrone filtering & 2 880 338 & 12.53\\
\hline
\end{tabular}
\label{cuts_pgs_recap}
\end{table*}

\section{Isochrone filtering: GALAH, LAMOST, and GSP-spec}
\label{isofit_other}

In the same fashion as in Sec.~\ref{validapo}, we briefly discuss the validation of the isochrone filtering method for GALAH, LAMOST and GSP-spec. Table~\ref{cuts_spectro_recap} summarises the sample size of each cross-match between PGS and spectroscopic surveys before and after applying the isochrone filtering method.

\begin{table}[ht]
    \caption{Sample size of the cross-matches between PGS and relevant spectroscopic catalogues.}
    \begin{tabular}[t]{ccc}
    \hline
    Survey & Before filtering & After filtering \\
     & (giants + dwarfs) &  (only giants) \\
    \hline
     APOGEE & 178 637 & 92 765\\
     GALAH & 42 424 & 14 424 \\
     LAMOST & 2 002 355 & 313 833\\
     \textit{Gaia} RVS (GSP-spec) &  2 078 139 & 1 101 704\\
    \hline
    \end{tabular}
    \tablefoot{The cross-matches are used for the validation of the isochrone filtering method described in Sec.~\ref{validapo}.}
    \label{cuts_spectro_recap}
    \end{table}

\begin{figure}
\centering
\begin{subfigure}{0.45\textwidth}
    \includegraphics[width=\textwidth]{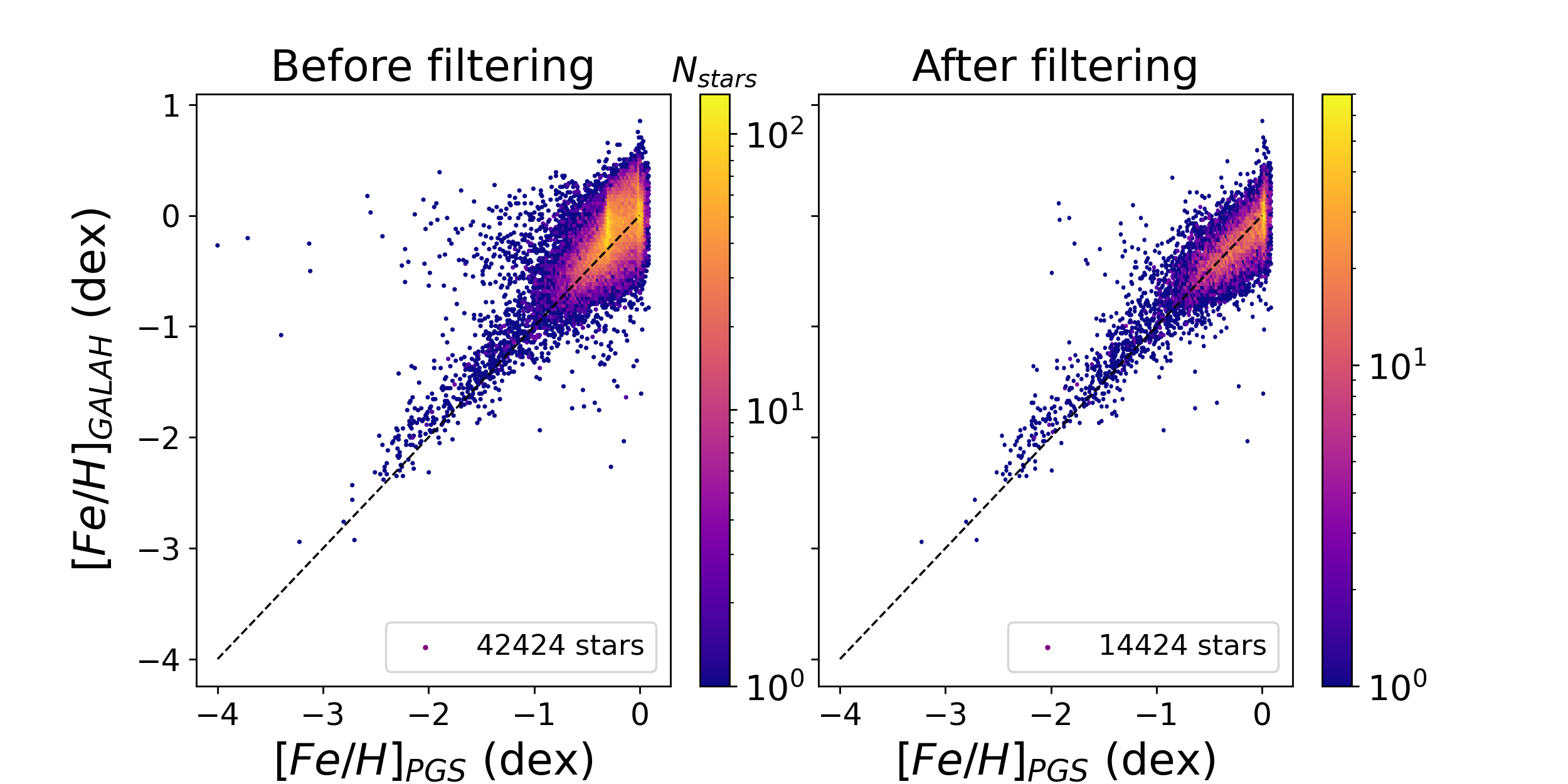}
    {\phantomsubcaption\label{fehfeh_gp}}
\end{subfigure}
\hfill
\begin{subfigure}{0.45\textwidth}
    \includegraphics[width=\textwidth]{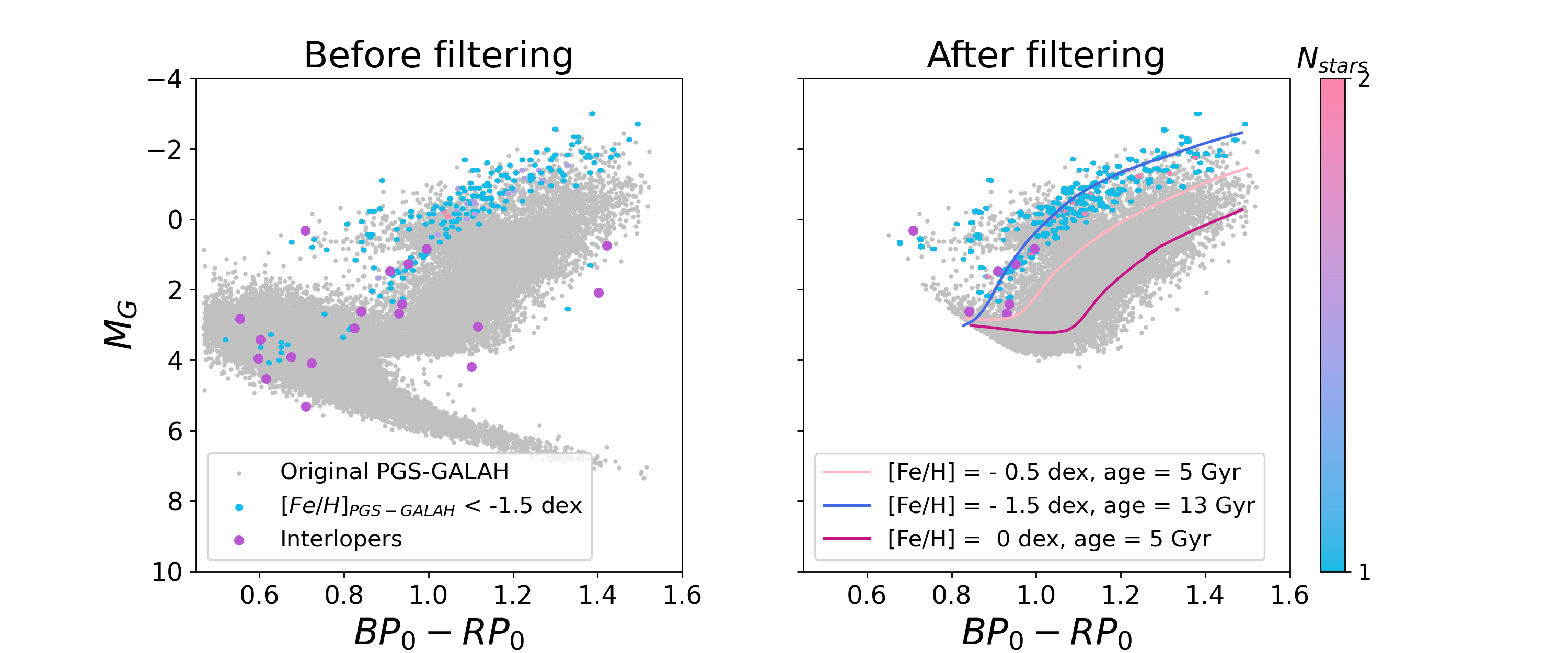}
    {\phantomsubcaption\label{gp_cmd}}
\end{subfigure}
\hfill
\caption{PGS - GALAH before (left) and after (right) applying the isochrone filtering method. Top: $\text{[Fe/H]}_{\text{GALAH}}$ versus $\text{[Fe/H]}_{\text{PGS}}$ (Fig.~\ref{fehfeh_gp}). Bottom: CMD (Fig.~\ref{gp_cmd}).}
\end{figure}

GALAH is the sample for which we have the lowest statistics. However, we are able to evidence in the left panel of Fig.~\ref{fehfeh_gp} the same horizontal sequence of outliers identified in the PGS-APOGEE cross-correlation. This sequence makes up for 13.5 $\%$ of the total number of PGS-GALAH metal-poor candidates. After filtering, 63 $\%$ of GALAH interlopers are removed. Although this score is less satisfying than with APOGEE, most of the horizontal sequence is effectively removed as seen in the top right panel of Fig.~\ref{fehfeh_gp}. Additionally, 5.8 $\%$ of PGS-GALAH metal-poor sample are also discarded.

\begin{figure}
\centering
\begin{subfigure}{0.45\textwidth}
    \includegraphics[width=\textwidth]{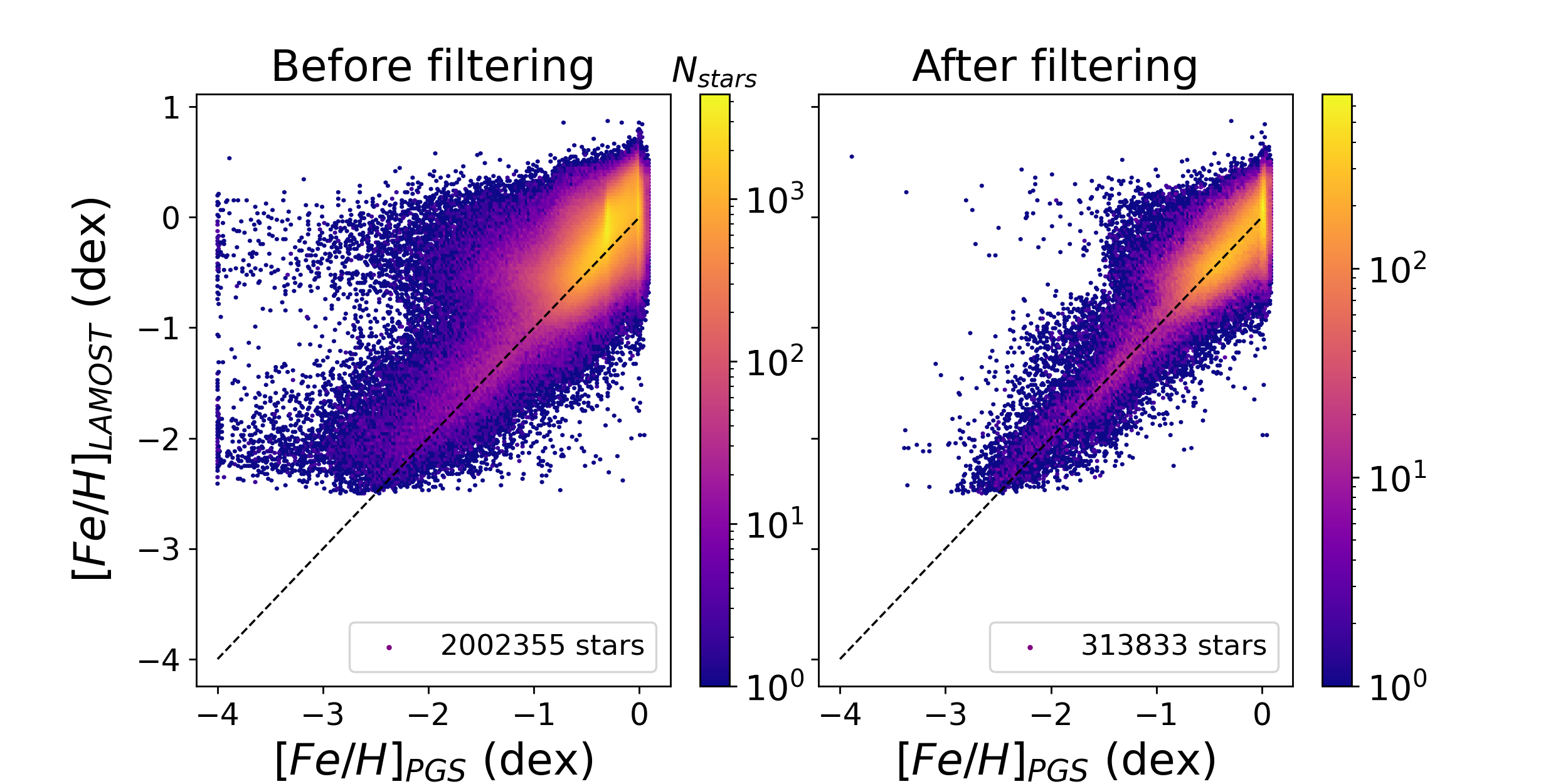}
    {\phantomsubcaption\label{fehfeh_lp}}
\end{subfigure}
\hfill
\begin{subfigure}{0.45\textwidth}
    \includegraphics[width=\textwidth]{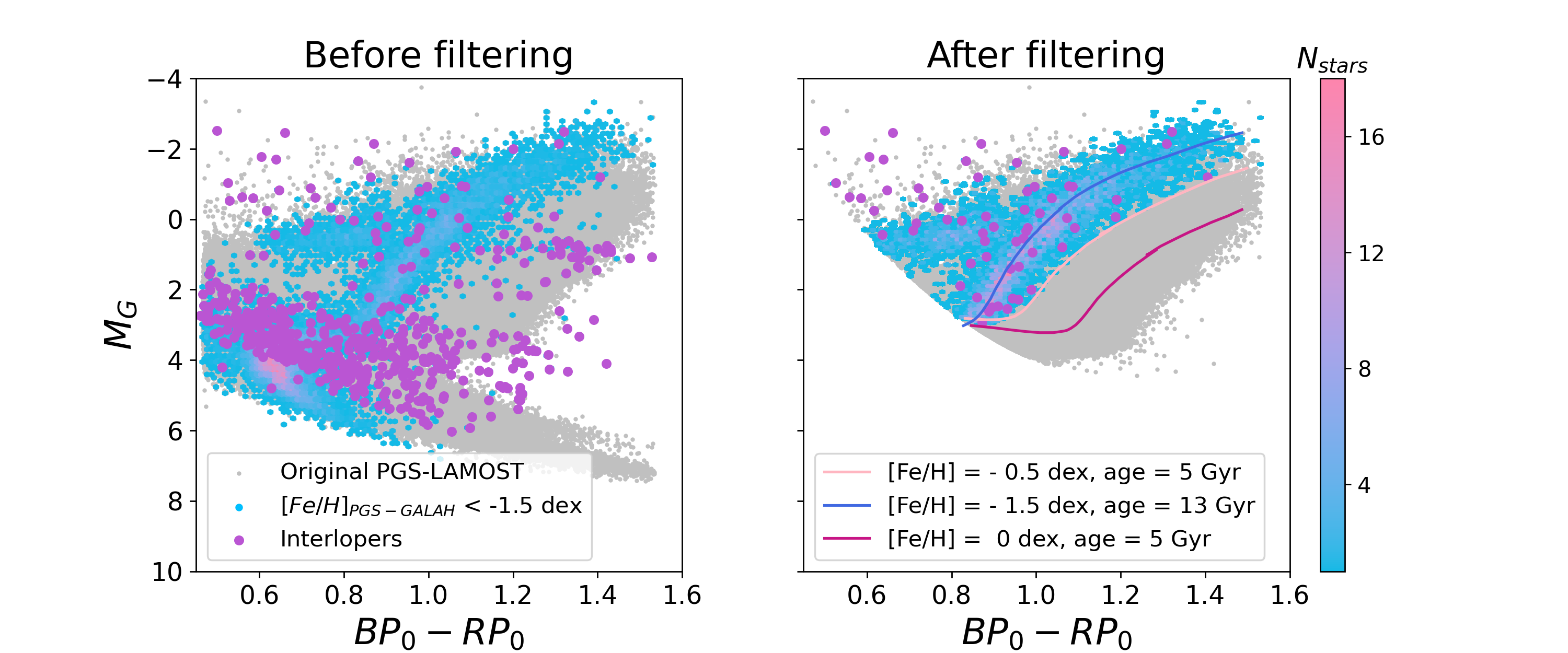}
    {\phantomsubcaption\label{lp_cmd}}
\end{subfigure}
\hfill
\caption{PGS - LAMOST before (left) and after (right) applying the isochrone filtering method. Top: $\text{[Fe/H]}_{\text{LAMOST}}$ versus $\text{[Fe/H]}_{\text{PGS}}$ (Fig.~\ref{fehfeh_lp}). Bottom: CMD (Fig.~\ref{lp_cmd}).}
\end{figure}

The LAMOST sample exhibits a bold horizontal sequence of interlopers in the left panel of Fig.~\ref{fehfeh_lp}. As seen in the left panel of Fig.~\ref{lp_cmd}, they are for the most part dwarf stars, which are removed as part of our selection function; nonetheless, the clump at $M_G$ $\sim$ 0 similar to the PGS-APOGEE cross-correlation is still present, but can be effectively removed with the isochrone filtering method. The interlopers make up for 11.2 $\%$ of the PGS-LAMOST metal-poor subsample. After filtering, 89 $\%$ of LAMOST interlopers are discarded, with an additional 56.4 $\%$ of the true PGS-LAMOST metal-poor stars (mostly dwarf stars) and effective removal of the interloper sequence in the right panel of Fig.~\ref{fehfeh_lp}.

\begin{figure}
\centering
\begin{subfigure}{0.45\textwidth}
    \includegraphics[width=\textwidth]{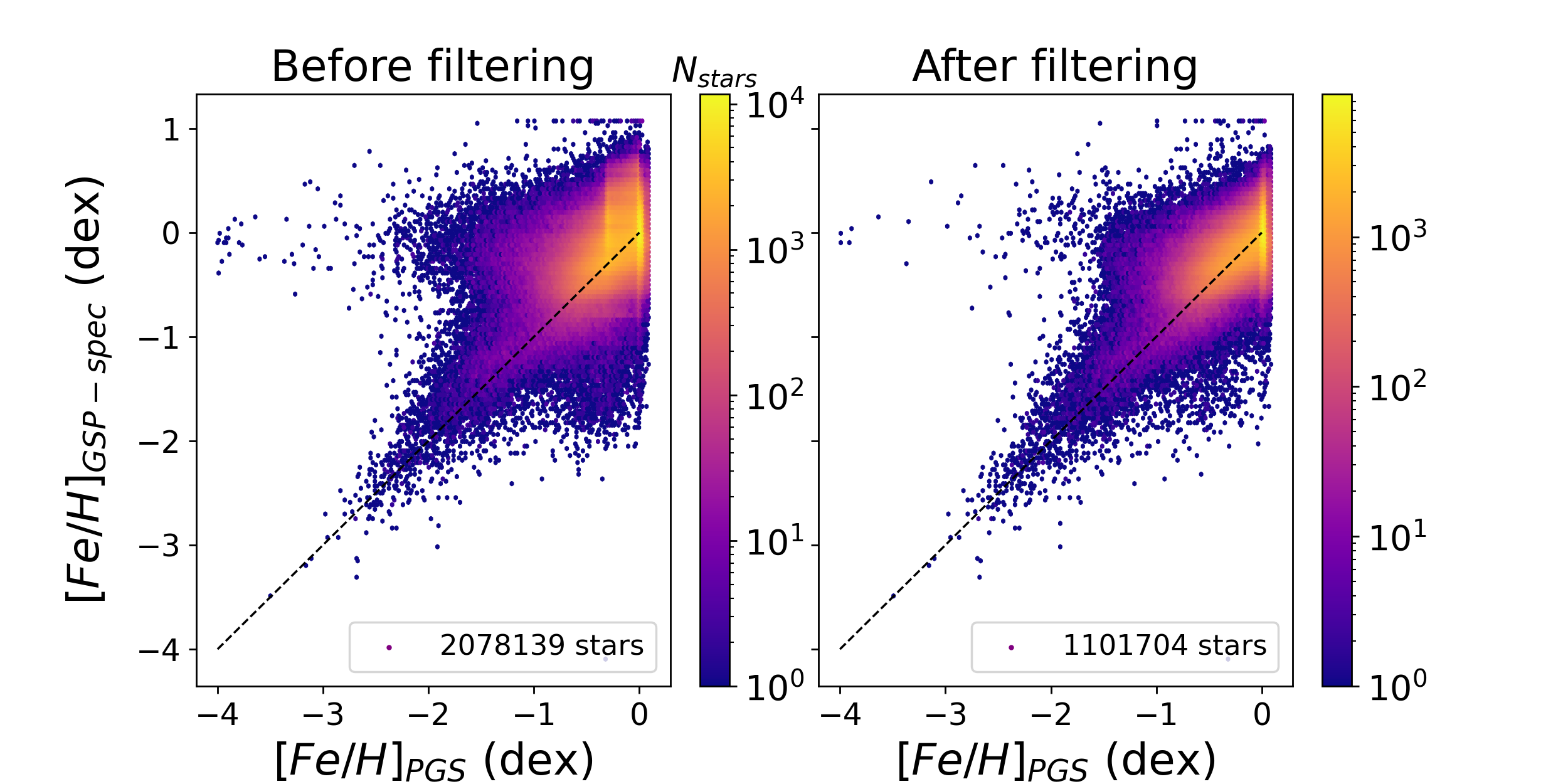}
    {\phantomsubcaption\label{fehfeh_gspp}}
\end{subfigure}
\hfill
\begin{subfigure}{0.45\textwidth}
    \includegraphics[width=\textwidth]{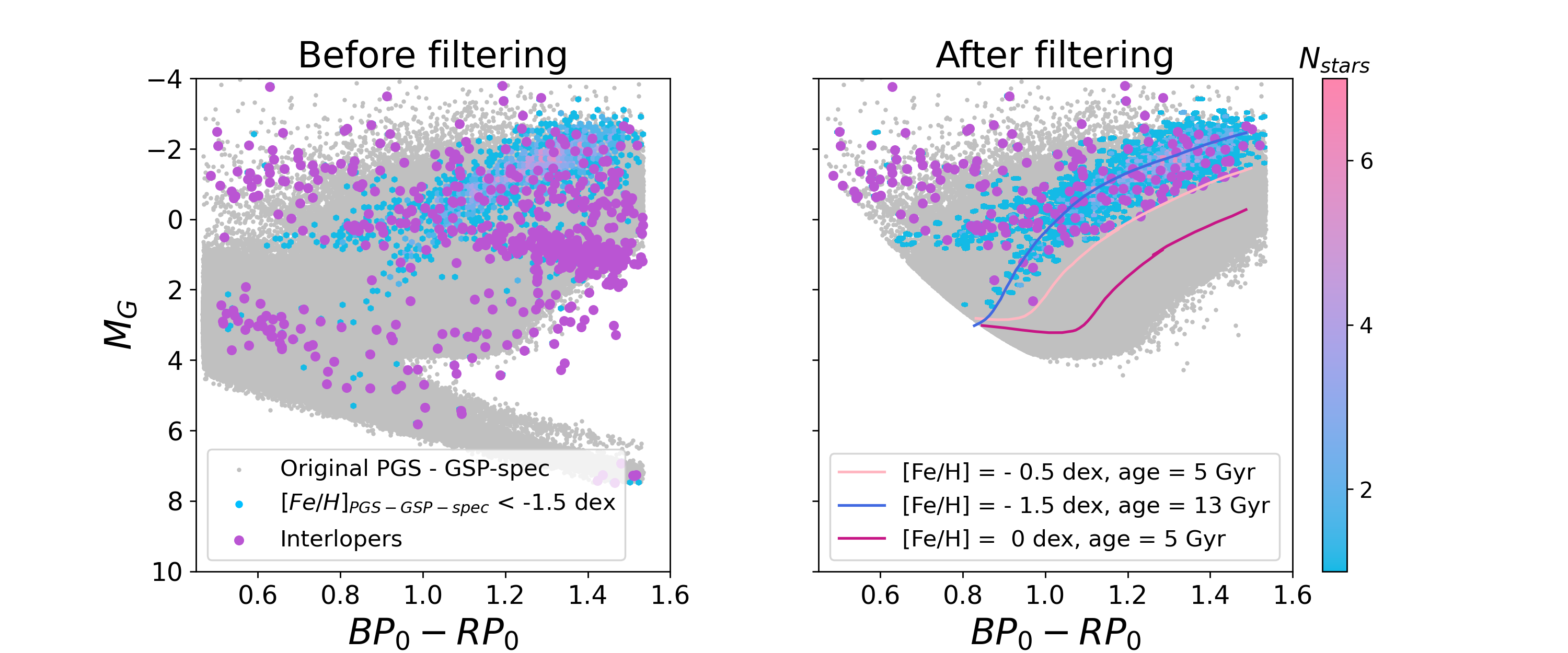}
    {\phantomsubcaption\label{gspp_cmd}}
\end{subfigure}
\hfill
\caption{PGS - GSP-spec before (left) and after (right) applying the isochrone filtering method. Top: $\text{[Fe/H]}_{\text{GSP-spec}}$ versus $\text{[Fe/H]}_{\text{PGS}}$ (Fig.~\ref{fehfeh_gspp}). Bottom: CMD (Fig.~\ref{gspp_cmd}).}
\end{figure}

The GSP-spec sample, in addition to the horizontal interloper sequence, displays a vertical interloper sequence towards solar [Fe/H] in the left panel of Fig.~\ref{fehfeh_gspp}, this time corresponding to metal-poor (most likely spectroscopic) interlopers. 
Interlopers are clumped in the reddest part of the CMD and are mainly giant stars, see left panel of Fig.~\ref{gspp_cmd}. They make up for 31 $\%$ of the PGS - GSP-spec metal-poor subsample. The isochrone filtering method discards both interloper sequences quite successfully. After filtering, 67 $\%$ of GSP-spec interlopers are discarded, with an additional 2.2 $\%$ of the true PGS - GSP-spec metal-poor stars.  

\section{GSP-spec flags quality selection}

Table~\ref{gspspec_flags} describes the flag and stellar parameter selection of our GSP-spec sample.

\begin{table}[h]
    \caption{Flag and stellar parameter selection of our GSP-spec sample.}
    \centering
    \begin{tabular}[t]{cc}
    \hline
    \textit{vbroadT} & 0,1 \\
    \textit{vradT} & 0,1 \\
    \textit{vbroadG} & 0,1 \\
    \textit{vradG} & 0,1 \\
    \textit{vbroadM} & 0,1 \\
    \textit{vradM} & 0,1 \\
    \textit{negFlux} & 0 \\
    \textit{KMgiantPar} & 0 \\
    \textit{extrapol} & 0,1,2 \\
    \textit{fluxNoise} & 0,1,2 \\
    \textit{nullFluxErr} & 0,1 \\
    \textit{nanFlux} & 0,1 \\
    \textit{emission} & 0,1 \\
    $T_{\text{eff}}$ (K) & [3800, 6000] \\
    RUWE & [0, 1.4] \\
    log \textit{g} (dex) & [0, 5] \\
    \hline
    \end{tabular}
    \tablefoot{For a more complete description of the flags, see Table 2 of \cite{gspspec_rvs}.}
    \label{gspspec_flags}
    \end{table}

\section{On the choice of spectroscopic contaminants}
\label{choice_spectro_cont}

\begin{figure*}
\centering
\begin{subfigure}{\textwidth}
    \includegraphics[width=\textwidth]{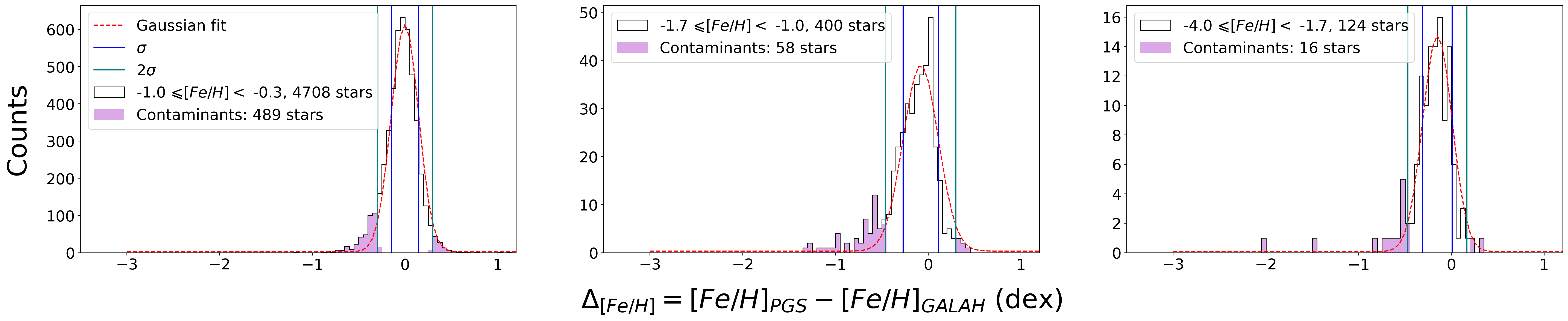}
    {\phantomsubcaption\label{gp_sigmadist}}
\end{subfigure}
\hfill
\begin{subfigure}{\textwidth}
    \includegraphics[width=\textwidth]{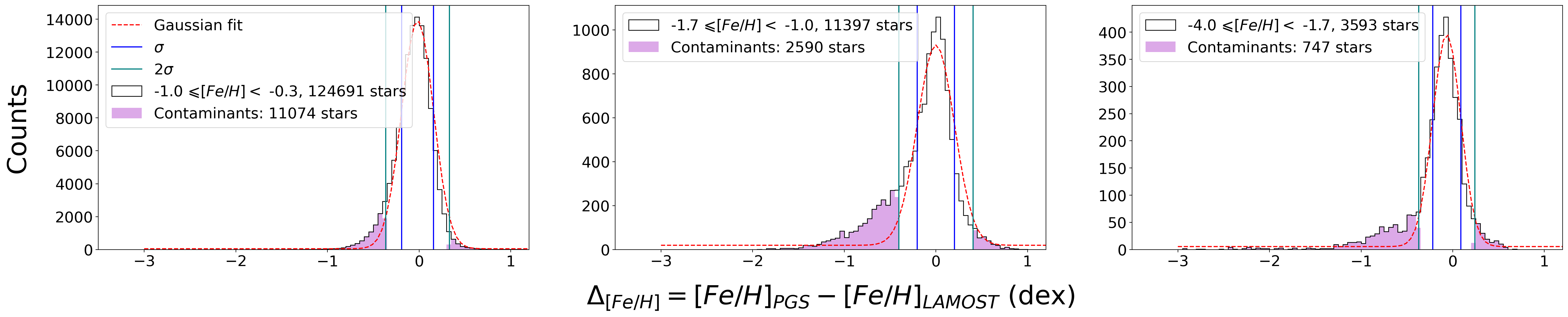}
    {\phantomsubcaption\label{lp_sigmadist}}
\end{subfigure}
\hfill
\begin{subfigure}{\textwidth}
    \includegraphics[width=\textwidth]{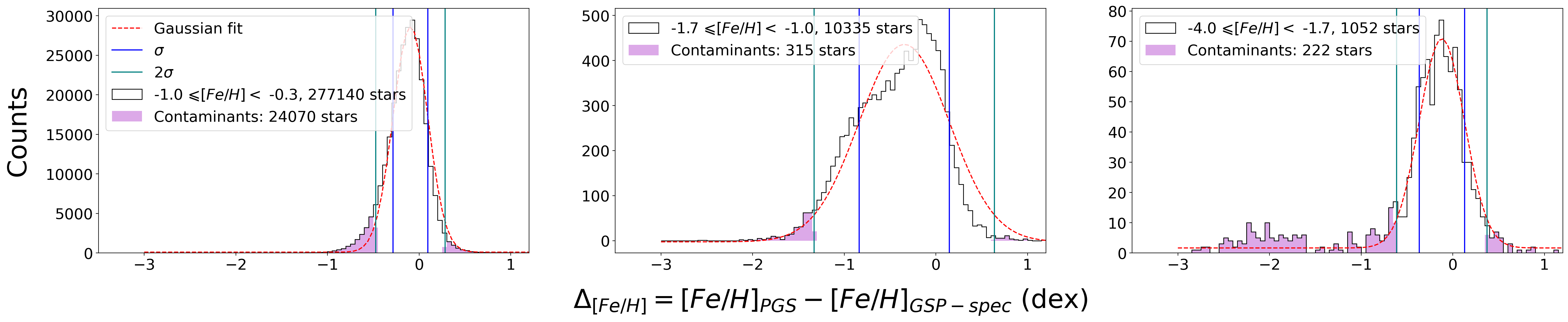}
    {\phantomsubcaption\label{gspp_sigmadist}}
\end{subfigure}
\hfill
\caption{$\Delta_{\text{[Fe/H]}} = \text{[Fe/H]}_{\text{PGS}} - \text{[Fe/H]}_{\text{survey}}$ distribution (black). The Gaussian fit to the distribution is overplotted in red dashed lines. $\sigma$ and 2$\sigma$ limits are respectively indicated in blue and teal vertical lines. The PGS-\textit{survey} contaminants, i.e. sources with $\Delta_{\text{[Fe/H]}} = \text{[Fe/H]}_{\text{PGS}} - \text{[Fe/H]}_{\text{survey}}$ > 2$\sigma$ are colour-coded in purple. 
Left panel: -1.0 < $\text{[Fe/H]}_{\text{PGS}}$ < -0.3. Middle panel: -1.7 < $\text{[Fe/H]}_{\text{PGS}}$ < -1.0. Right panel: -4.0 < $\text{[Fe/H]}_{\text{PGS}}$ < -1.7. From top to bottom: comparison with respect to GALAH, LAMOST and GSP-spec.}
\label{gauss_spectroconts}
\end{figure*}

\begin{figure}
    \centering
    \includegraphics[width=\linewidth]{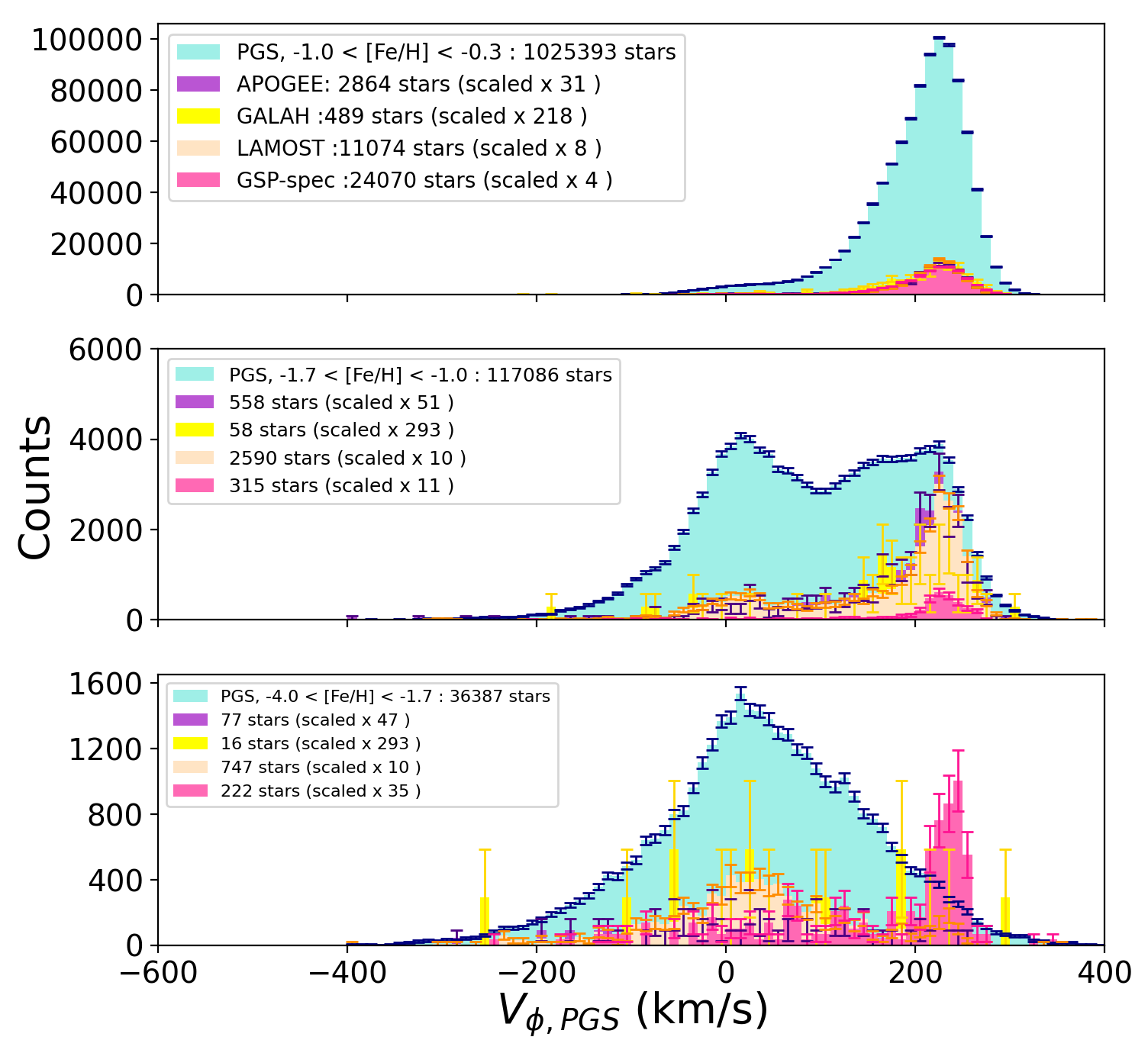}
    \caption{$V_\phi$ distribution of PGS, together with spectroscopic contaminants scaled by the size of PGS in each [Fe/H] subsample. Top panel: -1.0 < $\text{[Fe/H]}_{\text{PGS}}$ < -0.3. Middle panel: -1.7 < $\text{[Fe/H]}_{\text{PGS}}$ < -1.0. Bottom panel: -4.0 < $\text{[Fe/H]}_{\text{PGS}}$ < -1.7.}
    \label{vphi_pgs_allsurveys}
\end{figure}

To decide which surveys are the best fit for the spectroscopic decontamination of PGS, we investigated in Fig.~\ref{vphi_pgs_allsurveys} the $V_\phi$ distribution in the same [Fe/H] intervals defined in Fig.~\ref{vphidist_allbins}, and compared APOGEE, GALAH, LAMOST and GSP-spec. We refer the reader to Sec.~\ref{spectroconts} for a complete description of the procedure used to isolate spectroscopic contaminants, and Fig.~\ref{gauss_spectroconts} to visualise the effective selection relative to GALAH, LAMOST and GSP-spec. 

Between -1.0 and -1.7, and between -1.7 and -4.0, GSP-spec contaminants follow a trend distinct from the other spectroscopic surveys. 
In fact, below -1.7, the bulk of the contamination lies at $\sim$ 250 km.s$^{-1}$, whereas the contamination of the other surveys is centred around 0 km.s$^{-1}$ and spreads down to retrograde regions. In particular, GSP-spec contamination exceeds true PGS counts, which drives us to identify the cause for this overestimation. 

\begin{figure}[h]
    \centering
    \includegraphics[width=\linewidth]{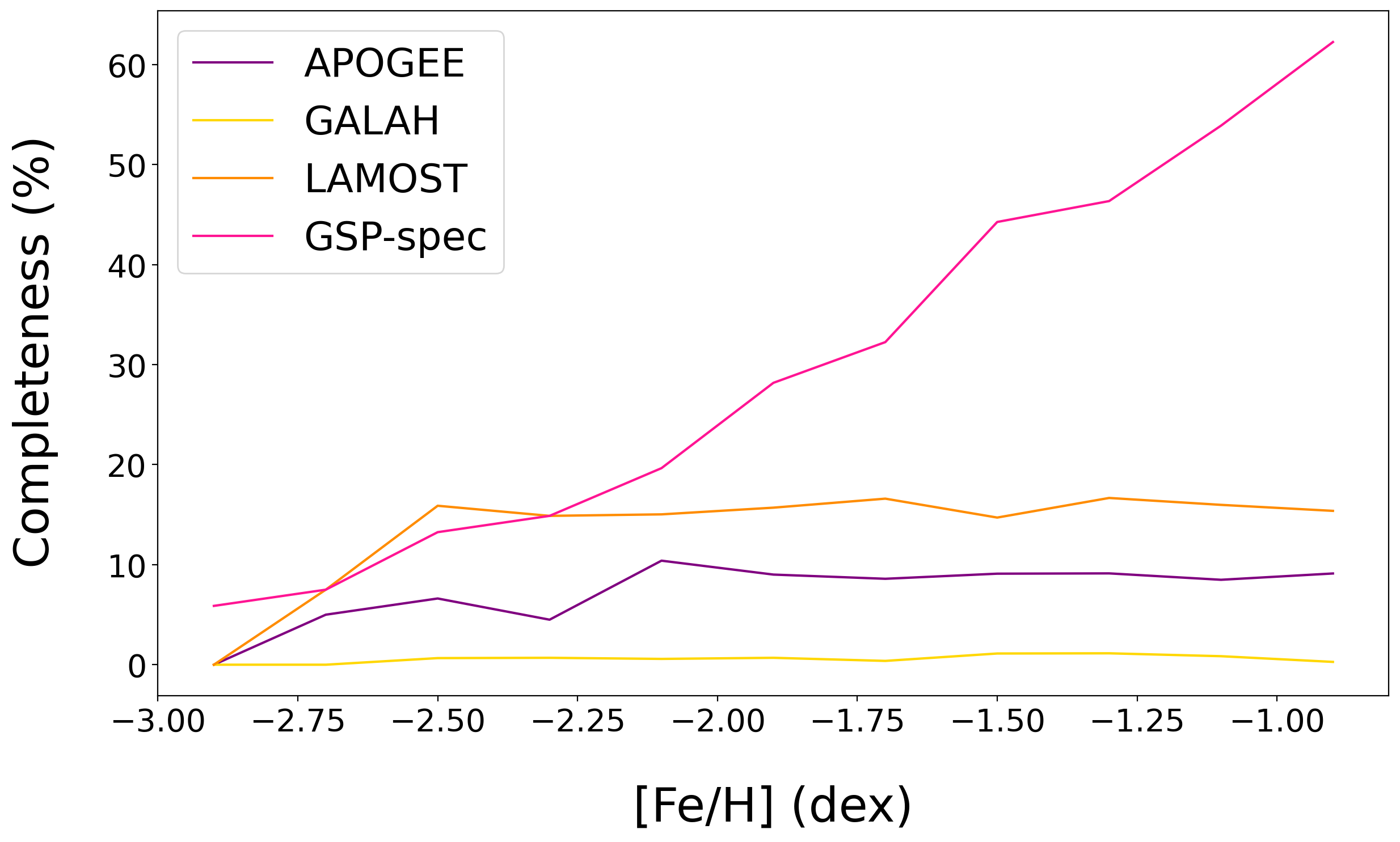}
    \caption{Completeness of halo stars for each spectroscopic survey as a function of [Fe/H].}
    \label{comp_surveys}
\end{figure}

We suspect that GSP-spec is incomplete in the metal-poor domain, due to the data selection itself; strictly following the quality flags advised in Table 2 of \cite{gspspec_rvs} can remove a significant number of metal-poor stars, due to lower accuracy in the retrieval of their atmospheric parameters and [Fe/H] (quality of the spectra). 
However, to be able to populate the contamination throughout the entire $V_\phi$ space, the cuts applied to our GSP-spec sample were loosened (see Table~\ref{gspspec_flags}).
Assuming that PGS is complete for halo (retrograde) bright stars (for which the Pristine pipeline performs best), we checked the completeness of APOGEE, GALAH, LAMOST and GSP-spec with respect to PGS, as a function of PGS metallicity in Fig.~\ref{comp_surveys}. We selected sources with $V_\phi$ < 80 km.s$^{-1}$ and \verb|phot_g_mean_mag| < 12 mag (to include the bulk of GSP-spec stars, see Fig.~2 of \citealt{gspspec_rvs}).

\begin{figure}[h]
    \centering
    \includegraphics[width=\linewidth]{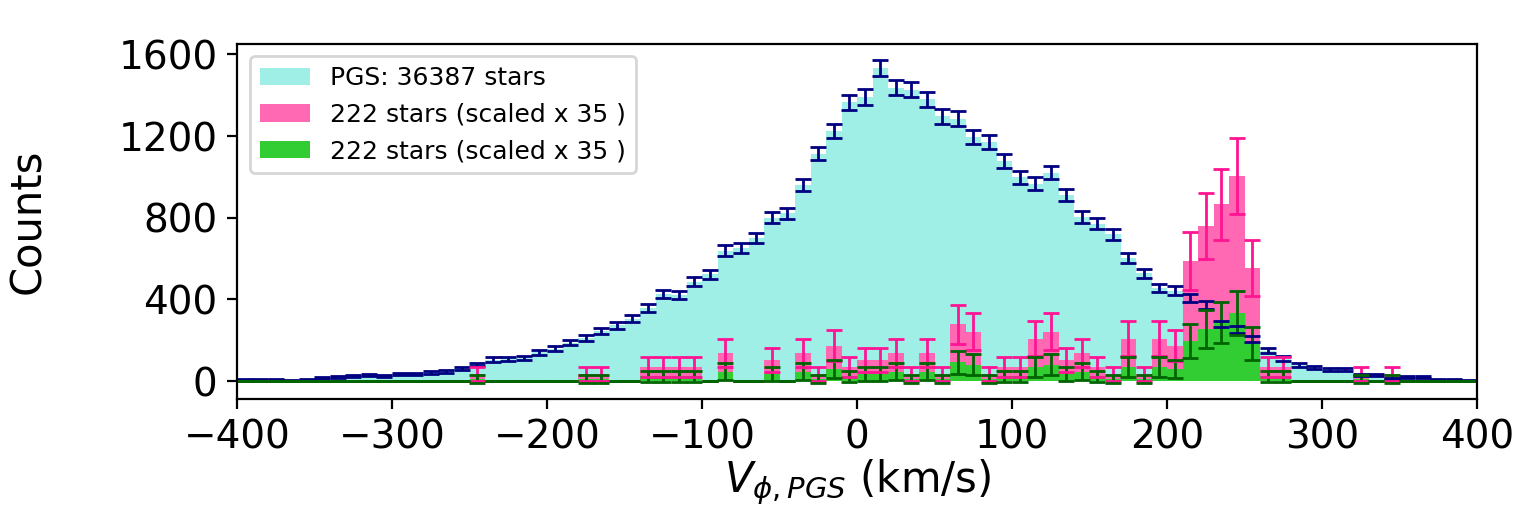}
    \caption{Same as Fig.~\ref{vphi_pgs_allsurveys}, but only with GSP-spec. Magenta: Original GSP-spec contaminants. Green: GSP-spec contaminants downgraded by the decrease factor in completeness of halo stars (see Fig.~\ref{comp_surveys}). The error bar is divided by the square root of the decrease factor.}
    \label{vphi_pgs_gspspec_downsized}
\end{figure}

We observe overall constant completeness with decreasing metallicity for APOGEE and LAMOST; still, the completeness of GSP-spec decreases with metallicity. Below -2.0 dex, it is downsized by a factor 3. Similarly to what is seen for retrograde stars, GSP-spec metal-poor sources might be missing in the high $V_\phi$ regions, leading to an overrepresentation of metal-rich contamination dominating the sample. 
Figure~\ref{vphi_pgs_gspspec_downsized} confirms that simply rescaling GSP-spec contaminants by that factor enables to recover values coherent with other spectroscopic trends as well as PGS itself, and that GSP-spec is incomplete at very low metallicities. For that reason, we did not use GSP-spec to decontaminate PGS.

Regarding GALAH contaminants, it is complex to infer a trend for completeness because it is almost constantly null, due to very small statistics available after cross-matching with PGS. Hence, we did not use GALAH to decontaminate PGS.

\section{Decontamination and chemodynamical analysis with LAMOST}
\label{lamost_discussion}

\begin{figure}
    \centering
    \includegraphics[width=0.8\linewidth]{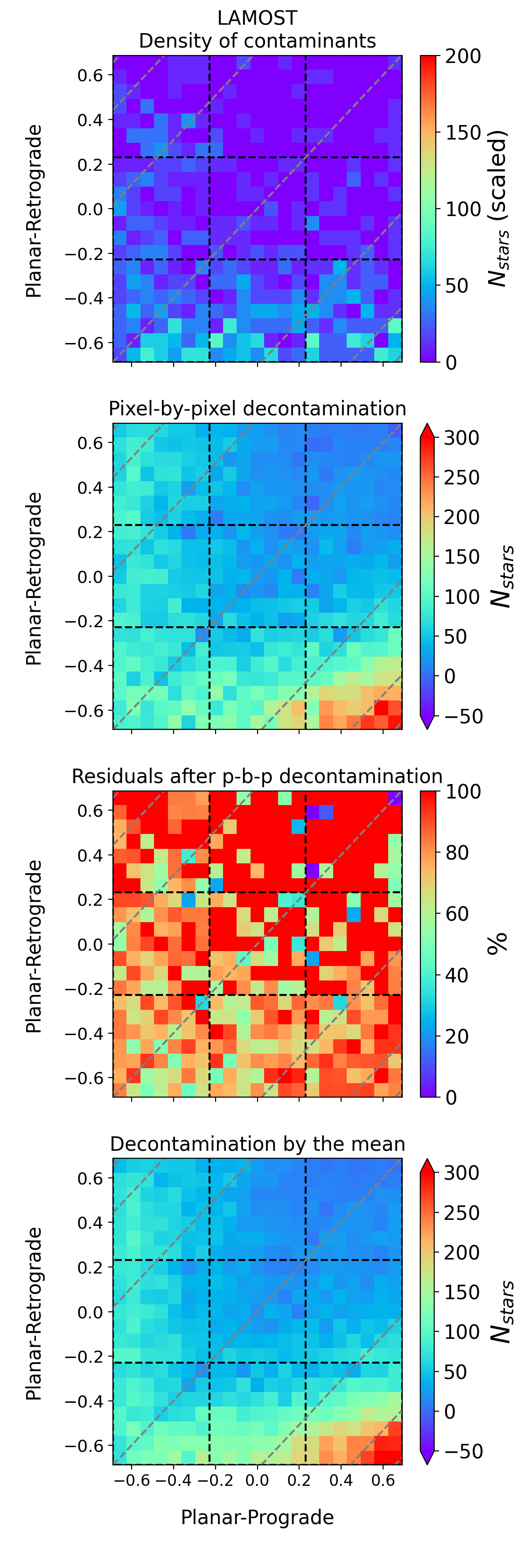}
    \caption{Densities in the action space. The grid corresponds to areas in the parameter space delimited in ($J_z - J_r$) and $L_z$. From top to bottom: first panel: density of LAMOST contaminants. Second panel: PGS decontaminated pixel-by-pixel, using LAMOST. Third panel: residuals percentages of the pixel-by-pixel PGS decontamination using LAMOST. Fourth panel: PGS decontaminated by the mean of the LAMOST contamination.}
    \label{actspacedecont_lamost}
    \end{figure}

\begin{figure}
    \centering
    \includegraphics[width=0.8\linewidth]{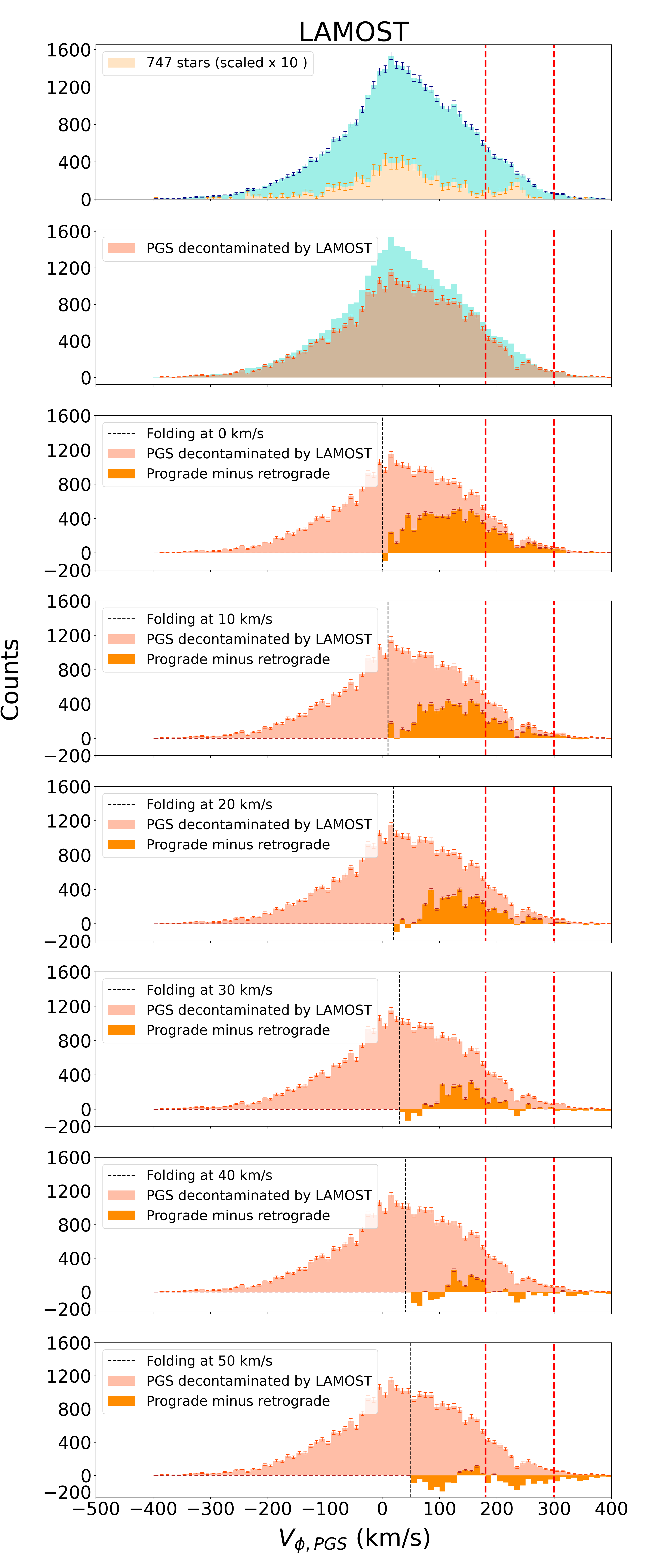}
    \caption{$V_\phi$ distribution for metallicities between -1.7 and -4.0. From top to bottom: panels 1 and 2: PGS decontamination by LAMOST. Panels 3-8: Folding of the retrograde counterparts over the prograde counterparts of the decontaminated PGS. The black dashed line indicates the location of our presumed halo. In panel 3, the folding is done assuming a theoretical halo. In the remaining panels, it is done assuming a prograde halo shifted by 10 km.s$^{-1}$ at every panel. The red dashed lines delimit the location of the population we aim to characterise. Same as Fig.~\ref{folding_vphi} of Sec.~\ref{discussion_folding}.}
    \label{folding_vphi_lamost}
\end{figure}

\begin{figure*}
    \sidecaption
        \includegraphics[width=12cm]
        {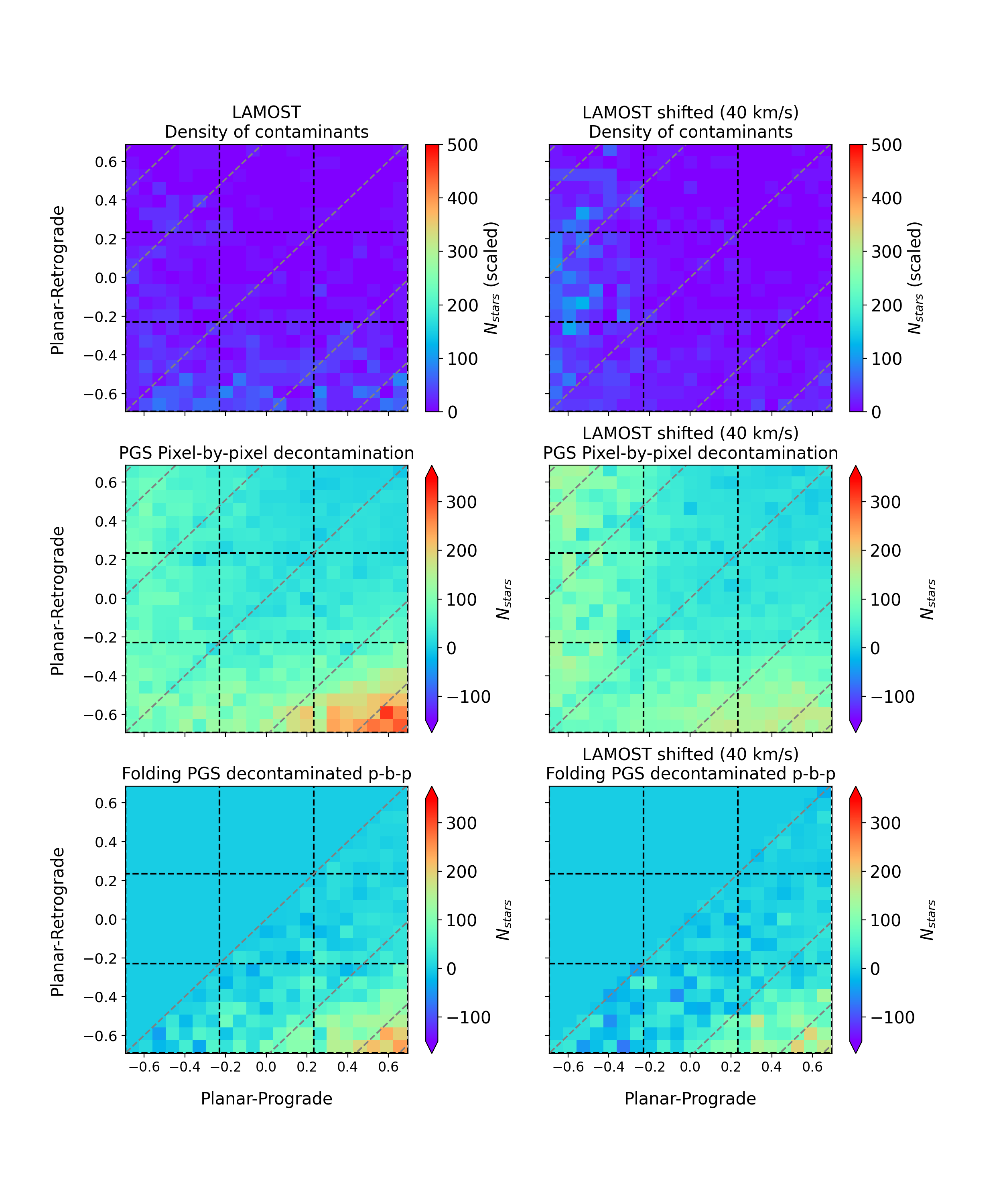}
        \caption{Rotated action space for metallicities between -4.0 $\leqslant$ [Fe/H] < -1.7. The grey dashed lines correspond to different values of $L_z$/$J_{\text{tot}}$: [-0.95, -0.8, -0.5, 0, 0.5, 0.8, 0.95]. Left column: decontamination process and folding using LAMOST contamination. Top row: density of spectroscopic contaminants. Middle row: pixel-by-pixel decontamination. Bottom row: folding of the retrograde components over the prograde components, for PGS decontaminated pixel-by-pixel. In this space, the folding axis of symmetry is the grey dashed line at $L_z$/$J_{\text{tot}}$ = 0. Right column: same as the left panels, but the three actions were computed shifting the observed $V_\phi$ by 40 km.s$^{-1}$. Same as Fig.~\ref{folding_actspace} of Sec.~\ref{discussion_folding}.}
        \label{folding_actspace_lamost}
\end{figure*}

We show in Fig.~\ref{actspacedecont_lamost} the decontamination of PGS using LAMOST, similarly to the methodology of Sec.~\ref{decontamination} with APOGEE. Figure~\ref{folding_vphi_lamost} and Fig.~\ref{folding_actspace_lamost} are relevant to the discussion of Sec.~\ref{discussion_folding}.

\end{appendix}

\end{document}